\documentstyle[11pt,makeidx]{book}

\makeindex

\newcommand{\bd}{\begin{displaymath}}
\newcommand{\ed}{\end{displaymath}}
\newcommand{\be}{\begin{equation}}
\newcommand{\ee}{\end{equation}}
\newcommand{\ba}{\begin{array}}
\newcommand{\ea}{\end{array}}
\newcommand{\bea}{\begin{eqnarray}}
\newcommand{\eea}{\end{eqnarray}}
\newcommand{\ben}{\begin{eqnarray*}}
\newcommand{\een}{\end{eqnarray*}}
\def\w42{$W^{(2)}_4$}
\def\wnl{$W^{(l)}_N$}
\def\p{\partial}

\begin{document}
\pagestyle{empty}
\begin{titlepage}
\vspace*{\fill}
\begin{center}

{\Huge \bf Toda-like Systems:}

\bigskip

{\Large \bf Solutions and Symmetries}

\vspace{0.15in}

%{\it Report on leave of post-doctoral station}

\vspace{3.0in} \it

\vspace{0.1in}\large\bf

Liu CHAO

\vspace{0.3in}

Institute of Modern Physics\\
Northwest University\\
Xian 710069, China

\vspace{0.2in}

\large\bf June 1995

\vspace*{1.5in}
\end{center}
\end{titlepage}

\newpage

\setcounter{page}{0}
\tableofcontents

\addcontentsline{toc}{chapter}{Abstract}
\chapter*{Abstract}
\pagestyle{empty}

This report consists of six independent chapters, each is summarized as
follows.

\paragraph{Chapter 1}
This chapter is an introduction of the whole report.
It  is mainly devoted to the introduction of infinite variaties of Toda
theories. It also briefly mentioned why and how this report is composed.

\paragraph{Chapter 2}
A free field representation for the two-extended principal conformal
Toda (2EPCT) theory based on simply-laced even-rank Lie algebras is given.
It is shown that the classical chiral exchange algebra
for such theories can be reconstructed from free chiral bosons via
Drinfeld-Sokolov linear systems and the local and periodic solutions of the
2EPCT fields can also be recovered. Most of
the results are in analogy to  the
standard Toda case and thus  are expected to carry  over to the quantum
case, whereas the exchange algebra is a bit more complicated due to some
additional $\delta$-function terms amd the more degrees of freedom.

\paragraph{Chapter 3}
A non-left-right symmetric conformal integrable Toda field theory
is constructed. It is found that the conformal algebra for this model
is the product of a left  chiral $W_{r+1}$ algebra and a right chiral
$W_{r+1}^{(2)}$ algebra. The general classical solution is
constructed out of the chiral vectors satisfying the so-called
classical exchange algebra. In addition, we derived an explicit
wronskian type solution in relation to the constrained WZNW theory.
We also showed that the $A_\infty$ limit of
this model is precisely the $(B_2, \ C_1)$
flow of the standard Toda lattice hierarchy.

\paragraph{Chapter 4}
A new class of integrable two-dimensional partial differential equations
is constructed from Bernoulli equation and called heterotic Liouville
systems due to the heterotic conformal symmetry. These systems are shown
to possess infinite many symmetries and are related to the surfaces
of non-constant Gauss curvatures in Euclidean three-space. The simplest
nontrivial extension of Liouville equation is just the heterotic Toda model
gauging the Witt algebra found recently.

\paragraph{Chapter 5}
Two-extended Toda fields associated with Saveliev-Vershik's continuum
Lie algebras are studied. Such fields satisfy three-dimensional
(integro-)differential equations. The structures such as
fundamental Poisson relation, classical Yang-Baxter equation, chiral
exchange algebra and dressing transformations are recovered, which are
in complete analogy to the two-dimensional case. The formal solution
is also considered using Leznov-Saveliev analysis, but this does not
lead to explicit solutions because we do not have enough knowledge
about the highest weight representations of the continuum Lie algebras.

\paragraph{Chapter 6}
General relationship between classical \wnl algebras and chiral
exchange algebra is established. As an example, \w42 algebra is
reconstructed from $A_3$ exchange algebra.

\newpage

\pagestyle{headings}
\chapter{Introduction}
\setcounter{page}{1}
\index{Introduction}

\section{What are Toda-like systems?}

Tradational (abelian) Toda systems \index{Toda}
are two-dimensional lattice integrable
models involving exponential-type potentials. These are straightforward
generalizations of the one-dimensional nonlinear dynamical model proposed by
M.Toda himself in the early 1970s. The basic features of such
theories can be summarized as follows:

\begin{itemize}
\item Interacting within a lattice (Toda lattice);\index{Toda ! lattice}
\item Exponential-type potentials;
\item 2d integrability, {\it i.e.} admitting zero-curvature formulism.
\end{itemize}

If the Toda lattice is infinite, the corresponding Toda theories are
described by the so-called Toda lattice hierarchy
\index{Toda ! lattice ! hierarchy}, which is the consequence
of the following infinite many zero-curvature \index{zero-curvature}
equations for sutably defined operators $B_n$, $C_m$,

\begin{eqnarray*}
& [ \partial_n - B_n,\; \partial_m - B_m ] = 0,&\\
& [ \partial_n - B_n,\; \partial_m - C_m ] = 0,&\\
& [ \partial_n - C_n,\; \partial_m - C_m ] = 0.&
\end{eqnarray*}

\noindent It is interesting to note that, in the simplest nontrivial case,
{\it i.e.} the $(B_1,\;C_1)$-case, the corresponding Toda equation
involves only exponential interactions, and the interactions take place only
between nearest neighbours on the Toda lattice. Another remarkable aspect is
that, the conventional conformal and affine (loop) Toda field theories
\index{Toda ! loop} \index{Toda ! affine} (these
are integrable lattice {\it field theories} with the corresponding lattice
identified with the Dynkin lattice of some Kac-Moody type Lie algebra
${\cal G}$) can both be obtained by appropriately reducing the
$(B_1,\;C_1)$-Toda model. Using the Lie algebra techniques one can easily
associate to each Kac-Moody Lie algebra (either being of the finite, loop or
affine type) a Toda field theory which also has the property that only
nearest neighbour interactions are involved. To achieve this one first
decomposes ${\cal G}$ according to the principal gradation
\index{principal ! gradation}, ${\cal G}=
\oplus_{n \in Z} {\cal G}^{(n)}$. Then choosing $\mu_\pm = \sum_{\alpha \;
{\rm simple}} E_{\pm \alpha}$ and defining the Lax pair

\begin{equation}
A_\pm = \pm \left[ \frac{1}{2} \partial_\pm \Phi + \exp \left( \mp
\frac{1}{2} {\rm ad} \Phi \right) \mu_\pm \right],~~~~
\Phi = \sum \phi^i H_i,
\end{equation}

\noindent it follows from the zero-curvature equation $ [ \partial_+ - A_+,\;
\partial_- - A_- ] = 0$ that

\[ \partial_+ \partial_- \Phi = [ \mu_+,\; \exp ({\rm ad} \Phi ) \mu_- ].\]

\noindent Since the Lax pair (1) take values in the subspaces
${\cal G}^{(0)} \oplus {\cal G}^{(\pm 1)}$ respectively, we shall denote
the corresponding Toda equation (2) as $(1,\;1)$-Toda fields.

The $(1,\;1)$-Toda field theory can be generalized along two directions. One
is to allow the theory to contain certain non-exponential interactions and
some non-nearest neighbour exponential interactions, the other is to consider
the non Kac-Moody gauge algebras, which can possibly depend on one or several
continuous parameter(s). Let us consider these two kinds of generalizations
in the sequence.

First let us keep the Kac-Moody type algebra ${\cal G}$ as gauge algebra
and generalize the notion of $(1,\;1)$-Toda fields to the case of
$(n,\;m)$-Toda fields. Formally this can be achieved by enlarging the
range of values of the Lax pair $A_\pm$ to the subspaces ${\cal G}^{(0)}
\oplus {\cal G}^{(1)} \oplus ... \oplus {\cal G}^{(n)}$ and ${\cal G}^{(0)}
\oplus {\cal G}^{(1)} \oplus ... \oplus {\cal G}^{(-m)}$ with $n,\;m \in
Z_+$. However in each distinguished case one need to pay some careful work
on choosing appropriate constants $\mu^{(n)}_+$ and $\mu^{(-m)}_-$ which are
analogues of the $\mu_\pm$ in the $(1,\;1)$-Toda case. Generally one can
choose as $\mu^{(n)}_+$ any element in ${\cal G}^{(n)}$ which has the
property that ${\rm Ker} ( {\rm ad} \mu^{(n)}_+ ) \cap {\cal G}_{-n} = \{ 0
\} $, where ${\cal G}_{-n}$ is the subalgebra of ${\cal G}$ consisted of
the subspaces ${\cal G}^{(i)}$ with $i \leq -n$. Similarly one can choose
$\mu^{(-m)}_-$. The only ambiguity in the above choice of $\mu^{(n)}_+$
($\mu^{(-m)}_-$, rep.) is a set of nonzero constant parameters of linear
combinations of the basic generators of ${\cal G}^{(n)}$
(${\cal G}^{(-m)}$, rep.), which play the role of coupling
constants in the corresponding Toda field theory. In the following we shall
cite two examples in which all the coupling constants are normalized to 1.

\begin{itemize}
\item The $(2,\;2)$-Toda field theory. In this case the Lax pair are
defined as

\[
A_\pm = \pm \left[ \frac{1}{2} \partial_\pm \Phi + \exp \left( \mp \frac{1}{2}
{\rm ad} \Phi \right) \left(\bar{\Psi}_\pm + \mu^{(\pm 2)}_\pm \right) \right]
\]

\noindent where
\begin{eqnarray*}
&\mu^{(\pm 2)}_\pm = \pm \frac{1}{2} \sum_{ \{ i,\;j:\; \alpha_i,\;
\alpha_j \;{\rm simple} \} } {\rm sign}(i-j)
 [ E_{\pm \alpha_i},\; E_{\pm \alpha_j} ] ,&\\
&\Psi_\pm = \sum_{ \{ i:\;\alpha_i \; {\rm simple} \} }
\psi_\pm^i E_{\mp \alpha_i},
{}~~~~\bar{\Psi}_\pm = \pm [ \mu^{(\pm 2)}_\pm,\; \Psi_\pm ].&
\end{eqnarray*}

\noindent Using the zero-curvature equation for $A_\pm$ one easily finds the
equations of motion, which read

\begin{eqnarray*}
&\partial_+ \partial_- \Phi - [ \bar{\Psi}_+,\; \exp ({\rm ad} \Phi )
\bar{\Psi}_- ] - [ \mu^{(2)}_+,\; \exp ({\rm ad} \Phi ) \mu^{(-2)}_- ] = 0,&\\
&\partial_- \Psi_+ = \exp \left( {\rm ad} \Phi \right) \bar{\Psi}_-,&\\
&\partial_+ \Psi_- = \exp \left( -{\rm ad} \Phi \right) \bar{\Psi}_+.&
\end{eqnarray*}

\noindent Expanding the above equations of motion into the component form,
we find that the fields $\psi_\pm^i$ interact non-exponontially, whereas
the fields $\phi^i$ interacts exponentially but the interacting fields
involves non-nearest neighbours. Further consideration would show that if the
Lie algebra ${\cal G}$ is of the finite or loop Kac-Moody type, the
corresponding $(2,\;2)$-Toda field theories can be obtained as reductions of
the $(B_2,\;C_2)$-member of the Toda lattice hierarchy.

\item The $(2,\;1)$-Toda fields. The Lax pair read

\begin{eqnarray}
&A_+ = \frac{1}{2} \partial_+ \Phi + \exp \left(- \frac{1}{2}
{\rm ad} \Phi \right) \left(\bar{\Psi}_+ + \mu^{(2)}_+ \right),&\nonumber \\
&A_- = - \left[ \frac{1}{2} \partial_- \Phi + \exp \left(
\frac{1}{2} {\rm ad} \Phi \right) \mu_- \right],&
\end{eqnarray}

\noindent which yield the equations of motion

\begin{eqnarray*}
&\partial_+ \partial_- \Phi = [ \bar{\Psi}_+,\; \exp ({\rm ad} \Phi )
\mu_- ],&\\
&\partial_- \Psi_+ = \exp \left( {\rm ad} \Phi \right) \mu_-.&
\end{eqnarray*}

\noindent Now the fields $\psi_+^i$ again interact non-exponentially with
the fields $\phi^i$, whilst the fields $\phi^i$ interact exponentially with
their nearest neighbours.
\end{itemize}

Next let us turn to consider the second way of generalizing the
$(1,\;1)$-Toda field theory, namely by gauging non Kac-Moody algebras. We
shall consider two different kinds of non Kac-Moody algebras: (i) the
Witt-Virasoro algebra and (ii) the so-called standard continuum
contragradient Lie algebras. In both cases we shall again have the
notation of $(n,\;m)$-Toda field theories \index{Toda ! $(n,~m)$-model},
formally with the form of the
Lax pairs  unchanged, but with the algebraic content
of the constants $\mu^{(\pm n)}$ and the fields $\Phi,\; \Psi_\pm$ replaced
appropriately.

\begin{itemize}
\item Witt-Virasoro gauge algebra. This kind of gauge algebras is naturally
$Z$-graded with each ${\cal G}^{(n)} \; (n \neq 0)$ being one dimensional.
So one can choose $\mu^{(n)}_+$ simply to be the generator $L_n$, and also
$\mu^{(-m)}_-$ to be $L_{-m}$. Therefore, by defining

\begin{eqnarray*}
&\Phi = \phi L_0,~~~~ \Psi_\pm = \psi_\pm L_{\mp 1}&\\
&\mu_\pm = L_{\pm 1},~~~~
\mu^{(\pm 2)}_\pm = L_{\pm 2},&
\end{eqnarray*}

\noindent we have the following $(1,\;1)$- $(2,\;1)$- and $(2,\;2)$-Toda
models for the Witt algebra,

\begin{enumerate}
\item $(1,\;1)$-model (Liouville model) \index{Liouville model}

$$
\partial_+ \partial_- \phi + 2 \exp (\phi ) =0;
$$

\item $(2,\;1)$-model

$$
\partial_+ \partial_- \phi - 6 \psi_+ \exp (\phi ) =0,
{}~~~~\partial_- \psi_+ - \exp (\phi ) = 0;
$$

\item $(2,\;2)$-model

\begin{eqnarray*}
&\partial_+ \partial_- \phi - 18 \psi_+ \psi_- \exp (\phi ) - 4
\exp (2 \phi )=0,&\\
&\partial_- \psi_+ = 3 \psi_- \exp (\phi ) = 0,
{}~~~~\partial_+ \psi_- = 3 \psi_+ \exp (\phi ) = 0.&
\end{eqnarray*}

\end{enumerate}

\noindent For the case of Virasoro gauge algebra, one needs only to replace
the definition of $\Phi$ by $\Phi = \phi L_0 + \sigma \ c $, where $c$ is the
center of the algebra, which lead to similar equations but with an additional
free field $\sigma$ (in the $(2,\;2)$-case), which is decoupled from $\phi$
and $\psi_\pm$.

\item Standard continuum contrgradient Lie algebras as gauge algebra. This
kind of Lie algebra usually depends on one or several continuous parameters.
The generating relations can be written as follows,
\index{continuum Lie algebra}

\begin{eqnarray*}
& [ h(t),\; h(t') ] = 0,
{}~~~~ [ h(t),\; e_\pm (t') ] = \pm K(t,\;t') e_\pm (t'),&\\
& [ e_+ (t),\; e_- (t') ] = \delta (t-t') h(t),&
\end{eqnarray*}

\noindent where $t$ is some $m$-dimensional continuous parameter, regarded as
the local coordinate on some smooth manifold ${\cal M}$ with
${\rm dim}{\cal M} = m$, $\delta (t-t') = \prod_{i=1}^{m} \delta (t_i -t_i')$.
The operator $K(t,\;t')$ is the Cartan operator, which is called {\it
symmetrizable} if there exists some function $v(t)$ such that $K(t,\;t')v(t')
=K(t',\;t)v(t)$.
\index{Cartan operator} \index{Cartan operator ! symmetrizable}

With the above gauge algebra in hand, one can now redefine the variables

\begin{eqnarray*}
&\Phi = \int {\rm d}t \ \phi (t) h(t),
{}~~~~\Psi_\pm = \int {\rm d}t \ \psi_\pm (t) e_\mp(t),&\\
&\mu_\pm = \int {\rm d}t \ e_\pm(t),
{}~~~~\mu^{(\pm 2)}_\pm = \pm \frac{1}{2} \int {\rm d}t \ {\rm d}t' \
\Omega (t,\;t') \ [ e_\pm (t),\; e_\pm (t') ],&
\end{eqnarray*}

\noindent where $\Omega (t,\;t')$ is some antisymmetric function of the
arguments. It then follows from the above standard procedure that for each
pair of given $K(t,\;t')$ and $\Omega (t,\;t')$ we have a complete series
of $(n,\;m)$-Toda theories \index{Toda ! $(n,~m)$-model},
the equations of which are usually
$(2+m)$-dimensional integro-differential equations
\index{integro-differential equations}.

Among various choices of the Cartan operator \index{Cartan operator}
$K(t,\;t')$, two special
cases seems to be of particular importance. The first case is that when
$K(t,\;t')$ is purely a differential polynomial of the $\delta$-functions
of $t_i - t_i'$, the other case is that when $K(t,\;t')$ is symmetrizable.
It can be easily seen that the first case corresponds to purely
differential $(2+m)$-dimensional Toda field equations, and the second case
implies that the corresponding Toda field models have a lagrangian
formulation because that, if $K(t,\;t')$ is symmetrizable, then the
corresponding continuum Lie algebra yields a symmetric bilinear form,

\begin{eqnarray*}
&\langle h(t),\; h(t') \rangle = K(t,\;t') v(t'),&\\
&\langle e_+ (t),\; e_- (t') \rangle = \delta (t-t') v(t').&
\end{eqnarray*}

\noindent This is in complete analogy to the case of Kac-Moody gauge
algebras. It should be remarked that, since the Witt-Virasoro algebra have no
nontrivial bilinear symmetric form, the corresponding Toda field models
do not have a lagrangian formulism.
\end{itemize}

To end this section, let us cite some of the symmetrized purely
differential Cartan operators \index{Cartan operator ! purely differential}
in the cases of ${\rm dim}{\cal M}= 1,\;2$,
and, in the case of ${\rm dim}{\cal M}= 1$, the first few of the
simplest 3-dimensional Toda field equations.

\begin{itemize}
\item Symmetrized purely differential Cartan operators.
\begin{enumerate}
\item The case of ${\rm dim}{\cal M} =1$.
\begin{eqnarray*}
&K(t,\;t') = \delta (t-t'),&\\
&K(t,\;t') = \partial_t^2 \delta (t-t'),&
\end{eqnarray*}
\item The case of ${\rm dim}{\cal M} =2$.
\begin{eqnarray*}
&K(t_1,\;t_2;\;t_1',\;t_2') = \delta (t_1-t_1')\delta (t_2-t_2'),&\\
&K(t,\;t') = \partial_{t_1}\delta (t_1-t_1')\delta (t_2-t_2')
-\delta (t_1-t_1') \partial_{t_2} \delta (t_2-t_2'),&\\
&K(t_1,\;t_2;\;t_1',\;t_2') = \partial_{t_1}\delta (t_1-t_1')
\partial_{t_2}\delta (t_2-t_2'),&\\
&K(t,\;t') = \partial_{t_1}^2 \delta (t_1-t_1')\delta (t_2-t_2')
+ \delta (t_1-t_1') \partial_{t_2}^2 \delta (t_2-t_2'),&\\
&K(t,\;t') = \partial_{t_1}^3\delta (t_1-t_1')\delta (t_2-t_2')
-\delta (t_1-t_1') \partial_{t_2}^3 \delta (t_2-t_2'),&\\
&K(t_1,\;t_2;\;t_1',\;t_2') = \partial_{t_1}^3\delta (t_1-t_1')
\partial_{t_2}\delta (t_2-t_2'),&\\
&K(t_1,\;t_2;\;t_1',\;t_2') = \partial_{t_1}^2\delta (t_1-t_1')
\partial_{t_2}^2\delta (t_2-t_2').&
\end{eqnarray*}
\end{enumerate}
\item 3-dimensional Toda field equations in the case of $K(t,\;t')
= \partial_t^2 \delta (t-t')$ and $\Omega (t,\;t') = t-t'$.
\begin{enumerate}
\item $(1,\;1)$-model
\begin{eqnarray*}
\partial_+ \partial_- \phi + \exp ( - \partial_t^2 \phi ) = 0;
\end{eqnarray*}
\item $(2,\;1)$-model
\begin{eqnarray*}
&\partial_+ \partial_- \phi - 2 \partial_t \psi_+ \exp (- \partial_t^2 \phi )
= 0,&\\
&\partial_- \psi_+ = \exp ( - \partial_t^2 \phi );&
\end{eqnarray*}
\item $(2,\;2)$-model
\begin{eqnarray*}
&\partial_+ \partial_- \phi - 4 \partial_t \psi_-
\partial_t \psi_+ \exp (- \partial_t^2 \phi ) + 2\exp (-2 \partial_t^2 \phi )
= 0,&\\
&\partial_- \psi_+ = 2\partial_t \psi_- \exp ( - \partial_t^2 \phi ),&\\
&\partial_+ \psi_- = 2\partial_t \psi_+ \exp ( - \partial_t^2 \phi ).&
\end{eqnarray*}
\end{enumerate}
\end{itemize}

We remark that these final examples of Toda theories can actually be
identified with the first few members of the so-called quasi-classical Toda
hierarchy or continuous Toda hierarchy
\index{Toda ! continuous hierarchy}, and they all have the property
of being conformally invarialt in the $(x_+,\;x_-)$-plane. Many structures
concerning the integrability of the usual Toda theories can be
generalized to these cases without any difficulty, however these theories
are not Lorentz invariant and thus are of interests mainly due their
integrabilities and possible applications in various branches of physics.

\section{Why study Toda-like systems}

The answer to this question is not very easy. Different authors have
different motivations on studying Toda-like systems. These systems are
so rich in both mathematical and physical structures that scholars
from quite a vast rainge of fields may all raise their interests to study
them. For examples, Toda-like systems are of great importance in the studies
of Hamiltonian reductions and $W$ algebras. They are also important
prototypes of integrable (two-dimensional) field theories. Toda systems
played an important role in two-dimensional gravity in the conformal gauge,
and are also closely related with matrix models, KdV and KP hierarchies
and string theories, etc. and etc.

To speak of my personal flavor, I would like to stress that Toda-like systems
are, among various integrable systems, most easy to generalize to higher
dimensions and to the case of more dynamical degrees of freedom. Toda-like
systems possess the very beautiful structure of chiral exchange algebras,
and such algebras are the key to construct both the explicit solutions
and the $W$ symmetry algebras of the systems themselves.
There is an infinite variaty of Toda-like systems. Among them, all have
their common properties and each has its own special features. I would like
to extract their common properties out of these special features.

\section{What are included}

Of cause one cannot expect everything concerning Toda-like systems
to be found here. The reasons are as follows. First, Toda(-like) systems
have been studied for quite a long time, there were already countless
literatures which reflect the most promising successes in the study of
Toda(-like) systems. Second, although there were so much progress in this
field, there still remains much more requirement for efforts to study
it more thoroughly and systematically, and this work is quite out of
my personal abilities. Last, as this report is only intended for a brief
summary of my research project during the last two years, I think it
should not include such materials as those were carried out by me but
before the last two years. Therefore the question ``what is included here''
makes sence.

For the time being, I am not intended to compose a completely new book
for this report. In stead, I chosed several major papers during the last
two years to fulfill this task. In principle, each chapter is an independent
paper, with only minor modifications in the text to make the appearance
of the document more fitful as a single chapter in a long report as this
one. The work of all the six chapters were carried out in collaboration
with other peoples, who's name should not be neglected here. They are

\begin{itemize}
\item Prof. Bo-yu Hou and Dr. Yan-Shen Wang (Chapter 2)
\item Prof. Bo-yu Hou (Chapter 3)
\item Dr. Chang-Zheng Qu (Chapter 4)
\item Drs. Xiang-Mao Ding and Yan-Shen Wang (Chapter 5)
\item Dr. Yan-Shen Wang (Chapter 6).
\end{itemize}

\noindent As a consequence of the way this report is composed, I gave
an abstract to each chapter in the ``Abstract'' section instead of
writing one for the whole report. To the point of view of myself,
the content of each chapter will be best
reflected in this way and not any other one.

\chapter{Free Field Representation of Two-Extended Principal Conformal Toda
Theory}

\index{principal ! conformal Toda}
\index{Toda ! two-extended}
\index{Toda ! two-extended ! free field representation of}

\section{Introduction}
One of  the basic techniques for  exploring the structures and properties
of conformal field  theories is via free field representation. For the
well-known standard Toda field theory,  this problem has been  studied
fully  by several authors both in the classical and quantum cases [1-5]. The
basic idea is that for each   Toda field  theory, there associated two
chiral Drinfeld-Sokolov \index{Drinfeld-Sokolov}
(DS) linear systems,  each lies in one  of the two
chiral sectors. The  DS potentials are consisted of purely chiral bosons,
thus starting  from the dynamics of free bosons one can recover most of
the basic features of Toda field theory, especially the local and periodic
solutions of Toda fields and the normalized chiral exchange algebras
which underlie the conformal and integrable properties of Toda field theory.
Moreover, the exchange algebras realized from these chiral bosons
have the specific property that they are completely chiral-splitted
except that the zero modes of the Toda fields  are shared by both
chiralities [1].

In this chapter we shall consider the parallel problem for the 2EPCT model.
We shall restrict ourselves to the classical case and leave the quantum
problem to the next chapter in the same series.

The 2EPCT model first come from the hamiltonian reduction of WZNW  theory
associated with second  order constraints [11], and is called bosonic
superconformal Toda model in some of our earlier chapters [11-13]. Now it
is fairly well understood that many features of the standard Toda theory
are shared by this model, such as the conformal invariance, chiral
exchange algebra and W-algebra symmetry, {\em etc} [12]. The difference lies
in that the former contains some extra fields of conformal dimensions
$\left(\frac{1}{2},\;0\right)$ and $\left(0,\;\frac{1}{2}\right)$
and thus the  theory admits a fractional generalization of the usual W$_N$
algebras, namely the W$_N^{(2)}$  algebras.

Let us specify the problem in a more explicit  fashion. Let ${\cal G}$
be a simply-laced finite dimensional Lie algebra with the standard
Chevalley generators $\left\{H_i,\;E_i,\;F_i\right\}_{i=1}^{r}$. ${\cal G}$
is equipped with a natural {\bf Z}-gradation ${\cal  G}=
{\cal G}_- \oplus {\cal G}_0 \oplus {\cal  G}_+, \;{\cal G}_\pm =
\oplus_{n \in {\bf Z}} {\cal G}^{(\pm n)}$,
called principal gradation \index{principal gradation},
with the grading operator ${\cal H}=\sum_{i,j=1}^{r}A^{ij}H_j$ where
$A^{ij}$ is the inverse of the Cartan  matrix $K_{ij}$ and the gradation
is  realized as ${\rm ad}({\cal H})E_i=E_i,\;{\rm ad}({\cal H})F_i=-F_i$.
The equations of motion for the 2EPCT model then takes the form

\begin{eqnarray}
& &\partial_+\partial_-\Phi + \left[ {\rm exp(ad} \Phi)\bar{\Psi}_-,\;
\bar{\Psi}_+\right]+\left[ {\rm exp(ad} \Phi){\cal E}_-,\;
{\cal E}_+\right]=0,\nonumber\\
& &\partial_+\Psi_-={\rm exp(-ad}\Phi)\bar{\Psi}_+,
{}~~~~\partial_-\Psi_+={\rm exp(ad}\Phi)\bar{\Psi}_-,\label{1}
\end{eqnarray}

\noindent where $x_\pm \equiv x \pm t$ are lightcone coordinates (here
$x,\;t$ are the usual cylindrical world-sheet coordinates), $\partial_\pm
\equiv \partial_{x_\pm}$,  and

\begin{eqnarray}
& &\Phi  = \sum_i \phi^i H_i \in {\cal G}_0,
{}~~~~\Psi_+ =\sum_i \psi_+^i F_i \in {\cal  G}^{(-1)},
{}~~~~\Psi_- =\sum_i \psi_-^i E_i \in {\cal  G}^{(1)},\nonumber\\
& &{\cal E}_+=\frac{1}{2}\sum_{ij}
{\rm sign}(i-j) [ E_i,\;E_j ] \in {\cal G}^{(2)},\nonumber\\
& &{\cal E}_-=\frac{1}{2}\sum_{ij}{\rm sign}(j-i)
[ F_i,\;F_j ] \in {\cal G}^{(-2)}, \label{2}\\
& &\bar{\Psi}_+ = [ {\cal E}_+,\;\Psi_+ ] \in {\cal G}^{(1)},
{}~~~~\bar{\Psi}_- = [ \Psi_-,\;{\cal E}_- ] \in {\cal G}^{(-1)}.\nonumber
\end{eqnarray}

The equations of motion can be  regarded as the  compatibility condition
of the following linear system,

\begin{eqnarray}
& &\partial_+ T = \left[ \frac{1}{2} \partial_+ \Phi + {\rm exp(}-\frac{1}{2}
{\rm ad}\Phi)(\bar{\Psi}_+ + {\cal E}_+) \right] T, \nonumber\\
& &\partial_- T = - \left[ \frac{1}{2}
\partial_- \Phi + {\rm exp(}\frac{1}{2}
{\rm ad}\Phi)(\bar{\Psi}_- + {\cal E}_-)\right] T.\label{3}
\end{eqnarray}

\noindent Let $\lambda_{max}^{(i)}$ and $\lambda_{max}^{(i)}-\alpha^i$
respectively denote the highest weight and next to highest weight in
the $i$-th foundamental representation of {\cal G}, with the corresponding
weight states denoted by $|\lambda_{max}^{(i)} \rangle$ and
$F_i|\lambda_{max}^{(i)} \rangle$. It was shown in [12] that the vectors
\footnote{The vectors $\xi_{1,2}^{(i)}$ and $\bar{\xi}_{1,2}^{(i)}$ were
denoted as $\xi^{(i)},\;\bar{\xi}^{(i)},\;\zeta^{(i)}$ and
$\bar{\zeta}^{(i)}$ in Ref.[12].}

\begin{equation}
\xi_1^{(i)} (x) \equiv \langle \lambda_{max}^{(i)} | {\rm exp(}
\frac{1}{2} \Phi) T(x),
{}~~~~\xi_2^{(i)} (x) \equiv \langle \lambda_{max}^{(i)}|E_i
{\rm exp(}\Psi_+) {\rm exp(} \frac{1}{2} \Phi) T(x)\label{4}
\end{equation}

\noindent and

\begin{equation}
\bar{\xi}_{1}^{(i)} (x) \equiv   T^{-1} (x) {\rm exp(}
\frac{1}{2} \Phi)| \lambda_{max}^{(i)} \rangle,
{}~~~~\bar{\xi}_2^{(i)} (x) \equiv  T^{-1}(x){\rm exp(} \frac{1}{2} \Phi)
{\rm exp(}\Psi_-) F_i | \lambda_{max}^{(i)} \rangle \label{5}
\end{equation}

\noindent are respectively chiral and antichiral,

\begin{displaymath}
\partial_+ \bar{\xi}_a^{(i)}(x) = \partial_-\xi_b^{(j)}(x) = 0,
{}~~~~ a,b = 1,2.
\end{displaymath}

\noindent Moreover they obay the following exchange relations
(throughout this chapter, all the chiral
quantities are  assumed to be evaluated at equal time $t=0$),

\begin{eqnarray}
& &\{ \xi_a^{(i)} (x) \otimes_, \ \xi_b^{(j)} (y) \}
= \xi_a^{(i)} (x) \otimes \xi_b^{(j)} (y)
\left( r_+ \theta (x-y) + r_- \theta (y-x) \right)\nonumber\\
& &~~~~+\delta_{a,2} \delta_{b,2} \{\psi_+^i (x) ,\;\psi_+^j (y) \}
\xi_1^{(i)} (x) \otimes \xi_1^{(j)} (y) ,\nonumber\\
& &\{ \xi_a^{(i)} (x) \otimes_, \ \bar{\xi}_b^{(j)} (y) \}
= - \left(\xi_a^{(i)} (x) \otimes 1 \right) r_-
\left( 1 \otimes \bar{\xi}_b^{(j)} (y) \right) ,\nonumber\\
& &\{ \bar{\xi}_a^{(i)} (x) \otimes_, \ \xi_b^{(j)} (y) \}
= - \left( 1 \otimes \xi_b^{(j)} (y) \right) r_+ \left(
\bar{\xi}_a^{(i)} (x) \otimes 1 \right) , \nonumber\\
& &\{ \bar{\xi}_a^{(i)} (x) \otimes_, \ \bar{\xi}_b^{(j)} (y) \} =
\left( r_- \theta (x-y) + r_+ \theta (y-x) \right)
\bar{\xi}_a^{(i)} (x) \otimes \bar{\xi}_b^{(j)} (y) \nonumber\\
& &~~~~+\delta_{a,2} \delta_{b,2} \{\psi_-^i (x) ,\;\psi_-^j (y) \}
\bar{\xi}_1^{(i)} (x) \otimes \bar{\xi}_1^{(j)} (y), \label{6}
\end{eqnarray}

\noindent where $r_\pm$ are solutions of the classical Yang-Baxter
equation, \index{Classical Yang-Baxter equation ! solution of}

\begin{eqnarray}
& &r_+ = C_0 + 2 \sum_{\alpha > 0} E_\alpha
\otimes F_\alpha,\nonumber\\
& &r_- = -( C_0 + 2 \sum_{\alpha > 0}
F_\alpha \otimes E_\alpha ),\label{7}
\end{eqnarray}

\noindent and we have introduced the notation

\begin{displaymath}
C_0 = \sum_{i,j} A^{ij}H_i \otimes H_j
\end{displaymath}

\noindent as it will appear repeatedly in the following context.

Notice that it is not an easy task to calculate the Poisson brackets
$\{\psi_\pm^i (x),\; \psi_\pm^j (y)\}$ explicitly for $i \ne j$ [12].
This is because that the fields $\psi_\pm^i$ are first order fields and the
corresponding canonical momenta
\index{canonical momentum} are just linear combinations of these
fields themselves. For example, the canonical momentum corresponding to the
field $\psi_+^i$ reads

\begin{eqnarray*}
& &\pi (\psi_+^i ) (x) \equiv \frac{1}{2}
\langle F_i,\; \bar{\Psi}_+ (x) \rangle \\
& &~~~~= \frac{1}{2} \sum_k {\rm sign} (k - i) \psi_+^k
A_{ki} \equiv \frac{1}{2} \sum_k \psi_+^k M_{ki}.
\end{eqnarray*}

\noindent So the Poisson brackets like $\{\psi_\pm^i(x),
\;\psi_\pm^j(y)$ must be treated as Dirac Poisson brackets.
\index{Dirac Poisson bracket} Moreover,
the matrix $M = (M_{ki})$ is not always invertible so that
one can easily derive the Dirac Poisson brackets out of the naive canonical
ones, say, $\{\pi (\psi_+^i) (x),\; \psi_+^j (y)\}
= \delta_{ij} \delta (x - y)$. The fact that in some cases $M$ is not
invertible implies that ${\rm rank} (M) < {\rm rank} ({\cal G})$ and
thus the number of independent conjugate momenta of the $\psi_+$ fields
is smaller than the number of these fields themselves, or say that the
theory involves some more constraints, which make the calculation of Dirac
Poisson brackets more complicated. Fortunately, one can check case by case
that for even-rank Lie algebras, the matrix $M$ is always invertible. Thus we
can get well defined Poisson brackets for all pairs of $\psi_+^i$ and
$\psi_+^j$. The same is the case for the fields $\psi_-^i$.
So in what follows we shall restrict ourselves to the even-rank Lie algebras.

Recall that although the fundamental fields $\Phi,\;\Psi_\pm$
are periodic in $x$, the vectors $\xi_a^{(i)}$ and $\bar{\xi}_a^{(i)}$
are not. They have the following nontrivial monodromy properties,
\index{monodromy ! properties}

\begin{displaymath}
\xi_a^{(i)} (x+2\pi)=\xi_a^{(i)} T,
{}~~~~\bar{\xi}_a^{(i)} (x+2\pi) = T^{-1} \bar{\xi}_a^{(i)} (x),
\end{displaymath}

\noindent where $T \equiv T(x+2\pi)T^{-1}(x) = T(2\pi)$, the latter
equality holds if we normalize $T(x)$ such that $T(0)=1$. These monodromy
behaviors are actually the origin of the nontrivial couplings between
the left and right moving sectors in eq.(\ref{6}). In fact, such couplings
arise from the zero mode problem and the zero modes are contained
in the monodromy matrix $T$. It can be easily verified that
\index{zero mode}

\begin{eqnarray*}
& &\{T \otimes, T\} = \left[ r_\pm,\;T \otimes T \right],\\
& &\{\xi_a^{(i)} \otimes, T\} = -\xi_a^{(i)} \otimes T r_-,\\
& &\{\bar{\xi}_a^{(i)} \otimes, T\} = (1 \otimes T) r_+
(\bar{\xi}_a^{(i)} \otimes 1).
\end{eqnarray*}

\noindent So it seems that the $\xi$-$\bar{\xi}$ basis of the exchange
algebra may not be the most viable one in considering the
conformal properties of the model.

Another problem concerns about the general solution
\index{Toda ! two-extended ! solution of}
\index{Solution} of the 2EPCT fields.
It is easy to see [14] that the 2EPCT fields can be related to the vectors
$\xi_a^{(i)}$ and $\bar{\xi}_a^{(i)}$ in such a way that

\begin{equation}
{\rm e}^{\phi^i (x)} = \xi_1^{(i)}(x) \bar{\xi}_1^{(i)}(x),
{}~~~~
\psi_+^i(x) = \frac{\xi_1^{(i)}(x) \bar{\xi}_2^{(i)}(x)}{\xi_1^{(i)}(x)
\bar{\xi}_1^{(i)}(x)},
{}~~~~\psi_-^i(x) = \frac{\xi_2^{(i)}(x) \bar{\xi}_1^{(i)}(x)}{\xi_1^{(i)}(x)
\bar{\xi}_1^{(i)}(x)}.\label{8}
\end{equation}

\noindent One may desire that these formulas play a role in constructing
the local and periodic solutions \index{solution ! local and periodic}
of the model. It certainly should be
the case. However, again due to the nontrivial couplings between the
chiral and antichiral vectors, they cannot be given independently and thus
causing difficulties in getting explicit solutions of the 2EPCT
fields from the above simple formulas. Such difficulties can be
overcome only after the chiralities are completely splitted.

It so happens that all of the above problems can be cured if one chooses
a free field realization of the chiral exchange algebra. For this purpose
we need the following DS linear systems written in a specific gauge,

\begin{eqnarray}
& &\partial_+Q_+ = L_+Q_+,
{}~~~~\partial_-Q_+=0,\label{9}\\
& &\partial_-Q_- = Q_-L_-,
{}~~~~\partial_+Q_-=0,\label{10}
\end{eqnarray}

\noindent where

\begin{equation}
L_\pm \equiv \partial_\pm K_\pm(x) + \bar{P}_\pm(x) + {\cal E}_\pm
\in {\cal G}_0 \oplus {\cal G}_\pm,
{}~~~~\partial_\pm L_\mp = 0,\label{11}
\end{equation}

\noindent $L_\pm$ are periodic in $x$, and

\begin{eqnarray}
&K_\pm (x) \equiv \sum_{i}k_\pm^i (x)H_i \in {\cal G}_0,&\nonumber\\
&P_+ (x) \equiv \sum_{i}p_+^i (x)F_i \in {\cal G}^{(-1)},
{}~~~~P_- (x) \equiv \sum_{i}p_-^i (x)E_i \in {\cal G}^{(1)},&\label{12}\\
&\bar{P}_+ (x) \equiv [ {\cal E}_+,\;P_+ (x) ] \in {\cal G}^{(1)},~~~~
\bar{P}_- (x) \equiv [ P_- (x),\;{\cal E}_- ] \in {\cal G}^{(-1)}.&\nonumber
\end{eqnarray}

Since $L_\pm$ are respectively chiral and antichiral, so is $K_\pm$ and
$P_\pm$. It may be appropriate to regard $K_\pm (x)$ respectively as
the chiral and antichiral part of the free bosonic field $K(x)$ with
the action

\begin{displaymath}
S [ K ] = -\frac{1}{2} \int {\rm d}^2x {\rm tr}(\partial_+K(x) \partial_-
K(x)),~~~~ K(x) = K_+(x) + K_-(x),
\end{displaymath}

\noindent while the fields $P_+ (x)$-$\bar{P}_+ (x)$ and
$P_- (x)$-$\bar{P}_- (x)$ are considered as
two independent sets of multicomponent $\beta$-$\gamma$ systems ({\it i.e.}
first order chiral fields),

\begin{displaymath}
S [ P_\pm ] = - \frac{1}{2} \int {\rm d}^2x {\rm tr}(\bar{P}_\pm (x)
\partial_\mp P_\pm (x)).
\end{displaymath}

\noindent The canonical Poisson brackets for the fields $K_\pm$ can be
easily obtained from the action $S[K]$, which turn out to read

\begin{eqnarray}
& &\{\partial_\pm K_\pm (x) \otimes, \;K_\pm (y)\} = \pm \delta (x-y)
C_0, \nonumber \\
& &\{K_\pm(x)\otimes,\;K_\pm(y)\} = 0. \label{a}
\end{eqnarray}

\noindent However, the Poisson brackets for the first order fields $P_\pm$
must be calculated using the standard Dirac procedure (see appendix), which
yield the following consistent Dirac brackets for even-rank underlying
Lie algebras,

\begin{eqnarray}
& &\{\bar{P}_+ (x) \otimes,\; P_+ (y)\} = \delta (x-y) \sum_{i} E_i
\otimes F_i,\nonumber\\
& &\{\bar{P}_- (x) \otimes,\; P_- (y)\} = -\delta (x-y) \sum_{i} F_i
\otimes E_i,\nonumber\\
& &\{\bar{P}_+ (x) \otimes,\; \bar{P}_+ (y)\} = \delta (x-y) \sum_{ij}
A_{ij} {\rm sign} (i-j) E_i \otimes E_i,\nonumber \\
& &\{\bar{P}_- (x) \otimes,\; \bar{P}_- (y)\} = -\delta (x-y) \sum_{ij}
A_{ij} {\rm sign} (i-j) F_i \otimes F_i,\nonumber\\
& &\{P_+ (x) \otimes,\; P_+ (y)\} = \delta (x-y) \sum_{ij} (M^{-1})_{ji} F_i
\otimes F_j, \nonumber\\
& &\{P_- (x) \otimes,\; P_- (y)\} = -\delta (x-y) \sum_{ij} (M^{-1})_{ji} E_i
\otimes E_j.\label{b}
\end{eqnarray}

\noindent It is of special importance to note that
there are no nontrivial Poisson brackets between the left movers and the
right movers in this free field formalism. So our task is reduced to
the construction of chiral exchange algebra in an appropriate basis
and obtain the local and periodic solution of the original 2EPCT fields
starting from the above free fields. The line we shall follow is almost
the same as Ref.[1] except that we are considering a different model
with more degrees of freedom.

\section{Exchange algebra: the ($\sigma$, $\rho$)-$(\bar{\sigma}$,
$\bar{\rho}$) basis} \index{exchange algebra}
In this section we shall reconstruct the chiral exchange algebra for the
2EPCT theory using the Poisson brackets given in the end of the
introduction. Before doing this, let us now give some basic notions
on the mode expansions of the free fields.

Since the Cartan or diagonal parts of the DS potentials $L_\pm$ are equal to
the chiral derivatives of the field $K(x)$ and that $L_\pm$ are
periodic in $x$, we can expand $K_\pm (x)$ as
\footnote{We regrete for the inconveniences that may be brought to the
readers by the incoincidence between our notations and that of Ref.[1].}

\begin{equation}
K_\pm (x) = {\cal K}_\pm + {\cal K}_\pm^{(0)} x + \sum_{n \ne 0}
\frac{{\cal K}_\pm^{(n)}}{in} {\rm e}^{inx} \label{14}
\end{equation}

\noindent where $ \displaystyle {\cal K}_\pm \equiv -\sum_{n \ne 0}
\frac{{\cal K}_\pm^{(n)}}{in} $
so that $K_\pm (0) =0$ (the reason for this choice of normalization will
soon become clear). Similarly the periodic fields $P_\pm (x)$ can be
expanded as

\begin{displaymath}
P_\pm (x) =\sum_{n} {\cal P}_n {\rm e}^{inx},
\end{displaymath}
\begin{equation}
\int {\rm d}x P_\pm (x) =
{\cal P}_\pm + {\cal P}_\pm^{(0)} x + \sum_{n \ne 0}
\frac{{\cal P}_\pm^{(n)}}{in} {\rm e}^{inx},
{}~~~~{\cal P}_\pm \equiv -\sum_{n \ne 0} \frac{{\cal P}_\pm^{(n)}}{in}.
\label{15}
\end{equation}

\noindent The normalization we have chosen is such that $Q_\pm (0) = 1$
and that $Q_\pm (x)$ do not contain ${\cal Q}_\pm,\;{\cal F}_\pm$, the
conjugate variables of ${\cal K}_\pm^{(0)}$ and ${\cal P}_\pm^{(0)}$.
We shall see in the due course that the mode expansions of the
fields $P_\pm$ are irrespective to our purpose, except the above
normalization. However, those for the
fields $K_\pm$ are very important since the zero modes ${\cal K}_\pm^{(0)}$
and their conjugate variables play the central role in constructing the
local and periodic solutions.\index{zero mode}

Using the Poisson brackets (\ref{a}-\ref{b}) we can get

\begin{eqnarray}
\displaystyle
&\{{\cal K}_\pm^{(n)} \otimes,\;{\cal K}_\pm^{(m)}\} = \mp \frac{in}{2\pi}
C_0,&\nonumber\\
\displaystyle
&\{\partial_\pm K_\pm (x) \otimes,\;{\cal K}_\pm\} = \mp
(\delta (x) - \frac{1}{2\pi}) C_0.&\label{16}
\end{eqnarray}

\noindent Similarly we can obtain the Poisson brackets between various modes
of the fields $P_\pm (x)$, but we shall not make use of them.

In order to construct the exchange algebra we have to calculate the Poisson
brackets $\{Q_\pm (x) \otimes,\; Q_\pm (y) \}$.
As usual, we shall present the detailed calculations only in
the left moving sector and suply the result from the other sector wherever
it is necessary.

Let us present with some lemmas first.

\noindent {\bf Lemma 2.2.1}
\begin{eqnarray*}
\displaystyle & &\{L_+ (x)\otimes,\; L_+ (y)\} = - \frac{1}{2} C_0
(\partial_x - \partial_y )\delta (x-y) \\
& &~~~~~+ \sum {\rm sign}(i-j) A_{ij}
E_i \otimes E_j \delta (x-y),\\
\displaystyle & &\{L_+ (x)\otimes,\; L_- (y)\} =0,\\
\displaystyle & &\{L_- (x)\otimes,\; L_- (y)\} =  \frac{1}{2} C_0
(\partial_x - \partial_y )\delta (x-y) \\
& &~~~~~+ \sum {\rm sign}(j-i) A_{ij}
F_i \otimes F_j \delta (x-y).
\end{eqnarray*}

\noindent Proof: Straightforward from the canonical Poisson brackets
(\ref{a}-\ref{b}).Q.E.D.

\noindent {\bf Lemma 2.2.2}
\begin{eqnarray*}
& &\int_0^x {\rm d}x' \int_0^y {\rm d}y' Q_+^{-1}(x') \otimes Q_+^{-1}(y')
\{L_+(x') \otimes,\; L_+(y')\} Q_+ (x') \otimes Q_+ (y') = \\
& & \; \; \;
= \frac{1}{2}\left\{\theta (x-y) \left[ r - Q_+^{-1}(y) \otimes Q_+^{-1}(y)
(r-C_0)Q_+(y) \otimes Q_+(y)\right] \right. \\
& &\; \; \;
+  \frac{1}{2}\left. \theta (y-x) \left[ r - Q_+^{-1}(x) \otimes Q_+^{-1}(x)
(r+C_0)Q_+(x) \otimes Q_+(x)\right] \right\},
\end{eqnarray*}

\noindent {\em where $0<x,\;y<2\pi$, and $r$ is either $r_+$ or $r_-$.}

\noindent Proof: From Lemma 2.1 we get
\begin{eqnarray}
& &Q_+^{-1}(x') \otimes Q_+^{-1}(y')
\{L_+(x') \otimes,\; L_+(y')\} Q_+ (x') \otimes Q_+ (y') =\nonumber\\
& &~~~~
= Q_+^{-1}(x') \otimes Q_+^{-1}(y') \left(- \frac{1}{2} C_0
(\partial_{x'} - \partial_{y'} )\delta (x'-y') \right.\nonumber\\
& &~~\left.
+ \sum {\rm sign}(i-j) A_{ij}
E_i \otimes E_j \delta (x'-y')\right) Q_+ (x') \otimes Q_+ (y')\nonumber\\
& &~~=
- \frac{1}{2} (\partial_{x'} - \partial_{y'} ) \left[\delta (x'-y')
Q_+^{-1}(x') \otimes Q_+^{-1}(y') C_0 Q_+ (x') \otimes Q_+ (y') \right]
\nonumber\\
& &~~ + Q_+^{-1}(x') \otimes Q_+^{-1}(y')
\left( \sum {\rm sign}(i-j) A_{ij}
E_i \otimes E_j \delta (x'-y')\right)\nonumber\\
& &~~\times Q_+ (x') \otimes Q_+ (y')\nonumber\\
& &~~ + \frac{1}{2} \delta (x'-y')
(\partial_{x'} - \partial_{y'} ) \left(
Q_+^{-1}(x') \otimes Q_+^{-1}(y') C_0 Q_+ (x') \otimes Q_+ (y') \right).
\nonumber\\
\label{17}
\end{eqnarray}

\noindent The last term can be rewritten

\begin{eqnarray}
& &{\rm Last \ Term} = \frac{1}{2} \delta (x'-y')
Q_+^{-1}(x') \otimes Q_+^{-1}(y') \nonumber\\
& &~~~~
\times \left( [ C_0,\; L_+ (x') \otimes 1 ]
- [ C_0,\; 1 \otimes L_+(y') ] \right)
Q_+ (x') \otimes Q_+ (y')\nonumber\\
& &~~~~
=- \frac{1}{2} \delta (x'-y')Q_+^{-1}(x') \otimes Q_+^{-1}(y')
 [ r,\; L_+ (x')\otimes 1 + 1 \otimes L_+(y') ] \nonumber\\
& &~~~~
- \sum {\rm sign} (i-j) A_{ij} ( E_i \otimes E_j )
Q_+ (x') \otimes Q_+ (y').\label{18}
\end{eqnarray}

\noindent In the last step, we have used the explicit form of $L_+$ and the
identities

\begin{eqnarray*}
& [ r_\pm,\; E_i \otimes 1 + 1 \otimes E_i ] =
H_i \otimes E_i - E_i \otimes H_i,&\\
& [ r_\pm,\; [ E_i,\; E_j ]  \otimes 1 + 1 \otimes [ E_i,\; E_j ] ] =
H_i \otimes [ E_i,\;E_j ] - [ E_i,\; E_j ] \otimes H_i& \\
& - 2A_{ij} ( E_i \otimes E_j - E_j \otimes E_i )&.
\end{eqnarray*}

\noindent Inserting eq.(\ref{18}) into (\ref{17}) we get

\begin{eqnarray}
& &Q_+^{-1}(x') \otimes Q_+^{-1}(y')
\{L_+(x') \otimes,\; L_+(y')\} Q_+ (x') \otimes Q_+ (y') =\nonumber\\
& &\; \; \;
=- \frac{1}{2} (\partial_{x'} - \partial_{y'} ) \left[ \delta (x'-y')
Q_+^{-1}(x') \otimes Q_+^{-1}(y') C_0 Q_+ (x') \otimes Q_+ (y') \right]
\nonumber\\
& &\; \; \;
- \frac{1}{2} \delta (x'-y')Q_+^{-1}(x') \otimes Q_+^{-1}(y')
\left( [ r,\; L_+ (x') \otimes 1 + 1 \otimes L_+(y') ] \right)\nonumber\\
& &\; \; \;
Q_+ (x') \otimes Q_+ (y').\label{19}
\end{eqnarray}

\noindent The proof ends up after one substitutes eq.(\ref{19}) into
(\ref{16}) and performs the integration. Q.E.D.

\noindent {\bf Proposition 2.2.3} {\em If $0<x,\;y<2\pi$, we have}
\begin{eqnarray}
& & \{Q_+(x) \otimes,\;Q_+(y)\}=\frac{1}{2} Q_+(x) \otimes Q_+(y) \nonumber\\
& &\times \left\{\theta (x-y) \left[ r - Q_+^{-1}(y) \otimes Q_+^{-1}(y)
(r-C_0)Q_+(y) \otimes Q_+(y)\right] \right. \label{20}\\
& &\left. +\theta (y-x) \left[ r - Q_+^{-1}(x) \otimes Q_+^{-1}(x)
(r+C_0)Q_+(x) \otimes Q_+(x)\right] \right\},\nonumber
\end{eqnarray}
\noindent {\em If $0<x<2\pi,\; y=2\pi$, we have}
\begin{eqnarray}
& & \{Q_+(x) \otimes,\;S\}=\frac{1}{2} Q_+(x) \otimes S \nonumber\\
& &~~~~\times \left\{ r - Q_+^{-1}(x) \otimes Q_+^{-1}(x)
(r+C_0)Q_+(x) \otimes Q_+(x) \right.\nonumber\\
& &~~~~\left.+ (1 \otimes S^{-1}) C_0 (1 \otimes S)\right\},
\label{21}
\end{eqnarray}
\noindent {\em If $x=y=2\pi$, we have}
\begin{eqnarray}
& & \{S \otimes,\;S\}=\frac{1}{2} S \otimes S \left\{ r - S^{-1} \otimes
S^{-1} r S \otimes S \right. \nonumber\\
& &\; \; \;
\left. - (S^{-1} \otimes 1) C_0 (S \otimes 1) +
(1 \otimes S^{-1}) C_0 (1 \otimes S)\right\},\label{22}
\end{eqnarray}
\noindent {\em where $S$ is the monodromy matrix \index{monodromy ! matrix}
of $Q_+ (x)$,}
\begin{displaymath}
Q_+(x+2\pi) = Q_+(x)S,~~~~ S= Q_+(2\pi).
\end{displaymath}
\noindent Proof: Eq.(\ref{20}) follow directly from the formula
\begin{eqnarray*}
& & \{Q_+(x) \otimes,\;Q_+(y)\}=Q_+(x) \otimes Q_+(y)
\int_0^x {\rm d}x' \int_0^y {\rm d}y'\\
& & \; \; \;
 Q_+^{-1}(x') \otimes Q_+^{-1}(y')
\{L_+(x') \otimes,\; L_+(y')\} Q_+ (x') \otimes Q_+ (y')
\end{eqnarray*}
\noindent and Lemma 2.2. As for eqs.(\ref{21}-\ref{22})
we only need to notice that
the $\delta$-function in eq.(\ref{19}) is periodic and it is consistence
to set
\begin{displaymath}
\int_0^y {\rm d}x f(x)\delta (x) = \left\{
\begin{array}{ccc}
&$$ \frac{1}{2} f(0)$$          & $$   0<y<2\pi  $$    \cr
&$$ 0  $$                       & $$  y=0        $$    \cr
&$$\frac{1}{2} (f(0)+f(2\pi))$$ & $$   y=2\pi    $$
\end{array}
\right.
\end{displaymath}
\noindent and also $\theta (0) = \frac{1}{2}.$ Q.E.D.

Notice that although we are considering a theory with the extra fields
$P_\pm(x)$, the resulting exchange relations (\ref{20}-\ref{22})
for $Q_+(x)$ and $S$
are the same as those in the standard Toda case except a different sign
which is caused by our initial definition of the Poisson brackets.

Now define

\begin{displaymath}
\sigma^{(i)} (x) = \langle \lambda_{max}^{(i)} | Q_+ (x),
\end{displaymath}

\noindent we have for $0<x,\;y<2\pi$

\noindent {\bf Proposition 2.2.4}
\begin{eqnarray*}
& &\{ \sigma^{(i)} (x)\otimes,\;\sigma^{(j)} (y)\} =
\frac{1}{2} \sigma^{(i)} (x)\otimes \sigma^{(j)} (y) \left[ r_+ \theta (x-y)
+ r_- \theta (y-x) \right],\\
& &\{ \sigma^{(i)} (x)\otimes,\;S\} =
\frac{1}{2} \sigma^{(i)} (x)\otimes S \left[ r_- + (1 \otimes S^{-1}) C_0
(1 \otimes S) \right],\\
& & \{S \otimes,\;S\}= - \frac{1}{2} \left\{ [ r,\; S \otimes S]
+ (1 \otimes S) C_0 (S \otimes 1) -
(S \otimes 1) C_0 (1 \otimes S)\right\}.
\end{eqnarray*}

Similarly define
\begin{displaymath}
\bar{\sigma}^{(i)} (x) = Q_- (x) | \lambda_{max}^{(i)} \rangle,
\end{displaymath}
\noindent we have

\noindent {\bf Proposition 2.2.5}
\begin{eqnarray*}
& &\{ \bar{\sigma}^{(i)} (x)\otimes,\;\bar{\sigma}^{(j)} (y)\} =
\frac{1}{2} \left[ r_- \theta (x-y) + r_+ \theta (y-x) \right]
\bar{\sigma}^{(i)} (x)\otimes \bar{\sigma}^{(j)} (y) ,\\
& &\{ \bar{\sigma}^{(i)} (x)\otimes,\;\bar{S}\} =
\frac{1}{2}  \left[ r_+ - (1 \otimes \bar{S}) C_0
(1 \otimes \bar{S}^{-1}) \right]\bar{\sigma}^{(i)} (x)\otimes \bar{S},\\
& & \{\bar{S} \otimes,\;\bar{S}\}=
 \frac{1}{2} \left\{ [ r,\; \bar{S} \otimes \bar{S}]
- (1 \otimes \bar{S}) C_0 (\bar{S} \otimes 1) +
(\bar{S} \otimes 1) C_0 (1 \otimes \bar{S})\right\},
\end{eqnarray*}

\noindent {\em where}

\begin{displaymath}
\bar{S} \equiv Q_- (2\pi),
{}~~~~\bar{\sigma}^{(i)} (x+2\pi) = \bar{S} \bar{\sigma}^{(i)} (x).
\end{displaymath}

\noindent {\em Moreover, all the cross Poisson brackets between the left
moving sector (which is consisted of the quantities without a ``bar'')
and the right moving sector (consisted of those quantites
with a ``bar'') vanish.}

Now let us consider the Poisson bracket $\{{\rm e}^{P_+ (x)}
 \otimes,\; Q_+ (y)\}$. We have

\noindent {\bf Lemma 2.2.6}
\begin{eqnarray*}
& &\{{\rm e}^{P_+ (x)} \otimes,\; Q_+ (y)\} = - \theta (y-x)
1 \otimes Q_+(y) Q_+^{-1}(x)\\
& &~~~~\times \sum_{l=0}^{\infty}\frac{1}{(l+1)!}
\left[ \sum_{i} ({\rm ad}P_+(x))^{l}
F_i \otimes E_i \right] {\rm e}^{P_+(x)} \otimes Q_+(x),\\
& &\{Q_+ (x) \otimes,\; {\rm e}^{P_+ (y)}\} =  \theta (x-y)
Q_+(x) Q_+^{-1}(y) \otimes 1\\
& &~~~~\times \sum_{l=0}^{\infty}\frac{1}{(l+1)!} \left[ \sum_{i} E_i
\otimes ({\rm ad}P_+(y))^{l} F_i \right] Q_+(y) \otimes {\rm e}^{P_+(y)},\\
& &\{{\rm e}^{P_+ (x)} \otimes,\; S \} = - 1 \otimes S Q_+^{-1}(x)\\
& &~~~~\times \sum_{l=0}^{\infty}\frac{1}{(l+1)!}
\left[ \sum_{i} ({\rm ad}P_+(x))^{l}
F_i \otimes E_i \right] {\rm e}^{P_+(x)} \otimes Q_+(x),\\
& &\{ S \otimes,\; {\rm e}^{P_+ (y)}\} = S Q_+^{-1}(y) \otimes 1\\
& &~~~~\times \sum_{l=0}^{\infty}\frac{1}{(l+1)!} \left[ \sum_{i} E_i
\otimes ({\rm ad}P_+(y))^{l} F_i \right] Q_+(y) \otimes {\rm e}^{P_+(y)}.
\end{eqnarray*}

\noindent Proof: We shall only prove the first formula. It is easy to see
that

\begin{displaymath}
\{P_+(x)\otimes,\; L_+(y)\} = -\sum F_i \otimes E_i \delta (x-y),
\end{displaymath}.

\noindent Using the formula

\begin{displaymath}
\{{\rm e}^A\otimes,\;B\} = \sum_{l=0}^{\infty} \frac{1}{(l+1)!}
\left[ ({\rm ad}A)^{l} \otimes 1 \{A \otimes,\; B\}\right] {\rm e}^A
\otimes 1,
\end{displaymath}

\noindent we can get

\begin{displaymath}
\{{\rm e}^{P_+ (x)} \otimes,\; L_+ (y)\} = -\delta (x-y)
\sum_{l=0}^{\infty}\frac{1}{(l+1)!} \left[ \sum_{i} ({\rm ad}P_+(x))^{l}
F_i \otimes E_i \right] {\rm e}^{P_+(x)} \otimes 1.
\end{displaymath}

\noindent Integrating the above equation the first equation in Lemma 2.6
follow. Q.E.D.

Now define the additional chiral basis vectors

\begin{displaymath}
\rho^{(i)}(x) = \langle \lambda_{max}^{(i)} | E_i
{\rm e}^{P_+(x)}Q_+(x),
{}~~~~\bar{\rho}^{(i)}(x) = Q_-(x){\rm e}^{P_-(x)}
F_i | \lambda_{max}^{(i)} \rangle
\end{displaymath}

\noindent and using the Poisson brackets

\begin{eqnarray*}
& &\{{\rm e}^{P_+ (x)} \otimes,\; {\rm e}^{P_+ (y)} \}
= \delta (x-y) \sum_{l=0}^{\infty}\frac{1}{(l+1)!}
\sum_{m=0}^{\infty}\frac{1}{(m+1)!} \\
& &~~~~\times ({\rm ad}P_+ (x))^{l} \otimes ({\rm ad}P_+ (y))^{m} \sum_{ij}
M^{-1})_{ji} ( F_i \otimes F_j )
{\rm e}^{P_+ (x)} \otimes {\rm e}^{P_+ (y)},
\end{eqnarray*}

\noindent we can obtain
\bigskip

\noindent {\bf Proposition 2.2.7}

\begin{eqnarray*}
& &\{ \rho^{(i)} (x)\otimes,\;\rho^{(j)} (y)\} =
\frac{1}{2} \rho^{(i)} (x)\otimes \rho^{(j)} (y) \left[ r_+ \theta (x-y)
+ r_- \theta (y-x) \right] \\
& &~~~~+ (M^{-1})_{ji} \delta (x-y) \sigma^{(i)} (x)\otimes \sigma^{(j)} (y)
,\\
& &\{ \sigma^{(i)} (x)\otimes,\;\rho^{(j)} (y)\} =
\frac{1}{2} \sigma^{(i)} (x)\otimes \rho^{(j)} (y) \left[ r_+ \theta (x-y)
+ r_- \theta (y-x) \right],\\
& &\{ \rho^{(i)} (x)\otimes,\;\sigma^{(j)} (y)\} =
\frac{1}{2} \rho^{(i)} (x)\otimes \sigma^{(j)} (y) \left[ r_+ \theta (x-y)
+ r_- \theta (y-x) \right],\\
& &\{ \rho^{(i)} (x)\otimes,\;S\} =
\frac{1}{2} \rho^{(i)} (x)\otimes S \left[ r_- + (1 \otimes S^{-1}) C_0
(1 \otimes S) \right].
\end{eqnarray*}
\noindent {\bf Proposition 2.2.8}
\begin{eqnarray*}
& &\{ \bar{\rho}^{(i)} (x)\otimes,\;\bar{\rho}^{(j)} (y)\}
\frac{1}{2} \left[ r_- \theta (x-y) + r_+ \theta (y-x) \right]
\bar{\rho}^{(i)} (x)\otimes \bar{\rho}^{(j)} (y) \\
& &~~~~- (M^{-1})_{ji} \delta (x-y) \bar{\sigma}^{(i)} (x)\otimes
\bar{\sigma}^{(j)} (y), \\
& &\{ \bar{\sigma}^{(i)} (x)\otimes,\;\bar{\rho}^{(j)} (y)\} =
\frac{1}{2} \left[ r_- \theta (x-y) + r_+ \theta (y-x) \right]
\bar{\sigma}^{(i)} (x)\otimes \bar{\rho}^{(j)} (y) ,\\
& &\{ \bar{\rho}^{(i)} (x)\otimes,\;\bar{\sigma}^{(j)} (y)\} =
\frac{1}{2} \left[ r_- \theta (x-y) + r_+ \theta (y-x) \right]
\bar{\rho}^{(i)} (x)\otimes \bar{\sigma}^{(j)} (y) ,\\
& &\{ \bar{\rho}^{(i)} (x)\otimes,\;\bar{S}\} =
\frac{1}{2}  \left[ r_+ - (1 \otimes \bar{S}) C_0
(1 \otimes \bar{S}^{-1}) \right] \bar{\rho}^{(i)} (x)\otimes \bar{S}.
\end{eqnarray*}

\noindent Proof of Proposition 2.7:
Follows from Lemma 2.6 and Proposition 2.3 and the fact

\begin{eqnarray*}
\langle \lambda_{max}^{(i)}| E_i \otimes 1 (r_- + C_0) = -
\langle \lambda_{max}^{(i)}| E_i \otimes 1 (2 F_i \otimes E_i),\\
1 \otimes \langle \lambda_{max}^{(i)} | E_i(r_+ - C_0) =
1 \otimes \langle \lambda_{max}^{(i)} | E_i (2 E_i \otimes F_i).
\end{eqnarray*}

\noindent Q.E.D.

\noindent These last two propositions complete the construction of chiral
exchange algebra \index{exchange algebra}
in the $(\sigma,\;\rho)$-$(\bar{\sigma},\;\bar{\rho})$
basis. We note that although the exchange relations for the $\sigma$-vectors
take the same form as those in the usual Toda field theory, there are
extra $\delta$-function terms in the Poisson brackets
$\{\rho^{(i)} (x) \otimes,\;\rho^{(j)} (y) \}$ and
$\{\bar{\rho}^{(i)} (x) \otimes,\;\bar{\rho}^{(j)} (y) \}$ (these
$\delta$-function terms vanish while $i = j$ because the matrix $M$ is
antisymmetric and so is $M^{-1}$). The additional $\rho$-$\bar{\rho}$
vectors and the extra $\delta$-function terms reflect the complexity of
our model comparing to the usual Toda theory.

\section{The $(\mu,\;\nu)$-$(\bar{\mu},\;\bar{\nu})$ basis and
diagonal monodromies}
In the last section we have reconstructed the chiral exchange algebra
using the $(\sigma,\;\rho)$-$(\bar{\sigma},\;\bar{\rho})$ basis. This
basis has the merit that all the exchange relations are very simple and
that the left and right moving sectors are decoupled. However, as is shown
in Propositions 2.7 and 2.8, the $(\sigma,\;\rho)$-$(\bar{\sigma},
\;\bar{\rho})$ basis has nontrivial couplings with the
left and right monodromy matrices,
\index{monodromy ! matrix} respectively. Moreover, these monodromy
matrices take values respectively in ${\rm exp} ({\cal G}_0 \oplus
{\cal G}_\pm )$, which makes their evaluations somewhat involved. These
complexities can be avoided by choosing other convenient basis for the chiral
exchange algebra, one of such basis is that with the monodromy matrices
diagonalized and equal to the $K_\pm$-zero modes, respectively.

Recall that the monodromy matrices in the $(\sigma,\;\rho)$-$(\bar{\sigma},
\;\bar{\rho})$ basis are defined as $S=Q_+(2\pi)$, $\bar{S}=Q_-(2\pi).$
According to the original form of the DS linear systems (\ref{9}-\ref{11})
and the mode expansion (\ref{14}) of the
fields $K_\pm(x)$, we can conclude that the diagonal
part of $S$ and $\bar{S}$ must be consisted of the zero modes of
$K_+$ and $K_-$ respectively. It follows that there exists a unique
choice of matrices $g$ and $\bar{g}$ such that they diagonalize $S$ and
$\bar{S},$

\begin{eqnarray*}
S=g^{-1} \kappa g,~~~~ \kappa = {\rm e}^{2\pi {\cal K}_+^{(0)}},\\
\bar{S}=\bar{g}^{-1} \bar{\kappa} \bar{g},
{}~~~~ \bar{\kappa} = {\rm e}^{- 2\pi {\cal K}_-^{(0)}}.
\end{eqnarray*}

\noindent We emphasize that the uniqueness of $g$ and $\bar{g}$ is
ensured by the condition that the diagonal parts thereof
are unit matrices. Since the $K_\pm$-zero modes Poisson commute with
themselves, we can in principal reduce the Poisson brackets
$\{S\otimes,\; S\}$ and $\{\bar{S}\otimes,\; \bar{S}\}$ into those for
$g$ and $\bar{g}$'s. This is already done in Ref.[1]. The results
read

\noindent{\bf Proposition 2.3.1}

\begin{eqnarray*}
& &\{g \otimes,\; g\} g^{-1}\otimes g^{-1} = \frac{1}{4} {\rm coth}
\left(\pi {\rm ad}_2 {\cal K}_+^{(0)} \right) \left[ g \otimes g (r + C_0)
g^{-1} \otimes g^{-1} \right.\\
& &~~~~- \left. r -C_0 - 2(1 \otimes g)C_0 (1 \otimes g^{-1}) + 2 C_0 \right]
\\
& &~~~~+ \frac{1}{4} {\rm coth}
\left(\pi {\rm ad}_1 {\cal K}_+^{(0)} \right) \left[ g \otimes g (r - C_0)
g^{-1} \otimes g^{-1} \right.\\
& &~~~~- \left. r + C_0 + 2(g \otimes 1)C_0 (g^{-1} \otimes 1) - 2 C_0
\right],\\
\end{eqnarray*}
\begin{eqnarray*}
& &\{\bar{g} \otimes,\; \bar{g}\} \bar{g}^{-1} \otimes \bar{g}^{-1} =
\frac{1}{4} {\rm coth} \left(\pi {\rm ad}_2 {\cal K}_-^{(0)} \right)
\left[ \bar{g} \otimes \bar{g} (r - C_0)
\bar{g}^{-1} \otimes \bar{g}^{-1} \right.\\
& &~~~~- \left. r + C_0 + 2(1 \otimes \bar{g})C_0
(1 \otimes \bar{g}^{-1}) - 2 C_0 \right]\\
& &~~~~+\frac{1}{4} {\rm coth} \left(\pi {\rm ad}_1 {\cal K}_-^{(0)}
\right) \left[ \bar{g} \otimes \bar{g} (r + C_0)
\bar{g}^{-1} \otimes \bar{g}^{-1} \right.\\
& &~~~~- \left. r - C_0 - 2(\bar{g} \otimes 1)C_0
(\bar{g}^{-1} \otimes 1) + 2 C_0 \right],\\
\end{eqnarray*}
\begin{eqnarray*}
& &\{g \otimes,\;\bar{g}\} =0,
\end{eqnarray*}

\noindent where the signs were adjusted in accordance with our convention.
Moreover, due to the normalization conditions $Q_\pm(0)=1$, the $K_\pm$-zero
modes must commute with $Q_\pm$ under Poisson brackets, and they trivially
Poisson commute with ${\rm e}^{P_\pm}$. Therefore, the Poisson brackets
between $S$ ($\bar{S}$, resp.) and the basis vectors $\sigma^{(i)},
\;\rho^{(i)}$ ($\bar{\sigma}^{(i)}, \;\bar{\rho}^{(i)}$ resp.) can also be
reduced, giving rise to

\noindent {\bf Proposition 2.3.2}

\begin{eqnarray*}
& &\{ \sigma^{(i)} (x) \otimes,\; g\} 1 \otimes g^{-1}
= \frac{1}{4} {\rm coth} \left(\pi {\rm ad}_2 {\cal K}_+^{(0)} \right)
\left[ \sigma^{(i)}(x) \otimes g (r_- + C_0)1 \otimes g^{-1}\right]\\
& &~~~~+\frac{1}{4} \sigma^{(i)}(x) \otimes g (r_- - C_0)1 \otimes g^{-1}
+ \frac{1}{2} \sigma^{(i)}(x) \otimes 1 C_0,
\end{eqnarray*}
\begin{eqnarray*}
& &\{\rho^{(i)}(x)\otimes,\; g\} 1\otimes g^{-1}
= \frac{1}{4} {\rm coth} \left(\pi {\rm ad}_2 {\cal K}_+^{(0)} \right)
\left[ \rho^{(i)}(x) \otimes g (r_- + C_0)1 \otimes g^{-1}\right]\\
& &~~~~+ \frac{1}{4} \rho^{(i)}(x) \otimes g (r_- - C_0) 1 \otimes g^{-1}
+ \frac{1}{2} \rho^{(i)}(x)\otimes 1 C_0,
\end{eqnarray*}

\noindent {\bf Proposition 2.3.3}

\begin{eqnarray*}
& &\{\bar{\sigma}^{(i)}(x)\otimes,\; \bar{g}\} 1\otimes \bar{g}^{-1}
=- \frac{1}{4} {\rm coth} \left(\pi {\rm ad}_2 {\cal K}_-^{(0)} \right)
\left[ 1 \otimes \bar{g} (r_+ - C_0)\bar{\sigma}^{(i)}(x) \otimes
\bar{g}^{-1}\right]\\
& &~~~~- \frac{1}{4} 1 \otimes \bar{g} (r_+ + C_0)
\bar{\sigma}^{(i)}(x) \otimes g^{-1} + \frac{1}{2} C_0
\bar{\sigma}^{(i)}(x) \otimes 1,
\end{eqnarray*}
\begin{eqnarray*}
& &\{\bar{\rho}^{(i)}(x)\otimes,\; \bar{g}\} 1\otimes \bar{g}^{-1}
= -\frac{1}{4} {\rm coth} \left(\pi {\rm ad}_2 {\cal K}_-^{(0)} \right)
\left[ 1 \otimes \bar{g} (r_+ - C_0)\bar{\rho}^{(i)}(x) \otimes
\bar{g}^{-1}\right]\\
& &~~~~- \frac{1}{4} 1 \otimes \bar{g} (r_+ + C_0)
\bar{\rho}^{(i)}(x) \otimes g^{-1} + \frac{1}{2} C_0
\bar{\rho}^{(i)}(x) \otimes 1.
\end{eqnarray*}

\noindent The procedure for proving these propositions is rather
tedious and we refer the readers to Ref.[1] for the standard method.

Now we are ready to define a new set of basis for the chiral exchange
algebra which has diagonal monodromy matrices. The basis is defined as

\begin{eqnarray*}
& &\mu^{(i)} (x) = \sigma^{(i)} (x) g^{-1},
{}~~~~\nu^{(i)} (x) = \rho^{(i)} (x) g^{-1},\\
& &\bar{\mu}^{(i)} (x) = \bar{g}\bar{\sigma}^{(i)} (x),
{}~~~~\bar{\nu}^{(i)} (x) = \bar{g}\bar{\rho}^{(i)} (x),
\end{eqnarray*}

\noindent and the monodromy properties are

\begin{eqnarray*}
& &\mu^{(i)} (x+2\pi) = \mu^{(i)} (x) \kappa,
{}~~~~\nu^{(i)} (x+ 2\pi) = \nu^{(i)} (x)\kappa,\\
& &\bar{\mu}^{(i)} (x+2\pi) =
\bar{\kappa}\bar{\mu}^{(i)} (x) ,
{}~~~~\bar{\nu}^{(i)} (x+ 2\pi) = \bar{\kappa}\bar{\nu}^{(i)} (x).
\end{eqnarray*}

\noindent Using the previous results, it is straightforward to
check the following

\noindent{\bf Proposition 2.3.4}

\begin{eqnarray*}
& &\{\mu^{(i)}(x)\otimes,\;\mu^{(j)}(y)\} = \mu^{(i)}(x)\otimes \mu^{(j)}(y)
\left\{\frac{1}{4} (r_+ - r_-) {\rm sign }(x-y) \right.\\
& &~~~~-\frac{1}{4}{\rm coth} \left(\pi {\rm ad}_1 {\cal K}_+^{(0)} \right)
\left[ r_+ - C_0 \right]\\
& &~~~~-\frac{1}{4}{\rm coth} \left(\pi {\rm ad}_2 {\cal K}_+^{(0)} \right)
\left[ r_- + C_0 \right]\\
& &~~~~+\frac{1}{2}\left[ {\rm coth} \left(\pi {\rm ad}_1 {\cal K}_+^{(0)}
\right)
-1\right] \left[(g\otimes 1) C_0 (g^{-1} \otimes 1) - C_0 \right]\\
& &~~~~-\left. \frac{1}{2}\left[ {\rm coth}
\left(\pi {\rm ad}_2 {\cal K}_+^{(0)} \right)
-1\right] \left[(1\otimes g) C_0 (1 \otimes g^{-1}) - C_0 \right]
\right\},
\end{eqnarray*}

\noindent{\bf Proposition 2.3.5}

\begin{eqnarray*}
& &\{\mu^{(i)}(x)\otimes,\;\nu^{(j)}(y)\} = \mu^{(i)}(x)\otimes \nu^{(j)}(y)
\left\{\frac{1}{4} (r_+ - r_-) {\rm sign }(x-y) \right.\\
& &~~~~-\frac{1}{4}{\rm coth} \left(\pi {\rm ad}_1 {\cal K}_+^{(0)} \right)
\left[ r_+ - C_0 \right]\\
& &~~~~-\frac{1}{4}{\rm coth} \left(\pi {\rm ad}_2 {\cal K}_+^{(0)} \right)
\left[ r_- + C_0 \right]\\
& &~~~~+\frac{1}{2}\left[ {\rm coth} \left(\pi {\rm ad}_1 {\cal K}_+^{(0)}
\right)
-1\right] \left[(g\otimes 1) C_0 (g^{-1} \otimes 1) - C_0 \right]\\
& &~~~~-\left. \frac{1}{2}\left[ {\rm coth}
\left(\pi {\rm ad}_2 {\cal K}_+^{(0)} \right)
-1\right] \left[(1\otimes g) C_0 (1 \otimes g^{-1}) - C_0 \right]
\right\},
\end{eqnarray*}

\noindent{\bf Proposition 2.3.6}

\begin{eqnarray*}
& &\{\nu^{(i)}(x)\otimes,\;\nu^{(j)}(y)\} = \nu^{(i)}(x)\otimes \nu^{(j)}(y)
\left\{\frac{1}{4} (r_+ - r_-) {\rm sign }(x-y) \right.\\
& &~~~~-\frac{1}{4}{\rm coth} \left(\pi {\rm ad}_1 {\cal K}_+^{(0)} \right)
\left[ r_+ - C_0 \right]\\
& &~~~~-\frac{1}{4}{\rm coth} \left(\pi {\rm ad}_2 {\cal K}_+^{(0)} \right)
\left[ r_- + C_0 \right]\\
& &~~~~+\frac{1}{2}\left[ {\rm coth} \left(\pi {\rm ad}_1 {\cal K}_+^{(0)}
\right)
-1\right] \left[(g\otimes 1) C_0 (g^{-1} \otimes 1) - C_0 \right]\\
& &~~~~-\left. \frac{1}{2}\left[ {\rm coth}
\left(\pi {\rm ad}_2 {\cal K}_+^{(0)} \right)
-1\right] \left[(1\otimes g) C_0 (1 \otimes g^{-1}) - C_0 \right]
\right\}\\
& &~~~~+ (M^{-1})_{ji} \delta (x-y) \mu^{(i)} (x)\otimes \mu^{(j)} (y),
\end{eqnarray*}

\noindent{\bf Proposition 2.3.7}

\begin{eqnarray*}
& &\{\bar{\mu}^{(i)}(x)\otimes,\;\bar{\mu}^{(j)}(y)\}
= \left\{- \frac{1}{4} (r_+ - r_-) {\rm sign }(x-y) \right.\\
& &~~~~-\frac{1}{4}{\rm coth} \left(\pi {\rm ad}_1 {\cal K}_-^{(0)} \right)
\left[ r_- + C_0 \right]\\
& &~~~~-\frac{1}{4}{\rm coth} \left(\pi {\rm ad}_2 {\cal K}_-^{(0)} \right)
\left[ r_+ - C_0 \right]\\
& &~~~~-\frac{1}{2}\left[ {\rm coth} \left(\pi {\rm ad}_1 {\cal K}_-^{(0)}
\right)
-1\right] \left[(\bar{g}\otimes 1) C_0 (\bar{g}^{-1} \otimes 1)
- C_0 \right]\\
& &~~~~+\left. \frac{1}{2}\left[ {\rm coth}
\left(\pi {\rm ad}_2 {\cal K}_-^{(0)} \right)
-1\right] \left[(1\otimes \bar{g}) C_0 (1 \otimes \bar{g}^{-1})
- C_0 \right]
\right\}\\
& &~~~~\times  \bar{\mu}^{(i)}(x)\otimes \bar{\mu}^{(j)}(y),
\end{eqnarray*}

\noindent{\bf Proposition 2.3.8}

\begin{eqnarray*}
\displaystyle
& &\{\bar{\mu}^{(i)}(x)\otimes,\;\bar{\nu}^{(j)}(y)\}
= \left\{- \frac{1}{4} (r_+ - r_-) {\rm sign }(x-y) \right.\\
\displaystyle
& &~~~~-\frac{1}{4}{\rm coth} \left(\pi {\rm ad}_1 {\cal K}_-^{(0)} \right)
\left[ r_- + C_0 \right]\\
\displaystyle
& &~~~~-\frac{1}{4}{\rm coth} \left(\pi {\rm ad}_2 {\cal K}_-^{(0)} \right)
\left[ r_+ - C_0 \right]\\
\displaystyle
& &~~~~-\frac{1}{2}\left[ {\rm coth} \left(\pi {\rm ad}_1 {\cal K}_-^{(0)}
\right)
-1\right] \left[(\bar{g}\otimes 1) C_0 (\bar{g}^{-1} \otimes 1)
- C_0 \right]\\
\displaystyle
& &~~~~+\left. \frac{1}{2}\left[ {\rm coth}
\left(\pi {\rm ad}_2 {\cal K}_-^{(0)} \right)
-1\right] \left[(1\otimes \bar{g}) C_0 (1 \otimes \bar{g}^{-1})
- C_0 \right]
\right\}\\
\displaystyle
& &~~~~\times \bar{\mu}^{(i)}(x)\otimes \bar{\nu}^{(j)}(y),
\end{eqnarray*}

\noindent and finally,
\bigskip

\noindent {\bf Proposition 2.3.9}

\begin{eqnarray*}
\displaystyle
& &\{\bar{\nu}^{(i)}(x)\otimes,\;\bar{\nu}^{(j)}(y)\}
= \left\{- \frac{1}{4} (r_+ - r_-) {\rm sign }(x-y) \right.\\
\displaystyle
& &~~~~-\frac{1}{4}{\rm coth} \left(\pi {\rm ad}_1 {\cal K}_-^{(0)} \right)
\left[ r_- + C_0 \right]\\
\displaystyle
& &~~~~-\frac{1}{4}{\rm coth} \left(\pi {\rm ad}_2 {\cal K}_-^{(0)} \right)
\left[ r_+ - C_0 \right]\\
\displaystyle
& &~~~~-\frac{1}{2}\left[ {\rm coth} \left(\pi {\rm ad}_1 {\cal K}_-^{(0)}
\right)
-1\right] \left[(\bar{g}\otimes 1) C_0 (\bar{g}^{-1} \otimes 1)
- C_0 \right]\\
\displaystyle
& &~~~~+\left. \frac{1}{2}\left[ {\rm coth}
\left(\pi {\rm ad}_2 {\cal K}_-^{(0)} \right)
-1\right] \left[(1\otimes \bar{g}) C_0 (1 \otimes \bar{g}^{-1})
- C_0 \right] \right\}\\
\displaystyle
& &\times \bar{\nu}^{(i)}(x)\otimes \bar{\nu}^{(j)}(y)\\
\displaystyle
& &~~~~- (M^{-1})_{ji} \delta (x-y) \bar{\mu}^{(i)} (x) \otimes
\bar{\mu}^{(j)} (y).
\end{eqnarray*}

\noindent Again, the cross Poisson brackets between both chiral sectors
vanish under this basis. The $(\mu,\;\nu)$-$(\bar{\mu},\;\bar{\nu})$ basis
is sometimes called Bloch wave basis.

\section{The conjugate variables of $K_\pm$-zero modes}
\index{zero mode ! conjugate variables of}
So far we have not considered the role of conjugate variables of the
$K_\pm$-zero modes ${\cal Q}_\pm$. In this section, we shall
derive various Poisson brackets involving these variables for the purpose of
future use.

By definition, the conjugate $K_\pm$-zero modes ${\cal Q}_\pm$ are such that

\begin{displaymath}
\{{\cal Q}_\pm \otimes,\;{\cal K}_\pm^{(0)}\} = \mp \frac{1}{2\pi} C_0.
\end{displaymath}

\noindent Denoting

\begin{displaymath}
\Theta_\pm = {\rm e}^{\pm{\cal Q}_\pm \mp {\cal K}_\pm},
\end{displaymath}

\noindent where ${\cal K}_\pm$ are those given right after eq.(14),
we have

\noindent {\bf Lemma 2.4.1}
\begin{eqnarray*}
& &\{Q_+(x) \otimes,\;\Theta_+\} = Q_+(x) \otimes \Theta_+
\int_0^x {\rm d}z \delta (z) Q_+^{-1}(z)\otimes 1 C_0 Q_+(z)\otimes 1\\
& &~~~~= \frac{1}{2} Q_+(x) \otimes \Theta_+ C_0
{}~~~~( \ {\rm if} \ 0<x<2\pi \ )
\end{eqnarray*}

\noindent Proof: Following eq.(16), we have

\begin{displaymath}
\{L_+(x)\otimes,\; {\rm e}^{-{\cal K}_+}\} =
\left( \delta (x) - \frac{1}{2\pi}\right) 1\otimes {\rm e}^{-{\cal K}_+}
C_0.
\end{displaymath}

\noindent Integrating the above equation we get

\begin{eqnarray}
& &\{Q_+(x)\otimes,\; {\rm e}^{-{\cal K}_+}\} =
Q_+(x) \otimes {\rm e}^{-{\cal K}_+} \nonumber\\
& &~~~~\times \int_0^x {\rm d}z
\left( \delta (x) - \frac{1}{2\pi}\right)
Q_+^{-1}(z)\otimes 1 C_0 Q_+(z)\otimes 1. \label{23}
\end{eqnarray}

\noindent Similarly, since

\begin{displaymath}
\{L_+(x)\otimes,\; {\rm e}^{{\cal Q}_+}\} =
\frac{1}{2\pi} 1\otimes {\rm e}^{{\cal Q}_+} C_0
\end{displaymath}

\noindent we have

\begin{eqnarray}
\{Q_+(x)\otimes,\; {\rm e}^{{\cal Q}_+}\} =
\frac{1}{2\pi} Q_+(x) \otimes {\rm e}^{{\cal Q}_+}
\int_0^x {\rm d}z Q_+^{-1}(z)\otimes 1 C_0 Q_+(z)\otimes 1.\label{24}
\end{eqnarray}

\noindent Combining eqs.(\ref{23}) and (\ref{24}) the Lemma 2.4.1 follow.
Q.E.D.

Using Lemma 2.4.1 and the fact that ${\cal Q}_+$ Poisson commutes with
${\rm e}^{P_+}$, we have from the definitions of the chiral vectors
$\sigma^{(i)}(x)$ and $\rho^{(i)}(x)$ that

\noindent {\bf Proposition 2.4.2}

\begin{eqnarray*}
&\{\sigma^{(i)}(x) \otimes,\;\Theta_+\} = \frac{1}{2}
\sigma^{(i)}(x) \otimes \Theta_+ C_0,&\\
&\{\rho^{(i)}(x) \otimes,\;\Theta_+\} = \frac{1}{2}
\rho^{(i)}(x) \otimes \Theta_+ C_0,&\\
&\{S \otimes,\;\Theta_+\} = \frac{1}{2}
S \otimes \Theta_+ \left( C_0 + S^{-1} \otimes 1 C_0 S\otimes 1\right).&
\end{eqnarray*}

\noindent Similarly, we have for the other chirality

\noindent {\bf Proposition 2.4.3}

\begin{eqnarray*}
&\{\bar{\sigma}^{(i)}(x) \otimes,\;\Theta_-\} = - \frac{1}{2} C_0
\bar{\sigma}^{(i)}(x) \otimes \Theta_- ,&\\
&\{\bar{\rho}^{(i)}(x) \otimes,\;\Theta_-\} = - \frac{1}{2} C_0
\bar{\rho}^{(i)}(x) \otimes \Theta_- ,&\\
&\{\bar{S} \otimes,\;\Theta_-\} = - \frac{1}{2}
\left( C_0 + \bar{S} \otimes 1
C_0 \bar{S}^{-1} \otimes 1\right) \bar{S} \otimes \Theta_-.&
\end{eqnarray*}

As did in the last section, we can calculate the Poisson brackets for
$g$ and $\bar{g}$ starting from those for $S$ and $\bar{S}$. Thus we have

\noindent {\bf Proposition 2.4.4}

\begin{eqnarray*}
\{g \otimes,\;\Theta_+\} g^{-1} \otimes 1 = \frac{1}{2} 1 \otimes \Theta_+
{\rm coth} \left( \pi {\rm ad}_1 {\cal K}_+^{(0)} \right)
\left[ g\otimes 1 C_0 g^{-1} \otimes 1 - C_0 \right],\\
\{\bar{g} \otimes,\;\Theta_-\} \bar{g}^{-1} \otimes 1 = \frac{1}{2}
1 \otimes \Theta_-
{\rm coth} \left( \pi {\rm ad}_1 {\cal K}_-^{(0)} \right)
\left[ \bar{g}\otimes 1 C_0 \bar{g}^{-1} \otimes 1 - C_0 \right].
\end{eqnarray*}

\noindent Moreover, using Propositions 2.4.2-2.4.4 and by straightforward
calculations we obtain

\noindent {\bf Proposition 2.4.5}

\begin{eqnarray*}
& &\{\mu^{(i)}(x) \otimes,\; \Theta_+\} = \frac{1}{2}
\mu^{(i)}(x) \otimes \Theta_+ \left[ g \otimes 1 C_0
g^{-1} \otimes 1 \right.\\
& &~~~~- \left. {\rm coth} \left( \pi {\rm ad}_1 {\cal K}_+^{(0)} \right)
\left( g\otimes 1 C_0 g^{-1} \otimes 1 - C_0 \right) \right] ,\\
& &\{\nu^{(i)}(x) \otimes,\; \Theta_+\} = \frac{1}{2}
\nu^{(i)}(x) \otimes \Theta_+ \left[ g \otimes 1 C_0
g^{-1} \otimes 1 \right.\\
& &~~~~- \left. {\rm coth} \left( \pi {\rm ad}_1 {\cal K}_+^{(0)} \right)
\left( g\otimes 1 C_0 g^{-1} \otimes 1 - C_0 \right) \right] ,
\end{eqnarray*}

\noindent {\bf Proposition 2.4.6}

\begin{eqnarray*}
& &\{\bar{\mu}^{(i)}(x) \otimes,\; \Theta_-\} = - \frac{1}{2}
\left[ \bar{g} \otimes 1 C_0
\bar{g}^{-1} \otimes 1 \right.\\
& &~~~~- \left. {\rm coth} \left( \pi {\rm ad}_1 {\cal K}_-^{(0)} \right)
\left( \bar{g}\otimes 1 C_0 \bar{g}^{-1} \otimes 1 - C_0 \right) \right] \\
& &~~~~\times \bar{\mu}^{(i)}(x) \otimes \Theta_- ,\\
& &\{\bar{\nu}^{(i)}(x) \otimes,\; \Theta_-\} = - \frac{1}{2}
\left[ \bar{g} \otimes 1 C_0
\bar{g}^{-1} \otimes 1 \right.\\
& &~~~~- \left. {\rm coth} \left( \pi {\rm ad}_1 {\cal K}_-^{(0)} \right)
\left( \bar{g}\otimes 1 C_0 \bar{g}^{-1} \otimes 1 - C_0 \right) \right] \\
& &~~~~\times \bar{\nu}^{(i)}(x) \otimes \Theta_- .
\end{eqnarray*}

Let us remark here that all the Poisson brackets from Section 2 downward
{\em will not be affected} if we impose the constraint condition

\begin{equation}
{\cal K}_+^{(0)} = {\cal K}_-^{(0)} \equiv {\cal K}^{(0)},
{}~~~~{\cal Q}_+ = - {\cal Q}_- \equiv {\cal Q}.\label{25}
\end{equation}

\noindent However, when eq.(\ref{25}) is valid, the cross Piosson brackets
such as $\{\mu^{(i)}(x) $ $\otimes,\;$ $\Theta_-\}$ {\em etc} will become
nonvanishing. Actually, they do not yield a simple form as the non-cross
ones (which are not affected by the condition (\ref{25}) because of the
contributions from the Poisson brackets like

\begin{eqnarray*}
\{Q_+(x)\otimes,\; {\rm e}^{ - {\cal Q}_-}\} =
\frac{1}{2\pi} Q_+(x) \otimes {\rm e}^{ - {\cal Q}_- }
\int_0^x {\rm d}z Q_+^{-1}(z)\otimes 1 C_0 Q_+(z)\otimes 1
\end{eqnarray*}

\noindent which follow from eqs.(\ref{24}) and (\ref{25}). We stress that
the integrations over $z$ as in the above equation are unavoidable
for the cross Poisson brackets just mentioned. In order to cure this,
we consider instead of $\Theta_+$ and $\Theta_-$ the
diagonal matrix

\begin{eqnarray*}
{\cal D} &\equiv& {\rm exp} ({\cal Q} - {\cal K}_+ + {\cal K}_- )\\
&=& \Theta_+ {\rm e}^{{\cal K}_-} = \Theta_- {\rm e}^{-{\cal K}_+}.
\end{eqnarray*}

\noindent We can easily show that

\bigskip
\noindent {\bf Proposition 2.4.7} {\em While eq.(\ref{25}) is imposed, all
the Poisson brackets given in Propositions 2.4.2-2.4.6 will remain valid
if we replace everywhere by ${\cal D}$ the original matrices $\Theta_+$
and $\Theta_-$.}

\bigskip
\noindent This last proposition not only assures the correctness of
Propositions 2.4.2-2.4.6 after the constraint (\ref{25}) is imposed but also
provide the explicit results for the nonvanishing cross Poisson brackets.

\section{Local and periodic solutions of the 2EPCT fields}
\index{solution ! local and periodic}
This section is devoted to the construction of local and periodic solutions
of the 2EPCT fields. To do this it may be convenient to rewrite
the equations of motion in the component form,

\be
\partial_+\partial_- \phi^j - \sum_{i,k}{\rm sign}(i-j){\rm sign}(k-j)
\psi_+^iA_{ij}\psi_-^kA_{kj}w^j + \sum_{i (i \ne j)}w^i w^j A_{ij} = 0,
\label{26}
\ee
\be
\partial_- \psi_+^j = \sum_{i} {\rm sign}(i-j) \psi_-^i A_{ij} w^j,
{}~~~~\partial_+ \psi_-^j = \sum_{i} {\rm sign}(i-j) \psi_+^i A_{ij} w^j,
\label{27}
\ee
\be
w^j = {\rm exp}(- \sum_{i} \phi^i A_{ij}) = \prod_{i} \left[
{\rm e}^{\phi^i(x)}\right]^{-A_{ij}}.\label{28}
\ee

\noindent We have

\noindent {\bf Proposition 2.5.1}
{\em The fields}
\begin{eqnarray}
& &{\rm e}^{\phi^i (x)}  \mu^{(i)} (x) {\cal D}
\bar{\mu}^{(i)}(x), \label{29}\\
& &\psi_+^i (x) = \frac{\nu^{(i)}(x) {\cal D} \bar{\mu}^{(i)}(x)}{
\mu^{(i)} (x) {\cal D} \bar{\mu}^{(i)}(x)},\label{30}\\
& &\psi_-^i (x) = \frac{\mu^{(i)}(x) {\cal D} \bar{\nu}^{(i)}(x)}{
\mu^{(i)} (x) {\cal D} \bar{\mu}^{(i)}(x)} \label{31}
\end{eqnarray}

\noindent {\em satisfy the following statements,}

{\em
\begin{enumerate}
\item they are solutions of the equations of motion for the 2EPCT fields,
\item they are periodic,
\item they are local, i.e. Poisson commute with themselves
\end{enumerate}
}

\noindent {\em provided the constraint condition (\ref{25}) is imposed.}

\noindent Proof:

1. We take eq.(\ref{26}) as an example. The formula (\ref{29})
can be rewritten

\begin{displaymath}
\phi^i (x) = {\rm ln}
\left[ \mu^{(i)} (x) \Theta_+ \Theta_- \bar{\mu}^{(i)}(x) \right].
\end{displaymath}

\noindent By direct calculation one can show that

\begin{eqnarray*}
\partial_+\partial_- \phi^i(x) =
\frac{
{\rm det} \left(
\begin{array}{ccc}
&$$\mu^{(i)} (x) \Theta_+ \Theta_- \bar{\mu}^{(i)}(x)$$ &
$$\mu^{(i)} (x) \Theta_+ \Theta_- \partial_-
\bar{\mu}^{(i)}(x)$$ \cr
&$$\partial_+\mu^{(i)} (x) \Theta_+ \Theta_- \bar{\mu}^{(i)}(x)$$ &
$$\partial_+ \mu^{(i)} (x) \Theta_+ \Theta_- \partial_-
\bar{\mu}^{(i)}(x)$$
\end{array}
\right)
}{\left( \mu^{(i)} (x) \Theta_+ \Theta_- \bar{\mu}^{(i)}(x) \right)^{2}}.
\end{eqnarray*}

\noindent Denote by $\Delta$ the determinant in the numerator
of the above equation and write

\begin{displaymath}
G = Q_+ g^{-1} \Theta_+ \Theta_- \bar{g} Q_-,
\end{displaymath}

\noindent eq.(\ref{26}) can be rewritten as ($\lambda^{(i)} \equiv
\lambda_{max}^{(i)}$)

\begin{eqnarray}
&\displaystyle \Delta - \sum_{i,k}{\rm sign}(i-j){\rm sign}(k-j) A_{ij}A_{kj}
\frac{
\langle \lambda^{(i)} |E_i  {\rm e}^{P_+} G
| \lambda^{(i)}\rangle }{
\langle \lambda^{(i)} | G | \lambda^{(i)}\rangle }  & \nonumber\\
&\displaystyle
\times \frac{
\langle \lambda^{(k)} | G {\rm e}^{P_-}
F_k | \lambda^{(k)}\rangle }{
\langle \lambda^{(k)} | G | \lambda^{(k)}\rangle }
\prod_{l (l \ne j)} \left[
\langle \lambda^{(l)} | G | \lambda^{(l)}\rangle
\right]^{-A_{lj}}& \label{32} \\
&\displaystyle
+ \sum_{i (i \ne j)} A_{ij}
\prod_{l (l \ne j)} \left[
\langle \lambda^{(l)} | G | \lambda^{(l)}\rangle
\right]^{-A_{lj}}
\prod_{m} \left[
\langle \lambda^{(m)} | G | \lambda^{(m)}\rangle
\right]^{-A_{mi}}
= 0.& \nonumber
\end{eqnarray}

\noindent In order to prove this equation, we have to calculate the
determinant $\Delta$ first. It follows that

\begin{eqnarray*}
\Delta &=& \langle \lambda^{(i)} | G | \lambda^{(i)}\rangle
\langle \lambda^{(i)} | L_+ G L_- | \lambda^{(i)}\rangle
- \langle \lambda^{(i)} | L_+ G | \lambda^{(i)}\rangle
\langle \lambda^{(i)} | G L_- | \lambda^{(i)}\rangle \\
& =& \langle \lambda^{(i)} | G | \lambda^{(i)}\rangle
\langle \lambda^{(i)} | \bar{P}_+ G \bar{P}_-
| \lambda^{(i)}\rangle
- \langle \lambda^{(i)} | \bar{P}_+ G | \lambda^{(i)}\rangle
\langle \lambda^{(i)} | G \bar{P}_- | \lambda^{(i)}\rangle \\
& +& \langle \lambda^{(i)} | G | \lambda^{(i)}\rangle
\langle \lambda^{(i)} | \bar{P}_+ G {\cal E}_-
| \lambda^{(i)}\rangle
- \langle \lambda^{(i)} | \bar{P}_+ G | \lambda^{(i)}\rangle
\langle \lambda^{(i)} | G {\cal E}_- | \lambda^{(i)}\rangle \\
& + &\langle \lambda^{(i)} | G | \lambda^{(i)}\rangle
\langle \lambda^{(i)} | {\cal E}_+ G \bar{P}_-
| \lambda^{(i)}\rangle
- \langle \lambda^{(i)} | {\cal E}_+ G | \lambda^{(i)}\rangle
\langle \lambda^{(i)} | G \bar{P}_- | \lambda^{(i)}\rangle \\
& +& \langle \lambda^{(i)} | G | \lambda^{(i)}\rangle
\langle \lambda^{(i)} | {\cal E}_+ G {\cal E}_-
| \lambda^{(i)}\rangle
- \langle \lambda^{(i)} | {\cal E}_+ G | \lambda^{(i)}\rangle
\langle \lambda^{(i)} | G {\cal E}_- | \lambda^{(i)}\rangle.
\end{eqnarray*}

\noindent Substituting the definitions of $\bar{P}_\pm$ and ${\cal E}_\pm$
into the above equation, we are lead to

\begin{eqnarray}
\Delta &=& \frac{1}{2} \sum_{i,k}{\rm sign}(i-j){\rm sign}(k-j)
p_+^iA_{ij}p_-^kA_{kj} \langle \Lambda^{j} | G\otimes G | \Lambda^{j}
\rangle \nonumber\\
&+ &\frac{1}{2} \sum_{i,k}{\rm sign}(i-j) {\rm sign}(j-k) p_+^i A_{ij}
\langle \Lambda^{j} | G\otimes G | \Xi^{j,k}
\rangle \nonumber\\
&+& \frac{1}{2} \sum_{i,k}{\rm sign}(j-i) {\rm sign}(k-j) p_-^k A_{kj}
\langle \Xi^{j,i} | G\otimes G | \Lambda^{j}
\rangle \nonumber\\
&+ &\frac{1}{2} \sum_{i,k}{\rm sign}(i-j) {\rm sign}(k-j)
\langle \Xi^{j,i} | G\otimes G | \Xi^{j,k}
\rangle, \label{33}
\end{eqnarray}

\noindent where we have defined

\begin{eqnarray}
| \Lambda^{j} \rangle &=& | \lambda^{(j)} \rangle \otimes F_j
| \lambda^{(j)} \rangle - F_j | \lambda^{(j)} \rangle \otimes
| \lambda^{(j)} \rangle \nonumber\\
| \Xi^{j,k} \rangle &=& | \lambda^{(j)} \rangle \otimes [ F_k,\;F_j]
| \lambda^{(j)} \rangle - [ F_k,\;F_j]  | \lambda^{(j)} \rangle \otimes
| \lambda^{(j)} \rangle \nonumber \\
&=& (F_k \otimes 1 + 1\otimes F_k ) |\Lambda^{j} \rangle \label{34}
\end{eqnarray}

\noindent and, consequently,

\begin{eqnarray*}
\langle \Lambda^{j} | &=& \langle \lambda^{(j)} | \otimes
\langle \lambda^{(j)} | E_j - \langle \lambda^{(j)} | E_j  \otimes
\langle \lambda^{(j)} | \\
\langle \Xi^{j,k} | &=& \langle \lambda^{(j)} | \otimes
\langle \lambda^{(j)} |  [ E_j,\;E_k]
- \langle \lambda^{(j)} |  [ E_j,\;E_k] \otimes  \langle \lambda^{(j)} | \\
&=& \langle \Lambda^{j} | (E_k \otimes 1 + 1\otimes E_k ).
\end{eqnarray*}

\noindent It is clear that only a few of $|\Xi^{j,k}\rangle$ are
nonvanishing. To identify which is vanishing and which is not, we
notice that for simply-laced Lie algebras, $A_{ij}$ is either equal to
$0$ or equal to $-1$ for $i \ne j$. Moreover, $ [ E_i,\; E_j ] $
does not vanish if and only if $A_{ij}$ does not. So we can multiply
a $-A_{jk}$ factor to the definition of $| \Xi^{j,k} \rangle$ without
changing the content of the state.

It was shown in Ref.[1] that $|\Lambda^{j} \rangle$ is the
highest weight state of some tensorial product representation of the Lie
algebra ${\cal G}$. Furthermore, it can be set equivalent to the
state $\sqrt{2} \bigotimes_{k \ne j} |\lambda^{(k)} \rangle^{\otimes
(-A_{kj})}$ since they yield the same highest weight,

\begin{eqnarray*}
 ( H_i \otimes 1 + 1 \otimes H_i ) |\Lambda^{j} \rangle
&=& ( 2 \delta_{ji} - A_{ji}) |\Lambda^{j} \rangle \\
&\equiv& \langle \Lambda^{j},\;\alpha^i \rangle |\Lambda^{j} \rangle,\\
 ( H_i \otimes 1 + 1 \otimes H_i )
\sqrt{2} \bigotimes_{k \ne j} |\lambda^{(k)} \rangle^{\otimes (-A_{kj})}
&=& ( 2 \delta_{ji} - A_{ji})
\sqrt{2} \bigotimes_{k \ne j} |\lambda^{(k)} \rangle^{\otimes (-A_{kj})}
\end{eqnarray*}

\noindent and the same norm,

\begin{eqnarray*}
\langle \Lambda^{j}|\Lambda^{j} \rangle
=\left(
\sqrt{2} \bigotimes_{k \ne j} \langle \lambda^{(k)} |^{\otimes (-A_{kj})}
\right)\left(
\sqrt{2} \bigotimes_{k \ne j} |\lambda^{(k)} \rangle^{\otimes (-A_{kj})}
\right) = 2.
\end{eqnarray*}

\noindent Accordingly, eq.(\ref{34}) shows that $|\Xi^{j,k}
\rangle$ is the next to highest weight state in the representation
of ${\cal G}$ characterized by the highest weight $\Lambda^{j}$,

\begin{eqnarray*}
& &|\Xi^{j,k} \rangle = (F_k \otimes 1 + 1 \otimes F_k )
\sqrt{2} \bigotimes_{k \ne j} |\lambda^{(k)} \rangle^{\otimes (-A_{kj})},\\
& &\langle \Xi^{j,i} | \Xi^{j,k} \rangle = 2 \delta_{ik} (2 \delta_{kj} -
A_{kj}).
\end{eqnarray*}

Following the above discussions, it is now easy to obtain

\begin{eqnarray}
& &\frac{1}{2} \langle \Lambda^{j} | G\otimes G | \Lambda^{j} \rangle
= \prod_{l (l \ne j)} \left[
\langle \lambda^{(l)} | G | \lambda^{(l)}\rangle
\right]^{-A_{lj}}, \label{35}\\
& &\frac{1}{2} \langle \Lambda^{j} | G\otimes G | \Xi^{j,k} \rangle
= - A_{jk} \frac{ \langle \lambda^{(k)} | G F_k | \lambda^{(k)}\rangle
}{
\langle \lambda^{(k)} | G | \lambda^{(k)}\rangle }
\prod_{l (l \ne j)} \left[
\langle \lambda^{(l)} | G | \lambda^{(l)}\rangle
\right]^{-A_{lj}} , \label{36}\\
& &\frac{1}{2} \langle \Xi^{j,i} | G\otimes G | \Lambda^{j} \rangle
= - A_{ji} \frac{
\langle \lambda^{(i)} | E_i G | \lambda^{(i)}\rangle
}{
\langle \lambda^{(i)} | G | \lambda^{(i)}\rangle
} \prod_{l (l \ne j)} \left[
\langle \lambda^{(l)} | G | \lambda^{(l)}\rangle
\right]^{-A_{lj}} , \label{37}\\
& &\frac{1}{2} \langle \Xi^{j,i} | G\otimes G | \Xi^{j,k} \rangle
= A_{ji} A_{jk} \frac{
\langle \lambda^{(i)} | E_i G | \lambda^{(i)}\rangle
}{
\langle \lambda^{(i)} | G | \lambda^{(i)}\rangle
}
\frac{
\langle \lambda^{(k)} | G F_k | \lambda^{(k)}\rangle
}{
\langle \lambda^{(k)} | G | \lambda^{(k)}\rangle
} \nonumber\\
& &~~~~\times \prod_{l (l \ne j)} \left[
\langle \lambda^{(l)} | G | \lambda^{(l)}\rangle
\right]^{-A_{lj}} ~~~~ ( i \ne k )  \label{38}
\end{eqnarray}

\noindent and

\begin{eqnarray}
& &\frac{1}{2} \langle \Xi^{j,k} | G\otimes G | \Xi^{j,k} \rangle
= ( A_{jk} )^2 \frac{
\langle \lambda^{(k)} | E_k G F_k | \lambda^{(k)}\rangle
}{
\langle \lambda^{(k)} | G | \lambda^{(k)}\rangle
} \prod_{l (l \ne j)} \left[
\langle \lambda^{(l)} | G | \lambda^{(l)}\rangle
\right]^{-A_{lj}} \nonumber\\
& &~~~~= ( A_{jk} )^2 \frac{
\langle \lambda^{(k)} | E_k G | \lambda^{(k)}\rangle
\langle \lambda^{(k)} | G F_k | \lambda^{(k)}\rangle
}{
\langle \lambda^{(k)} | G | \lambda^{(k)}\rangle ^2
} \prod_{l (l \ne j)} \left[
\langle \lambda^{(l)} | G | \lambda^{(l)}\rangle
\right]^{-A_{lj}} \nonumber\\
& &~~~~- A_{jk} \prod_{m} \left[
\langle \lambda^{(m)} | G | \lambda^{(m)}\rangle
\right]^{-A_{mk}} \prod_{l (l \ne j)} \left[
\langle \lambda^{(l)} | G | \lambda^{(l)}\rangle
\right]^{-A_{lj}} . \label{39}
\end{eqnarray}

\noindent Substituting eqs.(\ref{35}-\ref{39})
into the left hand side of eq.(\ref{32}) it
can be verified that the right hand side vanishes.

The equations of motion for the fields $\psi_\pm^i$ can be verified in
a similar way.

2. The periodicity \index{periodicity}
of the solution (\ref{29}-\ref{31}) is obvious
because that the vectors $\mu^{(i)},\; \nu^{(i)}$ and
$\bar{\mu}^{(i)},\;\bar{\nu}^{(i)}$
have respectively the diagonal monodromy matrix
\index{monodromy ! matrix ! diagonal} $\kappa$ and $\bar{\kappa}$,
and that eq.(\ref{25}) implies $\kappa \bar{\kappa} =1$.

3. The locality of the solution follows from the Poisson brackets

\begin{eqnarray*}
& &\{\tau_{a}^{(i)}(x) {\cal D} \bar{\tau}_{b}^{(i)} (x) \otimes,\;
\tau_{c}^{(i)}(y) {\cal D} \bar{\tau}_{d}^{(i)} (y)\} = 0,\\
& &a,\;b,\;c,\;d =1,\;2,~~~~ \tau_1^{(i)} = \mu^{(i)},
{}~~~~ \tau_2^{(i)} = \nu^{(i)}
\end{eqnarray*}

\noindent which can be obtained using the Propositions given in the last
two sections. Using the same method, we can recover all the
canonical Poisson brackets for the 2EPCT fields, the nontrivial ones
being

\begin{eqnarray}
& &\{\partial_0 \Phi_\pm (x) \otimes, \;\Phi_\pm (y)\} = \delta (x-y)
C_0, \nonumber\\
& &\{\bar{\Psi}_+ (x) \otimes,\; \Psi_+ (y)\} = \delta (x-y) \sum_{i} E_i
\otimes F_i,\nonumber\\
& &\{\bar{\Psi}_- (x) \otimes,\; \Psi_- (y)\} = -\delta (x-y) \sum_{i} F_i
\otimes E_i.\nonumber
\end{eqnarray}

\noindent Such calculations are not presented here because
they are considerably tedious and long. Q.E.D.

\section{The $(\chi,\;\omega)$-$(\bar{\chi},\;\bar{\omega})$ basis}
The exchange algebra given in Sections 2 and 3 do not contain the conjugate
$K_\pm$-zero modes ${\cal Q}_\pm$. In practice, however, it is possible
to reformulate the exchange algebra in terms of bases containing these
quantities. The $(\chi,\;\omega)$-$(\bar{\chi},\;\bar{\omega})$ basis
to be studied in this section is just one of such bases.

Let us define

\begin{eqnarray*}
\chi^{(i)} = \mu^{(i)} \Theta_+,&~~~~&
\bar{\chi}^{(i)} = \Theta_- \bar{\mu}^{(i)} ,\\
\omega^{(i)} = \nu^{(i)} \Theta_+,&~~~~&
\bar{\omega}^{(i)} = \Theta_- \bar{\nu}^{(i)} ,
\end{eqnarray*}

\noindent we can get, via straightforward calculations, the following

\noindent {\bf Proposition 2.6.1} {\em The exchange algebra reformulated in
terms of the $(\chi,\;\omega)$-$(\bar{\chi},\;\bar{\omega})$ basis
takes the form (while ${\cal K}_\pm^{(0)}$ are considered as independent)
}

\begin{eqnarray*}
\displaystyle
& &\{\chi^{(i)}(x) \otimes,\;\chi^{(j)}(y)\} = \frac{1}{4}
\chi^{(i)}(x) \otimes \chi^{(j)}(y) \left\{ (r_+ - r_-) {\rm sign} (x-y)
\right.\\
\displaystyle
& &~~~~- {\rm coth} \left( \pi {\rm ad}_1 {\cal K}_+^{(0)}\right) (r_+ - C_0)
- {\rm coth} \left( \pi {\rm ad}_2 {\cal K}_+^{(0)}\right)
\left. (r_- + C_0) \right\},\\
& &\{\chi^{(i)}(x) \otimes,\;\omega^{(j)}(y)\} = \frac{1}{4}
\chi^{(i)}(x) \otimes \omega^{(j)}(y) \left\{ (r_+ - r_-) {\rm sign} (x-y)
\right.\\
\displaystyle
& &~~~~- {\rm coth} \left( \pi {\rm ad}_1 {\cal K}_+^{(0)}\right) (r_+ - C_0)
- {\rm coth} \left( \pi {\rm ad}_2 {\cal K}_+^{(0)}\right)
\left. (r_- + C_0) \right\},\\
& &\{\omega^{(i)}(x) \otimes,\;\omega^{(j)}(y)\} = \frac{1}{4}
\omega^{(i)}(x) \otimes \omega^{(j)}(y) \left\{ (r_+ - r_-) {\rm sign} (x-y)
\right.\\
\displaystyle
& &~~~~- {\rm coth} \left( \pi {\rm ad}_1 {\cal K}_+^{(0)}\right) (r_+ - C_0)
- {\rm coth} \left( \pi {\rm ad}_2 {\cal K}_+^{(0)}\right)
\left. (r_- + C_0) \right\}\\
& &~~~~+ (M^{-1})_{ji} \delta (x-y) \chi^{(i)} (x)\otimes
\chi^{(j)} (y),
\end{eqnarray*}

\begin{eqnarray*}
\displaystyle
& &\{\bar{\chi}^{(i)}(x) \otimes,\;\bar{\chi}^{(j)}(y)\} = - \frac{1}{4}
\left\{ (r_+ - r_-) {\rm sign} (x-y) \right.\\
& &~~~~+ {\rm coth} \left( \pi {\rm ad}_1 {\cal K}_-^{(0)}\right) (r_- + C_0)
\\
\displaystyle
& &~~~~+\left. {\rm coth} \left( \pi {\rm ad}_2 {\cal K}_-^{(0)}\right)
(r_- + C_0) \right\} \bar{\chi}^{(i)}(x) \otimes \bar{\chi}^{(j)}(y),\\
& &\{\bar{\chi}^{(i)}(x) \otimes,\;\bar{\omega}^{(j)}(y)\} = - \frac{1}{4}
\left\{ (r_+ - r_-) {\rm sign} (x-y) \right.\\
& &~~~~+ {\rm coth} \left( \pi {\rm ad}_1 {\cal K}_-^{(0)}\right) (r_- + C_0)
\\
\displaystyle
& &~~~~+\left. {\rm coth} \left( \pi {\rm ad}_2 {\cal K}_-^{(0)}\right)
(r_- + C_0) \right\}\bar{\chi}^{(i)}(x) \otimes \bar{\omega}^{(j)}(y),\\
& &\{\bar{\omega}^{(i)}(x) \otimes,\;\bar{\omega}^{(j)}(y)\}= - \frac{1}{4}
\left\{ (r_+ - r_-) {\rm sign} (x-y) \right.\\
& &~~~~+ {\rm coth} \left( \pi {\rm ad}_1 {\cal K}_-^{(0)}\right) (r_- + C_0)
\\
\displaystyle
& &~~~~+\left. {\rm coth} \left( \pi {\rm ad}_2 {\cal K}_-^{(0)}\right)
(r_- + C_0) \right\}\bar{\omega}^{(i)}(x) \otimes \bar{\omega}^{(j)}(y)\\
& &~~~~- (M^{-1})_{ji} \delta (x-y) \bar{\chi}^{(i)} (x)\otimes
\bar{\chi}^{(j)} (y),
\end{eqnarray*}

\begin{eqnarray*}
& &\{\chi^{(i)}(x) \otimes,\;\bar{\chi}^{(j)}(y)\} = 0,\\
& &\{\chi^{(i)}(x) \otimes,\;\bar{\omega}^{(j)}(y)\} = 0,\\
& &\{\omega^{(i)}(x) \otimes,\;\bar{\chi}^{(j)}(y)\} = 0,\\
& &\{\omega^{(i)}(x) \otimes,\;\bar{\omega}^{(j)}(y)\} = 0.
\end{eqnarray*}

\noindent Notice that, while the condition (\ref{25}) is imposed, the
last few Poisson brackets (the cross ones) will nolonger vanish.
In order to evaluate these cross Poisson brackets, let us proceed to
replace the $\Theta_\pm$'s in the definition of $\chi^{(i)}$ and
$\omega^{(i)}$ by the constant matrix ${\cal D}$.
It can be easily seen that all the non-cross Poisson brackets
in Proposition 2.6.1 are not affected by this redefinition of basis, whereas,
for the cross Poisson brackets, we have

\noindent {\bf Proposition 2.6.2}

\begin{eqnarray*}
\displaystyle
& &\{\chi^{(i)}(x) \otimes,\;\bar{\chi}^{(j)}(y)\} =
\frac{1}{2} \chi^{(i)}(x) \otimes 1 \left\{ \left[
1 - {\rm coth} \left(\pi {\rm ad}_1 {\cal K}^{(0)} \right) \right] \right.\\
& &~~~~\times \left({\cal D}^{-1}g \otimes 1 C_0 g^{-1} {\cal D} \otimes 1
- C_0 \right) \\
\displaystyle
& &~~~~+ \left. \left[ 1 - {\rm coth}
\left(\pi {\rm ad}_2 {\cal K}^{(0)} \right) \right]
\left(1 \otimes {\cal D} \bar{g} C_0 1 \otimes \bar{g}^{-1} {\cal D}^{-1}
- C_0 \right) + 2C_0 \right\} 1\otimes \bar{\chi}^{(j)},\\
& &\{\chi^{(i)}(x) \otimes,\;\bar{\omega}^{(j)}(y)\} =
\frac{1}{2} \chi^{(i)}(x) \otimes 1 \left\{ \left[
1 - {\rm coth} \left(\pi {\rm ad}_1 {\cal K}^{(0)} \right) \right] \right.\\
& &~~~~\times \left({\cal D}^{-1}g \otimes 1 C_0 g^{-1} {\cal D} \otimes 1
- C_0 \right) \\
\displaystyle
& &~~~~+ \left. \left[ 1 - {\rm coth}
\left(\pi {\rm ad}_2 {\cal K}^{(0)} \right) \right]
\left(1 \otimes {\cal D} \bar{g} C_0 1 \otimes \bar{g}^{-1} {\cal D}^{-1}
- C_0 \right) + 2C_0 \right\} 1\otimes \bar{\omega}^{(j)},\\
& &\{\omega^{(i)}(x) \otimes,\;\bar{\chi}^{(j)}(y)\} =
\frac{1}{2} \omega^{(i)}(x) \otimes 1 \left\{ \left[
1 - {\rm coth} \left(\pi {\rm ad}_1 {\cal K}^{(0)} \right) \right] \right. \\
& &~~~~\times \left({\cal D}^{-1}g \otimes 1 C_0 g^{-1} {\cal D} \otimes 1
-C_0 \right) \\
\displaystyle
& &~~~~+ \left. \left[ 1 - {\rm coth}
\left(\pi {\rm ad}_2 {\cal K}^{(0)} \right) \right]
\left(1 \otimes {\cal D} \bar{g} C_0 1 \otimes \bar{g}^{-1} {\cal D}^{-1}
-C_0 \right) + 2C_0 \right\} 1\otimes \bar{\chi}^{(j)},\\
& &\{\omega^{(i)}(x) \otimes,\;\bar{\omega}^{(j)}(y)\} =
\frac{1}{2} \omega^{(i)}(x) \otimes 1 \left\{ \left[
1 - {\rm coth} \left(\pi {\rm ad}_1 {\cal K}^{(0)} \right) \right] \right.\\
& &~~~~\times \left({\cal D}^{-1}g \otimes 1 C_0 g^{-1} {\cal D} \otimes 1
- C_0 \right) \\
& &~~~~+ \left. \left[ 1 - {\rm coth} \left(\pi {\rm ad}_2
{\cal K}^{(0)} \right) \right]
\left(1 \otimes {\cal D} \bar{g} C_0 1 \otimes \bar{g}^{-1} {\cal D}^{-1}
- C_0 \right) + 2C_0 \right\} 1\otimes \bar{\omega}^{(j)}.
\end{eqnarray*}

Before ending this section it should be remarked that there can be an
infinite number of choices for the base vectors of the chiral exchange
algebra, and each choice has its own advantages and disadvantages.
The $(\sigma,\;\rho)$-$(\bar{\sigma},\;\bar{\rho})$ basis is the simplest
one for the calculation of exchange relations. The $(\mu,\;\nu)
$-$(\bar{\mu},\;\bar{\nu})$ basis is most appropriate for the chiral
splitting, and the $(\chi,\;\omega)$-$(\bar{\chi},\;\bar{\omega})$ basis
is the most viable one for studing the conformal transformations.
\footnote{This basis is the analogue of the $\psi-\bar{\psi}$ basis of the
exchange algebra of the standard Toda theory, which transforms covariantly
under the stress-energy tensor.}

\section{Relating the $(\xi$-$\bar{\xi})$ basis}
So far we have not pay a word on the relations between the exchange relations
and general solution of 2EPCT fields obtained from free fields and those
given in the introduction in terms of the $(\xi$-$\bar{\xi})$ basis. Now let
us give a brief discussion on such relations.

In Ref.[14] we proved that the $(\xi$-$\bar{\xi})$ basis can be expressed as

\begin{eqnarray*}
\displaystyle
\xi_1^{(i)}(x) &=& \langle \lambda_{max}^{(i)} | {\rm e}^{K_+(x)}M_+(x),\\
\displaystyle
\xi_2^{(i)}(x) &=& \langle \lambda_{max}^{(i)} | E_i {\rm e}^{P_+(x)}
{\rm e}^{K_+(x)}M_+(x),\\
\displaystyle
\bar{\xi}_1^{(i)}(x) &=& M_-^{-1}(x){\rm e}^{-K_-(x)}
| \lambda_{max}^{(i)} \rangle,\\
\displaystyle
\bar{\xi}_2^{(i)}(x) &=& M_-^{-1}(x){\rm e}^{-K_-(x)} {\rm e}^{P_-(x)}
F_i | \lambda_{max}^{(i)} \rangle,
\end{eqnarray*}

\noindent where $K_\pm \in {\cal G}_0,\; P_\pm \in {\cal G}_\mp^{(1)},\;
M_\pm \in {\rm exp} ({\cal G}_\pm )$ are respectively chiral and antichiral
vectors indicated by their subscripts, and

\begin{eqnarray*}
\displaystyle
\partial_+M_+(x)M_+(x)^{-1} &=& {\rm e}^{- {\rm ad} K_+(x)} (\bar{P}_+(x)
+ {\cal E}_+ ),\\
\displaystyle
M_-(x)\partial_-M_-^{-1}(x) &=& {\rm e}^{- {\rm ad} K_-(x)} (\bar{P}_+(x)
+ {\cal E}_+ ).
\end{eqnarray*}

\noindent The last equations can be gauge-transformed as $M_\pm \rightarrow
{\rm e}^{K_\pm} M_\pm$, yielding

\begin{eqnarray*}
\displaystyle
\partial_+({\rm e}^{K_+(x)} M_+(x))&=& (\partial_+ K_+(x) + \bar{P}_+(x)
+ {\cal E}_+ )({\rm e}^{K_+(x)} M_+(x)),\\
\displaystyle
\partial_-(M_-^{-1}(x){\rm e}^{-K_-(x)} )&=&(M_-^{-1}(x){\rm e}^{-K_-(x)} )
(\partial_- K_-(x) + \bar{P}_-(x)
+ {\cal E}_- ).
\end{eqnarray*}

\noindent These equations looks almost like the same as the DS systems
(\ref{9}-\ref{11}). However, there is a crucial difference,
namely the different normalizations. As is mentioned before, the $Q_\pm(x)$
in the DS systems (\ref{9}-\ref{10}) are normalized
as $Q_\pm(0) = 1$ so that they do not contain the conjugates of the zero
modes. But now ${\rm e}^{K_\pm(x)} M_\pm(x)$ are not so normalized, so they
must contain the conjugates of the zero modes. This means that the matrices
${\rm e}^{K_\pm(x)} M_\pm(x)$ differ from $Q_\pm(x)$ by the initial values
involving the conjugates of zero modes,

\begin{eqnarray*}
\displaystyle
Q_+(x) &=& {\rm e}^{K_+(x)} M_+(x)M_+^{-1}(0) {\rm e}^{- K_+(0)},\\
\displaystyle
Q_-(x) &=& {\rm e}^{K_-(0)} M_-(0)M_-^{-1}(x) {\rm e}^{- K_-(x)}.
\end{eqnarray*}

\noindent Consequently the $(\sigma,\;\rho)$-$(\bar{\sigma},\;\bar{\rho})$
basis is related to the $(\xi$-$\bar{\xi})$ basis as

\begin{eqnarray*}
\displaystyle
\sigma^{(i)}(x) &=& \xi_1^{(i)}(x)M_+^{-1}(0){\rm e}^{-K_+(0)},\\
\displaystyle
\rho^{(i)}(x) &=& \xi_2^{(i)}(x)M_+^{-1}(0){\rm e}^{-K_+(0)},\\
\displaystyle
\bar{\sigma}^{(i)}(x) &=& {\rm e}^{K_-(0)}M_-(0)\bar{\xi}_1^{(i)}(x),\\
\displaystyle
\bar{\rho}^{(i)}(x) &=& {\rm e}^{K_-(0)}M_-(0)\bar{\xi}_2^{(i)}(x).
\end{eqnarray*}

\noindent Moreover, the monodromy matrices $S$ and $\bar{S}$ are nothing but
\index{monodromy ! matrix}

\begin{eqnarray*}
\displaystyle
 S &=& {\rm e}^{K_+(2\pi)}M_+(2\pi)M_+^{-1}(0) {\rm e}^{- K_+(0)},\\
\displaystyle
 \bar{S}&=& {\rm e}^{K_-(0)}M_-(0)M_-^{-1}(2\pi) {\rm e}^{- K_-(2\pi)}.
\end{eqnarray*}

\noindent We thus see that the nontrivial couplings between the left and
right moving sectors under the $(\xi$-$\bar{\xi})$ basis is in fact the
consequence of different choice for the initial values of the chiral vectors.
For the sake of length we shall not fall into detailed calculations
on the relations of the exchange algebras under the
$(\sigma,\;\rho)$-$(\bar{\sigma},\;\bar{\rho})$
basis and the $(\xi$-$\bar{\xi})$ basis.

\section{Discussions}
In this chapter we constructed the exchange algebra and the local and periodic
solutions of the 2EPCT theory based on simply-laced
even-rank Lie algebras via free
field representation. More concretely, we have done the following:
\begin{enumerate}
\item Starting from the free chiral fields $K_\pm$ and $P_\pm$, we
constructed the exchange algebra under three different set of basis.
Only in the $(\mu,\;\nu)$-$(\bar{\mu},\;\bar{\nu})$ basis the chiralities
are completely splitted when the phase space is reduced by the condition
(\ref{25});
\item We obtained the local and periodic solutions of the 2EPCT fields
using the exchange algebra;
\item We established the connections between the bases used in the present
chapter and the $(\xi$-$\bar{\xi})$ basis considered earlier.
\end{enumerate}

\noindent Much of the results can be directly generalized to the
non-simply-laced case except that in proving the solution formulas we
encountered some complexities which remain to be resolved. The same
construction can also be generalized to the case of two-extended
principal conformal affine Toda models [13] based on simply-laced
affine Lie algebras.

In addition to what have been done we would like to point some
untouched problems.

The first problem is the converse of the classical free field
representations, {\it i.e.} the faithfulness problem of the free field
representation. In the standard Toda cases, it was shown in [2] that
not only one can obtain the exchange algebras and the local and periodic
solutions, but also all the dynamics of Toda theory. That means that
there is a one to one correspondence between the canonical
structures of Toda theory and that of free fields. For the 2EPCT fields
we hope the same is true but we have not worked that out yet.

The next problem is the problem of quantization. Since most of our results
are in analogy to the standard Toda case, and that there is already a
well established quantum theory for the Toda fields by going to the lattice
[3-5], we hope that our theory can also be quantized by defining a
consistence set of lattice exchange algebra. This work is now undertaken.

To end this chapter let us mention that both the standard Toda and the 2EPCT
theories are certain reductions of the WZNW model, therefore the similarity
between our results and those of Ref.[1] is reasonable. Actually,
the parallel problems of the classical and quantum exchange algebras in
the WZNW model have already been considered by
Balog {\em et al} [6], Faddeev [7], Alekseev-Shatashvilli [8]
and Fatteev-Lukuyanov [9-10]. So we believe that the exchange relations must be
the
most appropriate way of passing from the classical integrable systems
to the quantum analogues.

\section*{Appendix: Dirac procedure for calculating the Poisson brackets
for the fields $P_\pm$}
\addcontentsline{toc}{subsection}{Appendix}
\index{Dirac Poisson bracket}

In this appendix we shall show how we can get the Poisson brackets (\ref{b})
using the standard Dirac procedure. We take only the field $P_+$ as example.

{}From the action $S[P_+]$ we can get the Hamiltonian for the fields $p_+^i$,

\begin{eqnarray*}
H &=& \int {\rm d}x \left\{ \sum_{i} \pi (p_+^i) \partial_t p_+^i
- {\cal L}\right\}\\
&=& \int {\rm d}x \left\{ \sum_{i} \pi (p_+^i) \partial_t p_+^i
+ \frac{1}{2}\sum_{ij} M_{ij} p_+^i \partial_- p_+^j \right\},
\end{eqnarray*}

\noindent where $\pi (p_+^j) \equiv \frac{1}{2} \sum_i M_{ij} p_+^i$, which
yield the following {\em primary} constraints,

\begin{displaymath}
G_j \equiv \pi (p_+^j) - \frac{1}{2} \sum_i M_{ij} p_+^i \approx 0.
\end{displaymath}

\noindent The first class Hamiltonian

\begin{displaymath}
H_I \equiv H + \int {\rm d}x \sum_i \lambda^i (x) G_i (x)
\end{displaymath}

\noindent then implies that

\begin{eqnarray*}
\partial_t G_j &\equiv& \{ H_I,\; G_j \} \\
&=& \partial_x \pi (p_+^j) - \frac{1}{2} \sum_i M_{ij}
\partial_x p_+^i + \sum_{i} \lambda^i M_{ji} \approx 0,
\end{eqnarray*}

\noindent which simply determine the values of the Lagrangian multipliers
$\lambda^i$ and do not give rise to any secondary constraints. One can easily
calculate the ``matrix'' $\Delta (x,\;y)$ as follows,

\begin{eqnarray*}
\Delta_{ij} (x, \;y) = \{G_i(x),\; G_j(y)\} = M_{ji} \delta(x-y).
\end{eqnarray*}

\noindent Therefore, whenever $M$ is invertible, there will be no first class
constraints. This is precisely the case if $M$ is obtained as
$M_{ij} = A_{ij} {\rm sign} (i-j) $ and $A$ is the Cartan matrix of an
even-rank Lie algebra. It follows that in such cases the matrix
$\Delta$ is invertible and the inverse reads

\begin{eqnarray*}
(\Delta^{-1})_{ij} (x,\;y) = (M^{-1})_{ji} \delta (x-y).
\end{eqnarray*}

\noindent Moreover, the naive Poisson brackets between the
fields $p_+^i$, $\pi (p_+^i)$ and the constraints $G_j$ read

\begin{eqnarray*}
\{p_+^i(x),\; G_j(y)\} &=& - \delta_{ij} \delta (x-y),\\
\{\pi (p_+^i) (x),\; G_j(y)\} &=& - \frac{1}{2} M_{ij} \delta (x-y),\\
\{G_i(x),\; p_+^j(y)\} &=& \delta_{ij} \delta (x-y),\\
\{G_i(x),\; \pi (p_+^j) (y)\} &=& \frac{1}{2} M_{ji}  \delta (x-y).
\end{eqnarray*}

\noindent Thus by defining the Dirac bracket $\{\ , \ \}^{D}$ as

\begin{eqnarray*}
\{A,\; B\}^{D} = \{A,\; B\} - \int {\rm d}z {\rm d}w \sum_{ij}
\{A,\; G_i(z)\} (\Delta^{-1})_{ij} (z,\;w) \{G_j(w),\;B\}
\end{eqnarray*}

\noindent we find the following consistent results,

\begin{eqnarray*}
\{p_+^i(x),\;p_+^j(y)\}^{D} &=& (M^{-1})_{ji} \delta (x-y),\\
\{\pi (p_+^i)(x),\; p_+^j(y)\}^{D} &=&
\frac{1}{2} \delta_{ij} \delta (x-y),\\
\{\pi (p_+^i)(x),\; \pi (p_+^j) (y)\}^{D} &=&
\frac{1}{4} M_{ij} \delta (x-y).
\end{eqnarray*}

\noindent These last Poisson brackets, while rewritten in terms of the
matrix field $P_+ (x)$, give rise to those in eq.(\ref{b}) exactly,
but in the main context of this chapter we always write $\{\ ,\; \ \}$
instead of $\{\ ,\; \ \}^{D}$ for simplicity.

\chapter{Heterotic Toda Fields}

\section{Introduction}

Of all the conformal invariant integrable models Toda field theories
are the most interesting and extensively studied ones. Within the
framework of Toda  field theories, one finds the conformal Toda (CT)
\index{Toda ! conformal},
loop Toda \index{Toda ! loop}
(or affine Toda \index{Toda ! affine}, denoted as LT for short),
conformal affine Toda \index{Toda ! conformal affine}
(CAT) [1] and their various extensions,
especially the ``2-extensions'' \index{two-extension} [2-3,13]
studied by the authors sometime earlier. One of the major reason for why
conformal invariant Toda fields are so attractive is due to their nice
properties of yielding the W algebra symmetries \index{$W$ algebra}.
Nowadays it is becoming
a common practice to treat the conformally noninvariant Toda theories
as the result of appropriate deformations [4-5]
of the corresponding conformal
invariant ones. In view of this, the integrability of all Toda type
field theories is governed by the conformal (W algebra) symmetries of
the  undeformed Toda fields.

In the study of conformal invariant field theories, people are used
to treat only one-half of the complete model, namely one of the two
chiral sectors. This is because of the left-right symmetry of the model
under consideration. The left-right symmetry is in fact some kind of
``parity'' (P) or ``charge-parity'' (CP) invariance,
\index{CP invariance} which causes
an indistinguishability between both chiral sectors. To be specific,
let us consider the CT and the 2-extended (2-E) CT models. The
equations of motion for these two models can be written respectively as

\begin{displaymath}
\partial_+ \partial_- \varphi^i - \exp \left( -\sum_{j}K_{ij}
\varphi^j \right) =0  ~~~~ i, j=1, 2, \ldots, r
\end{displaymath}

\noindent and

\begin{displaymath}
\partial_+ \partial_- \varphi^j - \sum_{i, j}{\rm sign} (i-j)
{\rm sign} (k-j)
\psi_+^i K_{ij}\psi_-^k K_{kj} \omega^j + \sum_{i, (i \neq j)}\omega^i
\omega^jK_{ij}=0
\end{displaymath}
\begin{displaymath}
\partial_+ \psi_-^j=\sum_{i}{\rm sign} (i-j)\psi_+^iK_{ij}\omega^j
\end{displaymath}
\begin{displaymath}
\partial_- \psi_+^j=\sum_{i}{\rm sign} (i-j)\psi_-^iK_{ij}\omega^j
\end{displaymath}
\begin{displaymath}
\omega^i \equiv \exp \left( \sum_{j}K_{ij}
\varphi^j \right)  ~~~~ \left(i, j, k=1, 2, \ldots, r\right)
\end{displaymath}

\noindent where $K_{ij}$ are entries of the Cartan matrix of the
corresponding rank $r$ finite dimensional Lie algebra (now chosen as
simply-laced), and the signature function ${\rm sign} (i-j) $ is defined
such that it takes the value ``zero'' at equal arguments. One sees that
the left-right symmetry of CT is actually
the P invariance $x_+ \leftrightarrow x_-$, but for the 2-ECT case,
it is represented by the following ``CP invariance''

\begin{equation}
P: x_+ \leftrightarrow x_-, ~~~~ C: \psi_+^i \leftrightarrow
\psi_-^i.     \label{eq: cp}
\end{equation}

\noindent It is not difficult to check that the fields $\psi_\pm^i$
have respectively the conformal weights $\left(1/2, 0\right)$ and
$\left(0, 1/2\right)$, therefore the ``charge conjugation'' in
eq. (\ref{eq: cp}) is to be understood as the conjugation of conformal
charges instead of the normal (electronic) charges.

Despide of the elegant properties of the left-right symmetric models
like CT and 2-ECT mentioned above, there exists something in the
nature which is {\em not} left-right symmetric (such as the neutrinos).
Therefore, exploiting a conformal invariant model having no left-right
symmetry might be interesting. Such theories has already exist in
superstring theories, i.e. the heterotic string theory [6].
But now we are concerned about a Toda
type integrable theory which is also ``heterotic'' \index{heterotic}
(which we call heterotic Toda field theory, denoted HTFT for short).
We shall construct such a model by defining explicitly
its Lax pair \index{Lax pair}
representation, discuss its conformal properties (which
is represented by the product of a left chiral $W_{r+1}$ algebra and a right
chiral $W_{r+1}^{(2)}$ algebra---a mixture of CT and 2-ECT theories), and
consider the chiral exchange algebra, classical solution and the
relations to WZNW and Toda lattice hierarchies [8,5] as well.

\section{The heterotic Toda model}
\index{Toda ! heterotic}
Let us start by constructing explicitly the HTFT mentioned above.
Stressing its integrability, we begin with the Lax pair representation of
the model. The Lax pair is defined in the cylindrical space-time as follows,

\begin{eqnarray}
\partial_+ T &=& \left[\frac{1}{2} \partial_+
\Phi+\exp \left( -\frac{1}{2}
{\rm ad} \Phi \right)\left({\bar{\Psi}}_+ +\mu \right) \right]T,
\nonumber \\
\partial_- T &=& -\left[\frac{1}{2} \partial_-
\Phi+\exp \left( \frac{1}{2}
{\rm ad} \Phi \right) \nu \right]T,
\label {eq: lax}
\end{eqnarray}

\noindent where $x_\pm \equiv t \pm x$, $\partial_\pm \equiv
\partial_{x_\pm}$, and

\begin{eqnarray}
&\mu =\frac{1}{2} \sum_{i,j=1}^{r}{\rm sign}(i-j) [E_i,\ E_j],
{}~~~~\nu=\sum_{i=1}^{r}F_i, \nonumber \\
&\Phi=\sum_{i=1}^{r}\varphi^i H_i,~~~~ \Psi_+=\sum_{i=1}^{r}
\psi_+^i F_i, ~~~~ \bar{\Psi}_+=[\mu,\ \Psi_+] .
\nonumber
\end{eqnarray}

\noindent In the above definitions, $\{H_i, E_i, F_i\}$ denote
the Chevalley generators of
the rank $r$ finite dimensional Lie algebra {\tt g} (which will be restricted
to be the classical $A_r$ algebra for simplicity), and all the
component fields, $\varphi^i, \; \psi_+^i$, are assumed to be periodic
in the spacial  coordinate, $x$.

The equations of motion for the HTFT follow from the compatibility
conditions of the  Lax pair (\ref{eq: lax}). They turn out to be

\begin{eqnarray}
&\partial_+ \partial_{-} \Phi + [\nu, \exp (-{\rm ad} \Phi)\bar {\Psi}_{+}]
=0,\nonumber\\
&\partial_- \Psi_{+} - \exp ({\rm ad} \Phi) \nu=0. \label {eq: motion}
\end{eqnarray}

\noindent or, in component form,

\begin{eqnarray}
&\partial_+ \partial_{-} \varphi^i - \sum_{j} {\rm sign} (j-i)K_{ji}
\psi_{+}^{j} \omega^i=0, \nonumber \\
&\partial_{-}\psi_+^i - \omega^i=0, ~~~~ \omega^i
= \exp (-\sum_{j}K_{ji} \varphi^j) .
\label {eq: mtcomp}
\end{eqnarray}

\noindent One may wonder that how we can imagine such a complicated
model by direct construction. Indeed, the model
(\ref{eq: motion}-\ref{eq: mtcomp})
appears not to be very simple at a first glance, but from the point
of view of hamiltonian reductions of WZNW theories [9]
---which is now
a quite common way of constructing extended Toda models---the
origin of the model (\ref{eq: mtcomp}) can be made very clear:
it is nothing but the constrained WZNW theory under the following
constraints

\begin{equation}
\langle {\bf g}_-, \partial_+ gg^{-1} - \mu \rangle =0,
{}~~~~\langle {\bf g}_+, g^{-1}\partial_- g - \nu \rangle = 0,
\label{eq: wzwred}
\end{equation}

\noindent where ${\tt g_\pm}$ are respectively the maximal nilpotent
subalgebras consisted of positive and negative
step operators, and $\langle \ , \ \rangle $ is the
standard Killing form. Notice that the above constraint
equations are not of the left-right dual style, this is why the present
model is heterotic. Actually, if one perform the CP transform
(\ref{eq: cp}) on the model (\ref{eq: mtcomp}), he would arrive at
a {\it different} but {\it dual} model, with the roles of the
left and right chiral sectors interchanged.

There are two ways to identify the conformal invariance of the HTFT.
The first way is to study the conformal-preserving nature of the
WZNW $\rightarrow$ HTFT reduction \index{WZNW}.
In this way one can show that the
conformal algebra of the theory (\ref{eq: mtcomp}) is nothing but
the product of a left chiral (the $x_-$-depending sector) $W_{r+1}$ algebra
and a right chiral $W_{r+1}^{(2)}$ algebra. The second way of exploiting
the conformal invariance of the model (\ref{eq: mtcomp}) is to show
explicitly the behaviors of the equations of motion under conformal
change of spacetime variables. One can easily check that under the
conformal transformations of the coordinates

\[x_+ \rightarrow {\tilde x}_+ = f (x_+),
{}~~~~x_- \rightarrow {\tilde x}_- = h (x_-),\]

\noindent the equations of motion (\ref{eq: mtcomp}) is left invariant
provided the fields $\varphi^i, \; \psi_+^i$ transform as follows,

\[ \varphi^i \rightarrow \tilde \varphi^i = \varphi^i +
\sum_{j} \left( K^{-1}
\right)^{ji} \ln (f')^{1/2}h' \Rightarrow
\omega^i \rightarrow \tilde \omega^i =(f')^{-1/2}(h')^{-1} \omega^i,\]
\[ \psi_+^i \rightarrow \tilde \psi_+^i = (f')^{-1/2} \psi_+^i.\]

\noindent This shows that the fields $\psi_+^i$ and $\omega^i$
are respectively primary fields of conformal weights (0, 1/2) and
(1, 1/2). One may therefore expect that the conformal spectrum
of the left chiral sector is consisted of only integers, and that of the
right chiral sector may be consisted of integer and half-integers.
This observation is in agreement with the above-mentioned $(W_{r+1})_L
\otimes (W_{r+1}^{(2)})_R$ symmetry. \index{$W$ algebra symmetry !
$(W_{r+1})_L \otimes (W_{r+1}^{(2)})_R$}

Now let us write down the effective action \index{effective action:
heterotic} for the model
(\ref{eq: mtcomp}). It reads

\begin{eqnarray}
I[\Phi, \; \Psi_+] & = & \frac{1}{2}
\int {\rm d}^2 x \langle \partial_+ \Phi \partial_-
\Phi + \bar {\Psi}_+ \partial_- \Psi_+ - 2 \left( \exp (-{\rm ad} \Phi)
\bar {\Psi}_+ \right) \nu \rangle \nonumber\\
& = & \frac{1}{2}
\int {\rm d}^2 x \sum_{ij} \left[ \partial_+ \varphi^i K_{ij}
\partial_- \varphi^j + {\rm sign} (i-j) \psi_+^i K_{ij} \partial_-
\psi_+^j \right.
\label {eq: act}\\
& - & 2 \left. {\rm sign} (i-j) \psi_+^i K_{ij} \exp (- \sum
\varphi^l K_{lj} ) \right]. \nonumber
\end{eqnarray}

\noindent We see that the form of the kinematic terms in
eq.(\ref{eq: act}) are very similar to that of the heterotic string
theory [6]. The difference lies in that,
in the heterotic string theory,
the $\psi$ fields are fermionic and the full theory has a heterotic
supersymmetry, whilst in the present case, $\psi_+^i$ are bosonic
fields and therefore the theory has no supersymmetry (the curious
similarity between bosonic conformal algebras having the integer
half-integer conformal spectrum and the real superconformal algebras
is still an open area for further study, at least in the authors' own
view points).

Given the effective action (\ref{eq: act}), it is now ready to define the
conjugate momenta and the canonical Poisson brackets in the usual way,

\begin{eqnarray}
&\pi (\varphi_i) = \sum_{j} K_{ij} \partial_0 \varphi^j,
{}~~~~&\pi (\psi_+^i) = \frac{1}{2} \sum_{j} {\rm sign} (j-i) \psi_+^j K_{ji}
\nonumber \\
& \{ \pi (\varphi^i)(x), \; \varphi^j (y) \} = \delta^{ij} \delta (x-y),
& \{\pi (\psi_+^i)(x), \; \psi_+^j (y) \} = \delta^{ij} \delta (x-y),
\nonumber
\end{eqnarray}

\noindent where of cause the $\delta$-functions are also assumed to be
periodic due to the periodicity of the fundamental fields.
However, since the fields $\psi_+^i$ are of the first order in derivatives
in the action (\ref{eq: act}), one should treat the definitions of the
canonical momenta \index{canonical momentum}
of these fields as primary constraints and replace the
corresponding naive Poisson brackets by Dirac Poisson brackets. It follows
that provided the rank $r$ of the underlying Lie algebra $A_r$ is even,
all the above constraints are of the second class and there are no
further constraints in the theory. The final Dirac brackets for the
$\psi$-fields read (here we denote the Dirac brackets again by $\{\;,\;\}$)

\begin{displaymath}
\{\psi_+^i(x),\; \psi_+^j(y) \} = (M^{-1})_{ji} \delta(x-y),
\end{displaymath}

\noindent where $M_{ij} \equiv {\rm sign}(i-j) K_{ij}$, which is invertible
provided $r$ is even
\footnote{In the following context, as far as Poisson brackets are
concerned, we shall assume that $r$ is even. But the other results such as the
wronskian type solutions \index{solution ! Wronskian type}
{\em etc. } hold true without this restriction.}
. From the above Poisson brackets
we can calculate the fundamental
Poisson relation (FPR) for the spacial component of the Lax connection. Then,
upon integration with the initial value $T(0)=1$, the FPR for the
ultralocal transport matrix $T$ follow [7], \index{transport matrix}

\begin{equation}
\{ T(x) \otimes_, T(x) \} = [r_\pm, \; T(x) \otimes T(x) ],
\label{eq: fpr}
\end{equation}

\noindent where $r_\pm$ are the so-called classical r-matrices which
are well known in the standard CT theory,

\begin{eqnarray}
& r_+ = \frac{1}{2} \left \{ \sum \left( K^{-1}  \right) ^{ij} H_i \otimes
H_j +2 \sum_{ \alpha > 0} E_{ \alpha } \otimes F_{ \alpha } \right \} ,
\nonumber\\
& r_- = - \frac{1}{2} \left \{ \sum \left( K^{-1}  \right) ^{ij} H_i \otimes
H_j +2 \sum_{ \alpha > 0} F_{ \alpha } \otimes E_{ \alpha } \right \} .
\label{eq: r-matrix}
\end{eqnarray}

\noindent We mention that the FPR (\ref{eq: fpr}) together with
the r-matrices (\ref{eq: r-matrix}) are the characterizing properties
for the integrabilities of all the Toda type field theories. They are
also the starting points for studying the exchange algebras
[7,3] and dressing symmetries \index{dressing symmetries}
[10,3] for such theories. But now in this chapter we shall not
consider these issues in detail. Instead, we shall discuss briefly the
origin of the chiral exchange algebra \index{exchange algebra}
in HTFT with a specific
emphasis on its relations to the  classical solutions and the $W$ algebra
symmetries \index{$W$ algebra
symmetry} of the model. These discussions are part of the content
of the next section.

\section{Exchange algebra and Leznov-Saveliev analysis}
\index{Leznov-Saveliev analysis}

In this section we shall discuss some aspects connected with the chiral
vectors in the model. First let us show how there are chiral vectors
embedded in the present theory.

\subsection{Existence of chiral vectors}
As is well known, the Lax pair representation for integrable systems
admits a gauge freedom. In other words, the compatibility conditions
for a Lax pair are left invariant while the transport matrix $T$ is
shifted from the left by a gauge group element. Therefore, one can choose
different gauges for the Lax pair to obtain an optimized form
for the current usage. In the present case, we can choose the following
convenient gauges

\begin{eqnarray}
 \partial_+ T_L & = & \left \{ \partial_+ \Phi +  \bar {\Psi}_+
+ \mu \right \} T_L , \nonumber \\
 \partial_- T_L & = & - \left \{ \exp ( {\rm ad} \Phi ) \nu
\right \} T_L; \label{eq: laxl}
\end{eqnarray}

\noindent and

\begin{eqnarray}
 \partial_+ T_R & = & \left \{ \exp ( -{\rm ad} \Phi ) ( \bar {\Psi}_+
+ \mu ) \right \} T_R , \nonumber \\
 \partial_- T_R & = & - \left \{ \partial_- \Phi + \nu
\right \} T_R; \label{eq: laxr}
\end{eqnarray}

\noindent where

\begin{equation}
T_L=\exp \left( \frac{1}{2} \Phi \right) T, ~~~~ T_R=
\exp \left( - \frac{1}{2} \Phi \right) T. \label{eq: tltr}
\end{equation}

Assuming that $ |\lambda^i \rangle$ and $ \langle \lambda^i |$ are
respectively the highest weight and dual highest weight
vectors in $i$-th fundamental representation of {\tt g}, it follows
from eqs.(\ref{eq: laxl}) and (\ref{eq: laxr}) that the vectors

\begin{equation}
\xi^{(i)} (x_+) \equiv \langle \lambda^i | T_L ,
{}~~~~\bar{ \xi }^{(i)} (x_-) \equiv T^{-1}_R | \lambda^i \rangle
\label{eq: xi}
\end{equation}

\noindent are chiral,

\begin{displaymath}
\partial_- \xi^{(i)} (x_+) = 0,
{}~~~~\partial_+ \bar {\xi }^{(i)} (x_-) =0.
\end{displaymath}

\noindent Moreover, performing  another gauge transformation
$T_L \longrightarrow $ $\tilde {T_L}$ $ =\exp ( \Psi_+ ) T_L, \ $
we can show that the vectors

\begin{equation}
\zeta^{(i)} ( x_+ ) \equiv \langle \lambda^i - \alpha^{i} | \tilde {T_L}
=\langle \lambda - \alpha^{i} | \exp (\Psi_+ ) T_L \label{eq: zeta}
\end{equation}

\noindent are also chiral ( $\alpha^i$ being the $i$-th simple root)

\begin{displaymath}
\partial_- \zeta^{(i)} (x_+) = 0,
\end{displaymath}

\noindent These chiral vectors play the central role in the rest of
this chapter.

\subsection{Exchange algebra for the chiral vectors}
The chiral vectors obtained in the last subsection obey a very
nice exchange algebra. The method for obtaining such exchange
algebras is now quite familiar in the CT and 2-ECT theories. In the
present model, one can first calculate the Poisson brackets between
$\exp (\Phi ), \exp (\Psi_+ )$ and $T$, then,
using the definitions of the chiral vectors (\ref{eq: xi})
and (\ref{eq: zeta}) and by straightforward calculations, one obtains
the following exchange relations (throughout this chapter, all the chiral
quantities are  assumed to be evaluated at equal time $t=t_0$),
\index{exchange algebra}

\begin{equation}
\{ \xi^{(i)} (x) \otimes_, \ \xi^{(j)} (y) \}
= \xi^{(i)} (x) \otimes \xi^{(j)} (y)
\left( r_+ \theta (x-y) + r_- \theta (y-x) \right) ,\label{eq:zzzz1}
\end{equation}

\begin{equation}
\{ \xi^{(i)} (x) \otimes_, \ \bar{\xi}^{(j)} (y) \}
= - \left(\xi^{(i)} (x) \otimes 1 \right) r_-
\left( 1 \otimes \bar {\xi}^{(j)} (y) \right) ,\label{eq:zzzz2}
\end{equation}

\begin{equation}
\{ \bar{\xi}^{(i)} (x) \otimes_, \ \xi^{(j)} (y) \} = - \left( 1 \otimes
\xi^{(j)} (y) \right) r_+ \left( \bar{\xi}^{(i)} (x) \otimes 1 \right) ,
\label{eq:zzzz3}
\end{equation}

\begin{equation}
\{ \bar{\xi}^{(i)} (x) \otimes_, \ \bar{\xi}^{(j)} (y) \} =
\left( r_- \theta (x-y) + r_+ \theta (y-x) \right)
\bar{\xi}^{(i)} (x) \otimes \bar{\xi}^{(j)} (y)  ,\label{eq:zzzz4}
\end{equation}

\begin{eqnarray}
\{ \zeta^{(i)} (x) \otimes_, \ \zeta^{(j)} (y) \}
&=& \zeta^{(i)} (x) \otimes \zeta^{(j)} (y)
\left( r_+ \theta (x-y) + r_- \theta (y-x) \right) \nonumber\\
&+& (M^{-1})_{ji}\delta(x-y) \xi^{(i)}(x) \otimes \xi^{(j)}(y),
\label{eq:zzzz5}
\end{eqnarray}

\begin{equation}
\{ \xi^{(i)} (x) \otimes_, \ \zeta^{(j)} (y) \}
= \xi^{(i)} (x) \otimes \zeta^{(j)} (y)
\left( r_+ \theta (x-y) + r_- \theta (y-x) \right) ,\label{eq:zzzz6}
\end{equation}

\begin{equation}
\{ \zeta^{(i)} (x) \otimes_, \ \xi^{(j)} (y) \}
= \zeta^{(i)} (x) \otimes \xi^{(j)} (y)
\left( r_+ \theta (x-y) + r_- \theta (y-x) \right) ,\label{eq:zzzz7}
\end{equation}

\begin{equation}
\{ \zeta^{(i)} (x) \otimes_, \ \bar{\xi}^{(j)} (y) \} =
- \left(\zeta^{(i)} (x) \otimes 1 \right) r_- \left( 1 \otimes
\bar {\xi}^{(j)} (y) \right) ,\label{eq:zzzz8}
\end{equation}

\begin{equation}
\{ \bar{\xi}^{(i)} (x) \otimes_, \ \zeta^{(j)} (y) \} = - \left( 1 \otimes
\zeta^{(j)} (y) \right) r_+ \left( \bar{\xi}^{(i)} (x) \otimes 1 \right) .
\label{eq:zzzz9}
\end{equation}

\noindent Recalling that the fields $\Phi,\;\Psi_+$ are periodic, we have
the following monodromy properties \index{monodromy ! properties}
for the chiral vectors,

\begin{eqnarray*}
&\xi^{(i)}(x+2\pi ) = \xi^{(i)}(x) T&,\\
&\zeta^{(i)}(x+2\pi ) = \zeta^{(i)}(x) T&,\\
&\bar{\xi}^{(i)}(x+2\pi ) = T^{-1}\bar{\xi}^{(i)}(x)&,
\end{eqnarray*}

\noindent where $T \equiv T(2\pi)$. Moreover, thereis a set of nontrivial
Poisson  brackets between the above chiral vectors and the monodromy matrix,

\begin{eqnarray*}
&\{ T\;\otimes, T\} = \left[  r_{\pm},\;T\otimes T \right],&\\
&\{\xi^{(i)} (x)\;\otimes,T\}=- \left(\xi^{(i)}(x)\otimes 1 \right) r_-,&\\
&\{\zeta^{(i)} (x)\;\otimes,T\}=-
\left(\zeta^{(i)}(x)\otimes 1 \right) r_-,&\\
&\{\bar{\xi}^{(i)} (x)\;\otimes,T\}=  \left(1\otimes T\right)
r_+\left(\bar{\xi}^{(i)}(x)\otimes 1 \right) .&
\end{eqnarray*}

\noindent Notice that although the objects  $\xi,\;\zeta,\;\bar{\xi}$ are
chiral, there is a nontrivial coupling between  them.
This can happen only through the zero modes and the corresponding
conjugate variables. Actually, just  as in the usual  Toda context,
the degrees of freedom corresponding to the zero modes are contained
in the diagonal part of the monodromy matrix. Therefore in order to
choose an appropriate basis in which the chiralities are completely
splited we have  to first diagonalize the monodromy  matrix. This in
practice  is connected to  the so-called Drinfeld-Sokolov linear systems,
and we shall leave the task for skeching such procedures to subsection 3.4.

\subsection{Leznov-Saveliev analysis}
\index{Leznov-Saveliev analysis}

Let us now study the relations between the chiral vectors
(\ref{eq: xi}), (\ref{eq: zeta}) and the general solution of the
model. As will be shown below, the general solution of the model
can be represented by the products of these chiral vectors, which
can be rewritten in terms of appropriate matrix elements
in the fundamental representations of the underlying Lie algebra.
Such analysis were first carried out by Leznov and Saveliev in the
case of standard Toda theories, that is why the current subsection
is titled as above.

{}From eqs.(\ref{eq: xi}) and (\ref{eq: zeta}) we have

\begin{eqnarray}
\exp (\varphi^i) &=& \langle \lambda^i | \exp (\Phi ) | \lambda^i
\rangle = \langle \lambda^i | T_LT_R^{-1} | \lambda^i \rangle
= \xi^{(i)} (x_+) \bar{\xi}^{(i)} (x_-), \label{eq: phii} \\
\psi_+^i &=& \frac{\langle \lambda^i - \alpha^i| \exp (\Psi_+ )
T_L T_R^{-1} | \lambda^i \rangle} {\langle \lambda^i |
T_L T_R^{-1} | \lambda^i \rangle}
= \frac{\zeta^{(i)} (x_+) \bar{\xi}^{(i)} (x_-)}
{\xi^{(i)} (x_+) \bar{\xi}^{(i)} (x_-)}. \label{eq: psii}
\end{eqnarray}

\noindent Now recalling that the matrices $T_L$ and $T_R$ differ from
each other only by a diagonal part from the left (see eq.
(\ref{eq: tltr})), we can make the following Gauss decompositions,

\begin{equation}
T_L = {\rm e}^{K_+} N_- M_+, ~~~~ T_R = {\rm e}^{K_-} N_+ M_-,
\label{eq: gauss}
\end{equation}

\noindent where $K_\pm $ are respectively the diagonal parts of
$T_L$ and $T_R$, $N_+ , M_+$ and $N_- , M_-$ are upper- and
lower-triangular matrices with entries on the diagonal equal to one.
These upper and lower triangular matrices are intrinsically related
by the Gauss decompositions of the original
transport matrix $T$,

\begin{displaymath}
T = {\rm e}^{H_+} N_- M_+ = {\rm e}^{H_-} N_+ M_-,
\end{displaymath}

\noindent where $H_\pm $ are the diagonal part of $T$ under
each Gauss decomposition. Now substituting the Gauss decompositions
(\ref{eq: gauss}) into the definitions (\ref{eq: xi}) and
(\ref{eq: zeta}), it follows that

\begin{eqnarray}
& \xi^{(i)} (x_+) =  \langle \lambda^i | {\rm e}^{K_+} M_+,
{}~~~~ \bar{\xi}^{(i)} (x_-) = M_-^{-1} {\rm e}^{-K_-}
| \lambda^i \rangle, \label{eq: xi-km} \\
& \zeta^{(i)} (x_+) = \langle \lambda^i - \alpha^i |
{\rm e}^{K_+} \left( {\rm e}^{ -{\rm ad} K_+} {\rm e}^{\Psi_+} \right)
N_- M_+ . \label{eq: zeta-km}
\end{eqnarray}

\noindent Expanding the matrix $N_-$ into the form

\begin{equation}
N_- = \exp (\chi^{(-1)} ) \exp (\chi^{(-2)} )  ...
\label{eq: nexp}
\end{equation}

\noindent with $\chi^{(-i)}$ being a lower triangular matrix with nonzero
entries only in the $i$-th lower-diagonal, we can rewrite
eq.(\ref{eq: zeta-km}) into the  form,

\begin{eqnarray}
\zeta^{(i)} (x_+) &=& \langle \lambda^i - \alpha^i |
{\rm e}^{K_+} \left\{ 1+ {\rm e}^{ -{\rm ad} K_+} P_+  \right\} M_+ ,
\nonumber\\
P_+ &\equiv& \Psi_+ + {\rm e}^{ {\rm ad} K_+} \chi^{(-1)} .
\label{eq: zeta-km2}
\end{eqnarray}

\noindent The chirality of $\xi^{(i)}$ and $\bar{\xi}^{(i)}$ then implies
that the matrices $K_\pm $ and $M_\pm $ in eq.(\ref{eq: xi-km}) are
chiral,

\begin{displaymath}
\partial_\pm K_\mp = \partial_\pm M_\mp = 0.
\end{displaymath}

\noindent Consequently the chirality of $\zeta^{(i)} $ implies the
similar property of $P_+$,

\begin{displaymath}
\partial_- P_+ = 0.
\end{displaymath}

\noindent Substituting eqs.(\ref{eq: xi-km}) and (\ref{eq: zeta-km2})
into eqs.(\ref{eq: phii}-\ref{eq: psii}), it turns out that the general
solution of the model (\ref{eq: mtcomp})
is completely determined by the chiral quantities
$K_\pm, \ M_\pm \ $ and $P_+$,

\begin{eqnarray}
& {\rm exp}(\varphi^i) = \langle \lambda^i | {\rm e}^{K_+} M_+
M_-^{-1} {\rm e}^{-K_-} | \lambda^i \rangle, \nonumber \\
\displaystyle
& \psi_+^i = \frac{ \langle \lambda^i - \alpha^i |
{\rm e}^{K_+} \left( 1+ {\rm e}^{ -{\rm ad} K_+} P_+ \right)
M_+ M_-^{-1} {\rm e}^{-K_-} | \lambda^i \rangle }
{\langle \lambda^i | {\rm e}^{K_+} M_+
M_-^{-1} {\rm e}^{-K_-} | \lambda^i \rangle }. \label{eq: solution}
\end{eqnarray}

\noindent We have to point out that the chiral quantities
$K_\pm, \ M_\pm \ $ and $P_+$ are not all independent, they have to
obey some linear partial differential equations. To specify what
differential equations are satisfied by these objects, let us recall that
the gauge-transformed transport matrices $T_L$ and $T_R$ satisfy
the following equations,

\begin{equation}
\partial_+ T_L T_L^{-1} = \partial_+ \Phi + \bar{ \Psi}_+ + \mu,
{}~~~~\partial_- T_R T_R^{-1} = -( \partial_- \Phi + \nu).
\label{eq: nmnmnm}
\end{equation}

\noindent Substituting the Gauss decompositions \index{Gauss decompositions}
(\ref{eq: gauss}) into
eq.(\ref{eq: nmnmnm}) and projecting onto the upper- and lower-triangular
parts respectively, we have

\begin{eqnarray*}
\left[ N_- \left( \partial_+ M_+ M_+^{-1} \right) N_-^{-1} \right ]_+
= {\rm e}^{- {\rm ad} K_+ } ( \bar{\Psi}_+ + \mu ),
\end{eqnarray*}
\begin{eqnarray*}
\left[ N_+ \left(M_- \partial_- M_-^{-1} \right) N_+^{-1} \right]_-
= {\rm e}^{- {\rm ad} K_- } \nu .
\end{eqnarray*}

\noindent Considering the fact that $\mu $ has nonvanishing entries
only on the second upper-diagonal, $\nu $ has nonvanishing entries
only on the first lower-diagonal, and also remembering of the further
decomposition (\ref{eq: nexp}) of $N_-$ in terms $\chi^{(-i)} $,
we finally get

\begin{equation}
\partial_+ M_+ M_+^{-1}
= {\rm e}^{- {\rm ad} K_+ } ( \bar{P}_+ + \mu ),
{}~~~~M_- \partial_- M_-^{-1} = {\rm e}^{- {\rm ad} K_- } \nu, \label{eq: pd}
\end{equation}

\noindent with

\begin{equation}
\bar{P}_+ \equiv [ \mu, \ P_+ ]. \label{eq: pplus}
\end{equation}

\noindent Eq.(\ref{eq: solution}) together with (\ref{eq: pd}),
(\ref{eq: pplus}) constitute the general solution of the model
(\ref{eq: mtcomp}).

{\it Remark.} The construction we made in this subsection is in  some
sence a little formal because of the nontrivial couplings between
both chiralities. In order to reformulate the general solution of the
model in terms of free fields---which decouples from each other---we
again need to diagonalize the  monodromy matrix \index{monodromy ! matrix}
. This additional issue
also ensures that the general solution  obtained in this way  is
single-valued ({\it i.e.} periodic in $x$) and local ({\it i.e.}
Poisson commute).

\subsection{Sketch for a free field representation}
\index{free field representation ! heterotic Toda}

Let us now  give a brief sketch for the free field representation
of the general solution. The construction  is based on the following
Drinfeld-Sokolov (DS) \index{Drinfeld-Sokolov ! linear system}
linear systems,

\begin{displaymath}
\partial_+ Q_+ = ( \partial_+ K_+ + \bar{P}_+ + \mu ) Q_+ ,
{}~~~~ \partial_- Q_- = Q_- ( \partial_- K_- + \nu ) ,
\end{displaymath}

\noindent where $\partial_\pm K_\pm$ and $\bar{P}_+$ are the same as in
the last  subsection and are assumed to be periodic. Since in the above DS
systems everything is  chiral, we introduce the chiral vectors

\begin{eqnarray*}
&\sigma^{(i)}(x) =  \left\langle \lambda^i \right|Q_+  (x),&\\
&\bar{\sigma}^{(i)}(x) = Q_-(x) \left|\lambda^i \right\rangle,&\\
&\varsigma^{(i)}(x) =  \left\langle \lambda^i-\alpha^i \right|
{\rm e}^{P_+}Q_+  (x),&
\end{eqnarray*}

\noindent where the DS solutions $Q_\pm$ are nomalized as $Q_\pm(0)=1$.

It is obvious that these chiral vectors have the following
monodromy properties, \index{monodromy ! properties}

\begin{eqnarray*}
&\sigma^{(i)}(x+2\pi)=\sigma^{(i)}(x)S,
{}~~~~\varsigma^{(i)}(x+2\pi)=\varsigma^{(i)}(x)S,&\\
&\bar{\sigma}^{(i)}(x+2\pi)=\bar{S}\bar{\sigma}^{(i)}(x),&
\end{eqnarray*}

\noindent where $S  \equiv Q_+ (2\pi)$ and $\bar{S} \equiv Q_- (2\pi)$,
which are  respectively upper and lower triangular.

Now introducing the Poisson brackets

\begin{eqnarray*}
&\{\partial_\pm K_\pm(x)\; \otimes,\partial_\pm K_\pm(y)\}=\mp (\partial_x
-\partial_y)\delta(x-y)\sum_{i,j}(K^{-1})^{ij}H_i  \otimes H_j,&\\
&\{\bar{P}_+ (x)\;\otimes,P_+(y)\}  = \delta(x-y)\sum_{i}E_i\otimes F_i,&
\end{eqnarray*}

\noindent it can be proved that $\sigma, \varsigma$ and $\bar{\sigma}$
satisfy the following  exchange relations,

\begin{eqnarray*}
&\{ \sigma^{(i)} (x) \otimes_, \ \sigma^{(j)} (y) \} &
= \sigma^{(i)} (x) \otimes \sigma^{(j)} (y)
\left( r_+ \theta (x-y) + r_- \theta (y-x) \right) ,\\
&\{ \varsigma^{(i)} (x) \otimes_, \ \varsigma^{(j)} (y) \} &
= \varsigma^{(i)} (x) \otimes \varsigma^{(j)} (y)
\left( r_+ \theta (x-y) + r_- \theta (y-x) \right) ,\\
&~~~~&
+(M^{-1})_{ji}\delta(x-y) \sigma^{(i)}(x) \otimes \sigma^{(j)}(y),\\
&\{ \sigma^{(i)} (x) \otimes_, \ \varsigma^{(j)} (y) \} &
= \sigma^{(i)} (x) \otimes \varsigma^{(j)} (y)
\left( r_+ \theta (x-y) + r_- \theta (y-x) \right) ,\\
&\{ \bar{\sigma}^{(i)} (x) \otimes_, \ \bar{\sigma}^{(j)} (y) \} &
= \left( r_- \theta (x-y) + r_+ \theta (y-x) \right)
\bar{\sigma}^{(i)} (x) \otimes \bar{\sigma}^{(j)} (y) ,\\
&\{ \sigma^{(i)} (x) \otimes_, \ \bar{\sigma}^{(j)} (y) \} &
=\{ \varsigma^{(i)} (x) \otimes_, \ \bar{\sigma}^{(j)} (y) \}
=0.
\end{eqnarray*}

\noindent Moreover, it can be checked that the following expressions
are solutions of the equations of motion,

\begin{eqnarray*}
&{\rm exp}(\varphi^i) = \sigma^{(i)} U \bar{\sigma}^{(i)},&\\
\displaystyle
&\psi_+^i = \frac{\varsigma^{(i)}U \bar{\sigma}^{(i)}}
{\sigma^{(i)} U \bar{\sigma}^{(i)}}&,
\end{eqnarray*}

\noindent where $U$ is any constant matrix acting on the space
of the $i$-th fundamental representation of $A_r$. The problems  which are
still needed to be solved are that the above  solution have to be
periodic and local. These two requirements drastically reduces
the degrees of freedom  in choosing the constant matrix $U$ as will be
shown below.

Let us consider the periodicity. Inserting the monodromy
properties into  the above solutions and letting the fields $\varphi^i$
and $\psi_+^i$ to be periodic,  we have

\begin{equation}
SU\bar{S}=U.\label{eq: periodic}
\end{equation}

\noindent This requirement can be fulfilled as follows. Since the monodromy
matrix $S$ is upper triangular, it can be diagonalized by a unique
strictly upper triangular  matrix $g$,

\begin{displaymath}
S=g \kappa g^{-1},~~~~ \kappa = {\rm e}^{2\pi
\partial_+ K_+ (0)},
\end{displaymath}

\noindent where $\partial_+ K_+ (0)$ means $\partial_+ K_+ (x) |_{x=0}$,
which is obviously diagonal. Similarly we can diagonalize the monodromy
matrix $\bar{S}$ of the other chirality by a strictly lower triangular
matrix $\bar{g}$,

\begin{displaymath}
\bar{S}=\bar{g} \bar{\kappa} \bar{g}^{-1},
{}~~~~ \bar{\kappa} = {\rm e}^{- 2\pi
\partial_- K_- (0)},
\end{displaymath}

\noindent with the notation
$\partial_- K_- (0) = \partial_- K_- (x) |_{x=0}$.

\noindent It is then evident that the periodicity condition
(\ref{eq: periodic}) is satisfied if the constant matrix $U$ takes the form

\begin{displaymath}
U=g {\cal D} \bar{g}^{-1},
\end{displaymath}

\noindent together with a constraint  condition imposed on the positive and
negative zero modes,

\begin{displaymath}
\kappa \bar{\kappa} =  1
{}~~~~{\rm or}
{}~~~~\partial_+K_+(0)=\partial_-K_-(0).
\end{displaymath}

\noindent Actually, the above conditions simplyimply that the left and right
monodromies and the constant matrix $U$ canbe  diagonalized simultaneously
and the diagonal  parts of the left and right  monodromies must be
equal.

The remaining problem---the problem of locality of the general
solution---is far more difficult to prove. However, the general  principal
for this proof is rather simple [24]. Starting from the Poisson brackets
for $\partial_\pm K_\pm$ and $\bar{P}_+$, one can obtain well
defined exchange relations between each pair of the nomalized chiral fields
$\sigma^{(i)},\;\varsigma^{(i)},\;\bar{\sigma}^{(i)}$ and
the monodromy matrices $S$ and $\bar{S}$, then by direct calculations
using the general solution formula one can prove that the only
admissible diagonal matrix ${\cal D}$ satisfying the locality condition
is the one of the form

\begin{displaymath}
{\cal D} = \Theta \bar{\Theta},
{}~~~~\Theta = {\rm e}^{\Pi - \partial_+ K_+(0)},
{}~~~~\bar{\Theta}={\rm e}^{\bar{\Pi} + \partial_- K_-(0)}.
\end{displaymath}

\noindent where $\Pi$ and $\bar{\Pi}$ are respectively the conjugate
variables  of the zero modes $\partial_+K_+(0)$ and $\partial_-K_-(0)$.
As the explicit calculations are
considerably long and tedious, we prefer to publish them in a
seperate publication rather than present them in the present chapter.

Readers who  are smart enough might already have noticed that,
eq.(32), while appropriately gauge transformed, yields the following
DS{\it -like} linear systems,

\begin{displaymath}
\partial_+ V_+ = ( \partial_+ K_+ + \bar{P}_+ + \mu ) V_+ ,
{}~~~~ \partial_- V_- = V_- ( \partial_- K_- + \nu ) ,
\end{displaymath}

\noindent where $V_+ \equiv {\rm e}^{K_+} M_+,$ $V_- \equiv M_-^{-1}
{\rm e}^{-K_-}$. However, as the normalizations of $V_\pm$ and $Q_\pm$
are different, we cannot identify the last equations with the
standard DS-systems. In fact, these two objects differ from each other
by a right-shift with a constant matrix which results in different
monodromy  properties of the corresponding chiral  vectors.
To be more explicit, the chiral vectors $\xi^{(i)},\;\zeta^{(i)}$
and $\bar{\xi}^{(i)}$ described in subsection 3.1 are related to the
chiral vectors $\sigma^{(i)},\;\varsigma^{(i)}$ and $\bar{\sigma}^{(i)}$
as follows,

\begin{eqnarray*}
&\sigma^{(i)} = \xi^{(i)}V_+^{-1}(0) = \xi^{(i)}M_+^{-1}(0)
{\rm e}^{-K_+(0)},&\\
&\varsigma^{(i)} = \zeta^{(i)}V_+^{-1}(0)
= \xi^{(i)}M_+^{-1}(0) {\rm e}^{-K_+(0)},&\\
&\bar{\sigma}^{(i)} = V_-^{-1}(0)\bar{\xi}^{(i)} =
{\rm e}^{K_-(0)}M_-(0) \bar{\xi}^{(i)}.&
\end{eqnarray*}

\noindent Therefore, we can relate the general solution described in terms
of the chiral vectors $\xi^{(i)},\;\zeta^{(i)}$ and $\bar{\xi}^{(i)}$
to the one described by $\sigma^{(i)},\;\varsigma^{(i)}$ and
$\bar{\sigma}^{(i)}$ as

\begin{displaymath}
U=V_+ (0) V_-(0) = {\rm e}^{K_+(0)}M_+(0)M_-^{-1}(0) {\rm e}^{-K_-(0)}.
\end{displaymath}

\noindent So to ensure that the solution given by eqs.
(\ref{eq: phii})-(\ref{eq: psii}) to be
periodic and local, all what is needed is to choose appropriate
initial values for the fields $K_\pm$ and $M_\pm$.

{\it Remark.}

In this subsection we considered only the  zero  modes
of the  fields $K_\pm$. There are, however, zero mode problems for the
fields $P_+$  and $\bar{P}_+$. It is these zero modes that make the
chiral vectors $\sigma^{(i)}$ and $\varsigma^{(i)}$ interact
nontrivially. Since the zero modes for the fields $P_+$ and
$\bar{P}_+$ {\it  do not} lie in the diagonal, it is much more
difficult to disentangle them  than disentangling the zero modes of
$\partial_\pm K_\pm$. We hope to come  back to this point later.

\section{Special solution and WZNW model }
The general solution obtained in the last section involves matrix
elements in all the fundamental representations. Such a solution
is very useful while studying the symmetries of the model. But in
practice, one is often concerned about the explicit space-time
behaviors of the fundamental fields. In this case, the general
solution given above may appear not to be very helpful. In this
section, we shall give another attempt for deriving the explicit
solution of the model.

\subsection{Wronskian type special solution}
\index{solution ! Wronskian type}
Let us start from the chiral embedding of the light cone coordinates
$x_+, \ x_-$ into the product space $V = {\bf R}^{r+1} \otimes
{\bf R}^{2(r+1)}$. Such embedding are expressed by the identifications

\begin{eqnarray}
& \bar{X}^i = \bar{ \xi}^i (x_-),
{}~~~~X^i = \xi^i (x_+ ), ~~~~ Y^i = \zeta^i (x_+ ),& \nonumber\\
& i= 1, \ 2, \  ...  , r+1 &\label{eq: embed}
\end{eqnarray}

\noindent where $\bar{X}^i$ and $X^i, \ Y^i$ are respectively
coordinates of the left ${\bf R}^{r+1} $ and the right
${\bf R}^{2(r+1)} $ spaces, $\bar{\xi }^i, \ \xi^i$ and $\zeta^i$
are arbitrary functions of the arguments ({\em these functions must
obey some fixed monodromy properties in order to maintain the dynamics
since they are exactly the components of  the chiral vectors
$\xi^{(1)},\;\zeta^{(1)}$ and $\bar{\xi}^{(1)}$ as  we shall show later.
However, given the discussions in the last subsection, we are left with
no doubt in the fact that there {\em exist} physically meaningful solutions
for our model, and that is enough for our following discussions. Therefore
we do  not care about any explicit monodromy behaviors of the above chiral
embedding  functions}).

Now define two sets of
$(r+1)$ column-vectors $\bar{\bf f}_a$ and raw-vectors
${\bf f}_a$ as following,

\begin{eqnarray*}
& &\bar{f}_a^i (x_- ) = \partial_-^{a-1} \bar{\xi}^i (x_- ), \\
& &f_{2a-1}^i (x_+ ) = \partial_+^{a-1} \xi^i (x_+ ), \\
& &f_{2a}^i (x_+ ) = \partial_+^{a-1} \zeta^i (x_+ ), \\
& &i= 1, \ 2, \  ...  , r+1.
\end{eqnarray*}

\noindent From these vectors we can construct the $(r+1) \times (r+1)$
matrix of inner products

\begin{displaymath}
g_{ab} (x_-, \ x_+ ) \equiv {\bf f}_a {\bf f}_b
= \sum_{i=1}^{r+1} f_a^i (x_+ ) \bar{f}_b^i (x_- ),\  a,\  b=1,\ 2, \
 ...  , \ r+1.
\end{displaymath}

\noindent We also introduce the following notations,

\begin{eqnarray}
\Delta_a &\equiv& {\rm det} \left(
\begin{array}{ccc}
$$g_{11}$$ &  ...  & $$g_{1a}$$ \cr
\vdots & \ &   \vdots \cr
$$g_{a1}$$ &  ...  & $$g_{aa}$$
\end{array}
\right), ~~~~ a= 1, \ 2, \  ...  , r+1, \nonumber \\
\Delta_0 &\equiv& 1. \nonumber
\end{eqnarray}

\noindent Using these notations, we now define two sets of new vectors
${\bf e}_a$ and $\bar{\bf e}_a$,

\bea
& &e_a^i \equiv \frac{1}{\sqrt{ \Delta_{a-1} \Delta_{a} } }
\left|
\begin{array}{cccc}
$$g_{11}$$ &  ...  & $$g_{1,a-1}$$  & f_1^i \cr
\vdots & \ & \vdots & \vdots \cr
$$g_{a1}$$ &  ...  & $$g_{a,a-1}$$ & f_a^i
\end{array}
\right|,\nonumber\\
& &\bar{e}_a^i \equiv \frac{1}{\sqrt{ \Delta_{a-1} \Delta_{a} } }
\left|
\begin{array}{ccc}
$$g_{11}$$ &  ...  & $$g_{1,a}$$ \cr
\vdots & \ & \vdots \cr
$$g_{a-1,1}$$ &  ...  & $$g_{a-1,a}$$  \cr
$$\bar{f}_1^i$$ &  ...  & $$\bar{f}_a^i$$
\end{array}
\right|, \label{eq: e}
\eea

\noindent where ${\bf e}_a$ are $(r+1)$-raw vectors, and $\bar{\bf e}_a$
are $(r+1)$-column vectors. It can be easily proved
that the vectors ${\bf e}_a$ and $\bar{\bf e}_a$
are orthogonal to each other,

\begin{equation}
({\bf e}_a, \ \bar{\bf e}_b ) = \sum_{i=1}^{r+1} e_a^i \bar{e}_b^i
=\delta_{ab}. \label{eq: ortho}
\end{equation}

\noindent This is due to the following Laplacian expansions
\index{Laplacian expansion} of the
definition (\ref{eq: e}),

\begin{eqnarray}
e_a^i= \sqrt{ \frac{ \Delta_{a-1}}{\Delta_a} } \sum_{l=1}^{a}
\frac{ \Delta_{a} (l, \ a) }{\Delta_{a-1} } f_l^i (x_+ )
\equiv \sqrt{ \frac{ \Delta_{a-1}}{\Delta_a} } \sum_{l=1}^{a}
(A^{-1})_{al} f_l^i (x_+ ), \nonumber \\
\bar{e}_a^i= \sqrt{ \frac{ \Delta_{a-1}}{\Delta_a} } \sum_{l=1}^{a}
\frac{ \Delta_{a} (a, \ l) }{\Delta_{a-1} } \bar{f}_l^i (x_- )
\equiv \sum_{l=1}^{a} \bar{f}_l^i (x_- )
(C^{-1})_{la} \sqrt{ \frac{ \Delta_{a-1}}{\Delta_a} }, \label{eq: lap}
\end{eqnarray}

\noindent where $\Delta_a (i, \ j)$ are algebraic cominors of the entry
$(i, \ j)$. According to these Laplacian expansions, we can express
the derivatives of ${\bf e}_a$ and $\bar{\bf e}_a$ in terms of their
linear combinations,

\begin{eqnarray}
\partial_+{\bf e}_a = (\omega_+ )_a^b {\bf e}_b,
{}~~~~\partial_-{\bf e}_a = (\omega_- )_a^b {\bf e}_b, \nonumber \\
\partial_+ \bar{\bf e}_a = \bar{\bf e}_b (\bar{\omega}_+ )_a^b ,
{}~~~~\partial_- \bar{\bf e}_a = \bar{\bf e}_b (\bar{\omega}_- )_a^b ,
\nonumber
\end{eqnarray}

\noindent where as usual, the subscripts
$ _\pm $ specifies the upper- or
lower-triangularities of the corresponding $\omega$ matrices. The
orthogonality condition (\ref{eq: ortho}) now implies that
$\omega_\pm $ and $\bar{\omega}_\pm $ are not independent objects,

\begin{displaymath}
(\omega_\pm )_a^b = - (\bar{\omega}_\pm )_a^b.
\end{displaymath}

\noindent Further more, straightforward calculations show that
only a few of the matrix elements in $\omega_\pm $ are nonvanishing,
the nonvanishing elements being sited on the main diagonal and the
first and the second upper-/lower-diagonals. Explicitly, we have

\begin{eqnarray}
& &\partial_+{\bf e}_a = \frac{1}{2} \partial_+ \ln \left( \frac{
\Delta_a}{\Delta_{a-1}} \right) {\bf e}_a \nonumber\\
& &~~~~+ \sqrt{ \frac{\Delta_{a-1} \Delta_{a+1}}{\Delta_a^2}} \left(
\frac{\Delta_a (a-1, \ a)}{\Delta_{a-1}} - \frac{\Delta_{a+2}
(a+1, \ a+2)}{\Delta_{a+1}} \right) {\bf e}_{a+1} \nonumber \\
& &~~~~+ \sqrt{ \frac{\Delta_{a-1} \Delta_{a+1}}{\Delta_a^2}}
\sqrt{ \frac{\Delta_{a} \Delta_{a+2}}{\Delta_{a+1}^2}}
{\bf e}_{a+2}, \nonumber\\
& &\partial_-{\bf e}_a = - \frac{1}{2} \partial_- \ln \left( \frac{
\Delta_a}{\Delta_{a-1}} \right) {\bf e}_a \nonumber\\
& &~~~~- \sqrt{ \frac{\Delta_{a-2} \Delta_a}{\Delta_{a-1}^2}}
{\bf e}_{a-1}, \nonumber\\
& &a = 1, \ 2, \  ... , \ r+1, \nonumber\\
& &{\bf e}_{-1} = {\bf e}_{0} = {\bf e}_{r+2} = {\bf e}_{r+3} = 0.
\label{eq: e-delta}
\end{eqnarray}

\noindent Denoting

\begin{equation}
\varphi^a = \ln \Delta_a, ~~~~ \psi_+^a = -
\frac{ \Delta_{a+1} (a, \ a+1)}{\Delta_a },
\label{eq: explicit solution}
\end{equation}

\noindent eq.(\ref{eq: e-delta}) will become

\begin{eqnarray}
& &\partial_+{\bf e}_a = \frac{1}{2} \partial_+ ( \varphi^a -
\varphi^{a-1} ) {\bf e}_a \nonumber\\
& &~~~~+ (w^a )^{1/2} (\psi_+^{a-1} -\psi_+^{a+1} )
{\bf e}_{a+1} +(w^a w^{a+1} )^{1/2} {\bf e}_{a+2}, \nonumber\\
& &\partial_-{\bf e}_a = - \frac{1}{2} \partial_- ( \varphi^a -
\varphi^{a-1} ) {\bf e}_a \nonumber\\
& &~~~~- (w^{a-1} )^{1/2} {\bf e}_{a-1}, \nonumber\\
& &w^a \equiv \exp (\varphi^{a-1} + \varphi^{a+1} - 2\varphi^a ).
\label{eq: e-phipsi}
\end{eqnarray}

\noindent The compatibility condition $[\partial_+, \ \partial_- ]
{\bf e}_a = 0$ of eq.(\ref{eq: e-phipsi}) then yields

\begin{eqnarray}
& \partial_+ \partial_- \varphi^j -
(\psi_+^{j-1} - \psi_+^{j+1} )w^j =0,
{}~~~~ \partial_- \psi_+^j - w^j =0,
{}~~~~j=1, \  ... , \ r,& \nonumber\\
& \partial_+ \partial_- \varphi^{r+1} = 0,
{}~~~~\partial_- \psi_+^{r+1} =0. & \nonumber
\end{eqnarray}

\noindent Except for the nonvarnishing functions
$\varphi^{r+1}$ and $\psi_+^{r+1}$, the above equations are very
similar to the equations of motion (\ref{eq: mtcomp}) of the HTFT.
So if we can somehow fix the values of $\varphi^{r+1}$ and
$\psi_+^{r+1}$ to zero, then the above construction will really
result in an explicit solution to the system (\ref{eq: mtcomp}).

In order to fix the values of $\varphi^{r+1}$ and
$\psi_+^{r+1}$ to zero, let us mention that the $\Delta $ symbols
$\Delta_{r+1}$ and $\Delta_{r+2} (r+1, \ r+2)$ are simply
products of chiral objects,

\begin{displaymath}
\Delta_{r+1} = {\rm det} g = U \bar{U},
{}~~~~ \Delta_{r+2} (r+1, \ r+2) = V \bar{U},
\end{displaymath}

\noindent where $U, \ \bar{U}$ are respectively the determinants
of the matrices consisted of the functions $f_a^i$ and $\bar{f}_a^i$,
and $V$ is defined as follows,

\begin{displaymath}
V=\left|
\begin{array}{cccc}
f_1^1      & f_1^2      &  ...  & f_1^{r+1}     \cr
\vdots     & \vdots     & \      & \vdots        \cr
f_r^1      & f_r^2      &  ...  & f_r^{r+1}     \cr
f_{r+2}^1  & f_{r+2}^2  &  ...  & f_{r+2}^{r+1}
\end{array}
\right|.
\end{displaymath}

\noindent Now it is clear from eq.(\ref{eq: explicit solution}) that
if we set $\varphi^{r+1} = \psi_+^{r+1} =0 $ then the following
conditions on $U$ and $V$ have to be satisfied,

\begin{equation}
U=\bar{U}=1,~~~~ V=0. \label{eq: constraints}
\end{equation}

\noindent In terms of the original embedding functions,
these constraints are nothing but linear differential
equations for the $3(r+1)$ arbitrary functions. Such equations can
be easily solved  for any three of the embedding functions,
and that will finish the construction of the wronskian type solution.

\subsection{Connections with WZNW theory}
\index{WZNW theory ! connection with}
Readers who are familiar with the classical W-geometrical theory of
Gervais {\it et al} [12]
may have already recognized that the constructions
made above are almost based on the similar construction in the
conventional CT theory. That {\it is} the point. In the case of
standard CT theory, the wronskian type solution  is closely related
to the Drinfeld-Sokolov \index{Drinfeld-Sokolov ! gauge}
gauges of the corresponding $(W_{r+1})_L \otimes
(W_{r+1})_R$ symmetries of the model. Now we come to show that the
conformal algebra for the HTFT is $(W_{r+1})_L \otimes
(W_{r+1}^{(2)})_R$ in contrast to the CT ($(W_{r+1})_L \otimes
(W_{r+1})_R$ algebra) and 2-ECT ($(W_{r+1}^{(2)})_L \otimes
(W_{r+1}^{(2)})_R$ algebra) cases. The difference in conformal algebras
in the left and the right chiral sectors is the most crucial
property of the present model.

Remembering the definitions of the matrix elements $g_{ab}$, we have

\begin{eqnarray}
& \partial_+ g_{ab}  =  g_{a+2,b},
{}~~~~\partial_- g_{ab} = g_{a,b+1},& \nonumber \\
& a = 1, \ 2, \  ... , \ r-1,
{}~~~~b  =  1, \ 2, \  ... , \ r. &\nonumber
\end{eqnarray}

\noindent In matrix form, we have

\begin{equation}
\partial_+ g  = J_+ g,
{}~~~~\partial_- g  = g
J_-, \label{eq: wzw}
\end{equation}

\noindent where

\begin{equation}
J_- = \left(
\begin{array}{ccccc}
0        &        &        &         &*  \\
1        &0       &        &         &*  \\
	 &1       &\ddots  &         &\vdots  \\
	 &        &\ddots  &0        &*  \\
	 &        &        &1        &*
\end{array}
\right),
{}~~~~J_+ = \left(
\begin{array}{ccccccc}
&0        &0       &1       &        &         &   \\
&         &0       &0       &1       &         &   \\
&         &        &\ddots  &\ddots  &\ddots   &   \\
&         &        &        &0       &0        &1  \\
&*        &*       & ...   &*       &*        &*  \\
&*        &*       & ...   &*       &*        &*
\end{array}
\right), \label{eq: j+-}
\end{equation}

\noindent and the non-zero non-constant entries ``*'' in (\ref{eq: j+-})
are to be determined. Eq.(\ref{eq: wzw}) is nothing but a constrained
version of the well known WZNW equations, with $J_\pm$ being the
constrained WZNW currents. We recognize that this form of the constrained
WZNW currents is written in the well-known Drinfeld-Sokolov gauges
\index{Drinfeld-Sokolov ! gauge}
of the $(W_{r+1})_L$ and $(W_{r+1}^{(2)})_R$ algebras, where the ``*'' entries
are just the W algebra generators. It is straightforward to check that
the vectors ${\bf f}_a$ and $\bar{\bf f}_a$ also solve eq.(\ref{eq: wzw}),

\begin{equation}
\partial_+ {\bf f}_a = \sum_{b=1}^{r+1} ( J_+ )_{ab} {\bf f}_b,
{}~~~~\partial_- \bar{\bf f}_a = \sum_{b=1}^{r+1} \bar{\bf f}_b ( J_- )_{ba}.
\label{eq: fwzw}
\end{equation}

\noindent hence the matrices consisted of
the functions $f_a^i$ and $\bar{f}_a^i$
are respectively the chiral components of the WZNW field $g$,
and there is no problem in concluding that all the ``*'' in eq.
(\ref{eq: j+-}) are chiral objects.

Now let us go one step further to determine the relations between the
W algebra generators and the embedding functions (\ref{eq: embed}).

Substituting the relations $\partial_+ {\bf f}_a = {\bf f}_{a+2}, \
\partial_- \bar{\bf f}_b = {\bf f}_{b+1}$ into eq.(\ref{eq: fwzw}),
we have

\begin{equation}
\sum_{c=1}^{r+1} ( J_+ )_{ac} {\bf f}_c = {\bf f}_{a+2},
{}~~~~\sum_{c=1}^{r+1} \bar{\bf f}_c ( J_- )_{cb} = \bar{\bf f}_{b+1}.
\label{eq: fwzw2}
\end{equation}

\noindent Solving the above linear system of equations for the
variables $(J_+ )_{ac}$ and $(J_- )_{cb}$ with $ a= r, \ r+1 $
and $b = r+1$, it results in

\begin{eqnarray}
& (J_+)_{r,c} = - \frac{R_c (c, \ r+2)}{U},
{}~~~~(J_+)_{r+1,c} = - \frac{S_c (c, \ r+2)}{U},& \nonumber \\
& (J_-)_{r+1,c} = - \frac{\bar{R} (c, \ r+2)}{U},&
\label{eq: wgenerator}
\end{eqnarray}

\noindent where $R_c, \ \bar{R}_c$ and $S_c$ are given as follows,

\begin{eqnarray}
& & R_c = \left|
\begin{array}{cccc}
f_1^1        &  ...  & f_1^{r+1}      & f_1^{c}       \cr
\vdots       & \      & \vdots         & \vdots        \cr
f_{r+1}^1    &  ...  & f_{r+1}^{r+1}  & f_{r+1}^{c}   \cr
f_{r+2}^1    &  ...  & f_{r+2}^{r+1}  & f_{r+2}^{c}
\end{array}
\right| = 0,\nonumber\\
& & S_c = \left|
\begin{array}{cccc}
f_1^1        &  ...  & f_1^{r+1}      & f_1^{c}       \cr
\vdots       & \      & \vdots         & \vdots        \cr
f_{r+1}^1    &  ...  & f_{r+1}^{r+1}  & f_{r+1}^{c}   \cr
f_{r+3}^1    &  ...  & f_{r+3}^{r+1}  & f_{r+3}^{c}
\end{array}
\right| = 0, \nonumber \\
& & \bar{R}_c = R_c ({\rm with} \ f_a^i \leftrightarrow \bar{f}_a^i) .
\nonumber
\end{eqnarray}

\noindent Eq.(\ref{eq: wgenerator}) determines the W algebra generators
completely in terms of the embedding functions (\ref{eq: embed}). It is
interesting to notice that constraint conditions in
(\ref{eq: constraints}) implies that

\begin{displaymath}
(J_+)_{r,r+1}=0, ~~~~ (J_+)_{r,r}+(J_+)_{r+1, r+1} =0,
{}~~~~ (J_-)_{r+1, r+1} =0.
\end{displaymath}

\noindent This not only ensures the tracelessness of the $A_r$ WZNW
currents (\ref{eq: j+-}) but also makes $J_\pm$ fit in the framework
of eq.(\ref{eq: wzwred}).

Now let us give some brief remarks on the aspects which are not
mentioned above.

\begin{enumerate}
\item Using the standard method, we can also reduce eq.(\ref{eq: fwzw})
into the Gelfand-Dickey equations for the W algebras in the left and
right chiral sectors. The one for the $(W_{r+1})_L$ is a scalar
differential equation, and the one for $(W_{r+1}^{(2)})_R$ is a matrix one.
The matrix Gelfand-Dickey equation corresponding to the $W_{r+1}^{(2)}$
algebra was first given in Ref.[13].
\item The matrices $A, \ C$ defined in eq.(\ref{eq: lap})
have very profound meanings in the WZNW setting of the model. In fact,
from (\ref{eq: lap}), we can write

\begin{equation}
{\bf f}_a = \sum_{l=1}^{a}
A_{al} \sqrt{ \frac{ \Delta_{l}}{\Delta_{l-1}} } {\bf e}_l,
{}~~~~\bar{\bf f}_a = \sum_{l=1}^{a}
\bar{\bf e}_l \sqrt{ \frac{ \Delta_{l}}{\Delta_{l-1}} } C_{la} ,
\label{eq: ffac}
\end{equation}

\noindent from which we have

\begin{displaymath}
g_{ab} = {\bf f}_a \bar{\bf f}_b = ABC,
{}~~~~(B)_{ab} \equiv \left[ \exp (\Phi ) \right]_{ab} =
\frac{ \Delta_a}{ \Delta_{a-1}} \delta_{ab}.
\end{displaymath}

The last equation shows that the matrices $A, \ C$ are exactly the
lower- and upper-triangular parts in the Gauss decomposition
of the constrained WZNW field $g$. Moreover, it can be shown that
$A, \ C$ satisfy the following equations,

\begin{displaymath}
A^{-1} \partial_- A = \exp ({\rm ad} \Phi ) \nu,
{}~~~~\partial_+ C C^{-1} = \exp (-{\rm ad} \Phi )(\bar{\Psi}_+ + \mu ).
\end{displaymath}

\noindent These equations are precisely the constrained WZNW equations
re-expressed in terms of the Gauss components $A$ and $C$.

\item The chiral embedding functions (\ref{eq: embed}) are just the
components of the chiral vectors $\bar{\xi}^{(1)}, \ \xi^{(1)}$ and
$\zeta^{(1)}$ corresponding to the defining representation of the
Lie algebra $A_r$. This statement makes the discussions in
this and the last sections finally unified, and it follows that the
W algebras $(W_{r+1})_L$ and $(W_{r+1}^{(2)})_R$ are related to the
simple exchange algebra (\ref{eq:zzzz1}-\ref{eq:zzzz9}) by
eq.(\ref{eq: wgenerator}). So it seems that the exchange algebra
(\ref{eq:zzzz1}-\ref{eq:zzzz9}) is more fundamental than the
W symmetry algebras.

\indent Now let us spend some more words on the identification of chiral
embedding functions with the chiral vectors
(\ref{eq: xi}) and (\ref{eq: zeta}). Notice that while the constraints
(\ref{eq: constraints}) are imposed, eq.(\ref{eq: e-phipsi})
will become exactly the Lax pair (\ref{eq: lax}) of $A_r$ HTFT, with the
transport matrix $T$ defined as

\begin{displaymath}
T_{ai} \equiv {\bf e}_a^i, ~~~~ T_{ia}^{-1} \equiv
\bar{\bf e}_a^i.
\end{displaymath}

\noindent This observation enables us to rewrite eq.(\ref{eq: ffac})
into

\begin{equation}
f_a^i = \sum_{l=1}^{a}
A_{al} \left[ \exp \left( \frac{1}{2} \Phi \right) \right]_{ll} T_{li},
{}~~~~\bar{f}_a^i = \sum_{l=1}^{a}
\left(T^{-1} \right)_{il} \left[ \exp \left( \frac{1}{2} \Phi \right)
\right]_{ll} C_{la}.
\label{eq: fftt}
\end{equation}

\noindent On the other hand, the definitions of $A$ and $C$ implies that

\begin{equation}
A_{aa} =1, ~~~~ A_{a+1,a} = \psi_+^a,
{}~~~~C_{aa} =1. \label{eq: aacc}
\end{equation}

\noindent Substituting eq.(\ref{eq: aacc}) into (\ref{eq: fftt}) and
remember that $\xi^i = f_1^i, \ \zeta^i = f_2^i, \ \bar{\xi}^i =
\bar{f}_1^i \ $, we finally get

\begin{eqnarray}
& \bar{\xi}^i = \left( T^{-1} \right)_{i1} \left[ \exp
\left( \frac{1}{2} \Phi \right) \right]_{11} C_{11}
= \left\{ T^{-1} \exp \left( \frac{1}{2} \Phi \right) \right\}_{i1}, &
\label{eq: barxiixii}\\
& \xi^i = A_{11}  \left[ \exp
\left( \frac{1}{2} \Phi \right) \right]_{11} \left( T \right)_{1i}
= \left\{ \exp \left( \frac{1}{2} \Phi \right) T \right\}_{1i}, &
\label{eq: xiixii} \\
& \zeta^i = \left\{ \exp (\Psi_+ ) \exp \left( \frac{1}{2} \Phi \right)
T \right\}_{2i}. & \label{eq: zetazeta}
\end{eqnarray}

\noindent Eqs.(\ref{eq: barxiixii}-\ref{eq: zetazeta}) are exactly the
definitions (\ref{eq: xi}) and (\ref{eq: zeta}) rewritten in the
defining representation of the Lie algebra $A_r$.
\end{enumerate}

To end the present section, let us mention that the wronskian type
solution (\ref{eq: explicit solution}) is only a special solution.
The key point in relating this special solution to the  general
solution given in the last section is to perform {\it chiral} gauge
transformations starting from the Drinfeld-Sokolov type gauges
(\ref{eq: j+-}). In terms of the embedding functions, such gauge
transformations are expressed by replacing these chiral functions by
their arbitrary linear combinations, and that is all about the
W-geometrical picture about the HTFT.

\section{Connections with Toda lattice hierarchy}
\index{Toda ! lattice hierarchy}
In this section, we are going to consider the connections between HTFT
and the well known Toda lattice hierarchy (TLH). This
is a totally different view point in contrast to the discussions
made in the former sections.

First let us briefly recall the usual description of TLH
(here we are using the notations of Ref.[5]). Let
${\cal O}$ be the space of the unitary hermitian states
$\{ |n\rangle\}$ equipped with the metric

\begin{displaymath}
\langle m | n \rangle = \delta_{mn}.
\end{displaymath}

\noindent Let $\Lambda$ be the shifting operator acting on ${\cal O}$,

\begin{displaymath}
\Lambda |n \rangle = |n+1 \rangle,
{}~~~~\langle n | \Lambda = \langle n-1 |.
\end{displaymath}

\noindent It follows that any operator $W$ acting on the space
${\cal O}$ can be expressed as

\begin{displaymath}
W = \sum_{j \in {\bf Z}} W_j \Lambda^j,
{}~~~~W_{j} = \sum_{n} |n \rangle W_j (n) \langle n |.
\end{displaymath}

\noindent Moreover, each operator $W$ admits a unique factorization

\begin{displaymath}
W= W_+ + W_-, ~~~~ W_+ =
\sum_{j \geq 0} W_j \Lambda^j,
{}~~~~W_- =\sum_{j < 0} W_j \Lambda^j.
\end{displaymath}

\noindent Now consider two special operators of the form

\begin{displaymath}
L = \Lambda + \sum_{n=0}^{\infty } u_n \Lambda^{-n},
{}~~~~M= \sum_{n=-1}^{\infty } v_n \Lambda^n
\end{displaymath}

\noindent and defining

\begin{displaymath}
B_n \equiv (L^n)_+, ~~~~ C_n \equiv (M^n)_-,
\end{displaymath}

\noindent the TLH is then determined by the following evolution
equations for the operators $L$ and $M$,

\begin{eqnarray}
& \partial_n L = [B_n, \ L],&
{}~~~~\bar{\partial}_n L = [C_n, \ L], \nonumber\\
& \partial_n M = [B_n, \ M],&
{}~~~~\bar{\partial}_n M = [C_n, \ M],
\nonumber
\end{eqnarray}

\noindent or, equivalently, by the compatibility conditions

\begin{eqnarray}
\left[ \partial_n - B_n, \ \partial_m - B_m \right] = 0, \nonumber \\
\left[ \bar{\partial}_n - C_n, \ \bar{\partial}_m - C_m
\right] = 0, \nonumber \\
\left[ \partial_n - B_n, \ \bar{\partial}_m - C_m \right] = 0.
\label{eq: zerocurvature}
\end{eqnarray}

The claim of this section is that the $A_{\infty }$  limit
of the HTFT (\ref{eq: mtcomp}) is just the $(B_2, \ C_1)$ flow
in the hierarchy (\ref{eq: zerocurvature}).
To justify this claim let us recall that $B_2$ and $C_1$ must
be of the form

\begin{eqnarray}
& B_2= \sum_{s} \left( |s\rangle \langle s|\Lambda^2
+ |s\rangle b_1 (s) \langle s|\Lambda
+ |s\rangle b_0 (s) \langle s| \right), & \nonumber \\
& C_1 = \sum_{s} |s\rangle c_1 (s) \langle s|\Lambda^{-1}&.
\label{eq: b2c1}
\end{eqnarray}

\noindent Substituting eq.(\ref{eq: b2c1}) into the second
member of (\ref{eq: zerocurvature}), it follows that

\begin{eqnarray}
& \partial_+ c_1 (a) + c_1 (a) b_0 (a+1) - b_0 (a) c_1 (a) = 0,& \nonumber\\
& \partial_- b_1 (a) - c_1 (a) + c_1 (a-2) = 0,&\nonumber\\
& \partial_- b_0 (a) - c_1 (a) b_1 (a+1) + b_1 (a) c_1 (a-1)
=0,& \label{eq: mtbbc}
\end{eqnarray}

\noindent where $\partial_+ \equiv \partial_2, \ \partial_- \equiv
\bar{\partial}_1.\ $ Changing the variables $b_0 (a) \rightarrow
\partial_+ (\varphi^{a-1} - \varphi^a ), \  b_1 (a) \rightarrow
\psi_+^{a} -\psi_+^{a-2}, \ $ eq.(\ref{eq: mtbbc}) will become

\begin{eqnarray}
&\partial_+ \partial_- \varphi^a -(\psi_+^{a-1} - \psi_+^{a+1} )
\exp (\varphi^{a-1} +\varphi^{a+1} - 2 \varphi^a) =0,&\nonumber \\
& \partial_- \psi_+^a -
\exp (\varphi^{a-1} +\varphi^{a+1} - 2 \varphi^a) =0, &\nonumber\\
& a = -\infty, \  ... , \ \infty. & \label{eq: ainftyeq}
\end{eqnarray}

\noindent We see that eq.(\ref{eq: ainftyeq}) is just the
$A_{\infty } \ $ version of eq.(\ref{eq: mtcomp}). Therefore, the model
we are considering is really a Toda type theory.

Recently, it is of increasing interests to study the so-called
continuous Toda field theories [15-19].
Such theories have intimate relations
with the  W-infinity  algebras and self-dual
gravities. So it seems worthwhile
to remark here that the continuous limit of the present model will become
a (1+2)-dimensional heterotic Toda theory. Such a theory is expected
to be related to the heterotic W-infinity gravities. We hope to come
back to this point later.

\section{Summary and Discussions}
Let us now summarize  the whole chapter and make some more discussions.

In this chapter, we constructed a heterotic Toda field theory, studied
its chiral exchange algebra and gave the classical general solution and a
wronskian type  special solution. We also showed that the wronskian
type solution is closely related to the W algebra symmetries of the
model. The chiral vectors $\xi^{(i)}, \
\bar{\xi}^{(i)}$ and $\zeta^{(i)}$ are the cross-point of all these
different aspects. Besides all these, we showed that the HTFT is really
a member in the TLH. This final statement may be of interesting in
the theories of  W-infinity  algebras and self-dual gravities.
It can be expected that in the
continuous limit, the HTFT will yield two sets of  W-infinity  algebras,
one of them will be the usual $(w_\infty)_L$ algebra, and the other
will be a generalized $(w_\infty)_L$ algebra which may be denoted
$(w_\infty^{(2)})_R$. The possibility for the existence of
$(w_\infty^{(2)})_R$ type algebras was first proposed by the authors
in Ref.[3]. Detailed study on the  W-infinity  algebras and  W-infinity
gravities in the framework of HTFT will be presented elsewhere.

It is also worth mentioning that here we only worked on
the Lie algebra $A_r$. We can also construct similar models based on
other finite dimensional Lie  algebras as well as loop and affine
Lie algebras. In the latter cases, we shall arrive at heterotic
analogies of LT and CAT theories. Then the soliton behaviors
in these heterotic loop Toda and conformal affine Toda theories
may become interesting subjects of further study. Moreover, as
the complex affine (loop) Toda theories are receiving considerable
attentions [20-23]
because they give physically meaningful (real)
energy-momentum tensors, it may also be interesting to study the case of
complex coupling constants of these heterotic models.

\chapter{Heterotic Liouville systems from Bernoulli equation}

Liouville equation has attracted the attentions of both theoretical
physicists and mathematicians for quite a long history [1], and even after
over a centry's investigations, the importance thereof is still not faded
in both physics and mathematics. Physically, Liouville system is closely
related with the theories of two-dimensional gravity, strings and
conformal fields [2]. Mathematically, \index{Liouville model}
Liouville equation is the characteristic
equation of two-dimensional surfaces of constant Gauss curvatue lying in
Euclidean three-space [3]; a simple reduction of
Liouville equation leads to the
well known Ricatti equation and thus provides a prototype of integrable
nonlinear partial differential equations, {\it etc.} The recent development
of the theories of quantum groups and quantum conformal fields [4]
shows that Liouville system is a very interesting field of study
which calls for a re-union of physics and mathematics.

In this chapter, we will consider a large class of ``heterotic'' extensions
of Liouville equation, which contain one extra ``dependent'' variable
$\psi$ besides the original Liouville variable $\varphi$. These extensions
are called heterotic because a simple check shows that they possess
heterotic conformal symmetry except in the simplest nontrivial case, {\it
i.e.} the Liouville case, in which the heterotic conformal symmetry becomes
the normal one. We will show that these heterotic Liouville systems possess
interesting features in addition to the heterotic conformal symmetry.

Recall that the Liouville equation $\partial_z \partial_{\bar{z}} \varphi
=2 {\rm e}^{\varphi} $ can be reduced into Ricatti equation
\index{Recatti equation}
$ \partial_z w - w^2 + f(z) =0 $ by the substitution $\varphi = {\rm ln}
\partial_{\bar{z}} \psi$ and performing integration with respect to $\bar{z}$
twice. Conversely Liouville equation can be reconstructed from Ricatti
equation using the same substitution and differentiating with respect to
$\bar{z}$ twice. Inspired by this observation, let us
start by considering the Bernoulli equation \index{Bernoulli equation}

\begin{equation}
\partial_z \psi = \psi^n + f(z) \psi + g(z)
\end{equation}

\noindent with $n \geq 2$, $f(z)$, $g(z)$ are functions of $z$ only
whilst $\psi$ is considered to depend on both $z$ and $\bar{z}$.
Differentiate (1) with respect to $\bar{z}$ once, we get

\begin{eqnarray*}
\partial_z \partial_{\bar{z}} \psi = n \psi^{n-1} \partial_{\bar{z}}
\psi + f(z) \partial_{\bar{z}} \psi.
\end{eqnarray*}

\noindent Devided by $\partial_{\bar{z}} \psi$ and differentiate with
respect to $\bar{z}$ again, we obtain

\begin{eqnarray*}
\partial_{\bar{z}}\left(\frac{\partial_z \partial_{\bar{z}} \psi}{
\partial_{\bar{z}} \psi} \right) = n(n-1) \psi^{n-2} \partial_{\bar{z}}
\psi.
\end{eqnarray*}

\noindent Now using the substitution

\begin{eqnarray*}
\varphi = {\rm ln} \partial_{\bar{z}} \psi,
\end{eqnarray*}

\noindent we obtain

\begin{eqnarray}
\partial_z \partial_{\bar{z}} \varphi &=&
n(n-1)\psi^{n-2} {\rm e}^{\varphi}, \\
\partial_{\bar{z}} \psi &=& {\rm e}^{\varphi}.
\end{eqnarray}

\noindent This system of equations is just what have been called heterotic
Liouville system, among which the simplest case of $n=2$ corresponds to the
standard Liouville equation \index{Liouville model ! heterotic}

\begin{eqnarray*}
\partial_z \partial_{\bar{z}} \varphi = 2 {\rm e}^{\varphi}.
\end{eqnarray*}

The heterotic conformal symmetry of the system (2,3) can be easily verified
as follows. Let the ``space-time'' coordinates $z$, $\bar{z}$ undergo
the following conformal transformation

\begin{eqnarray*}
z \rightarrow \tilde{z}=\xi(z),
{}~~~~\bar{z} \rightarrow \tilde{\bar{z}}=\bar{\xi}(\bar{z}),
\end{eqnarray*}

\noindent and requiring that the system (2,3) is left invariant, we find
that the variables $\varphi$ and $\psi$ must transform as

\begin{eqnarray*}
\varphi \rightarrow \tilde{\varphi} &=& \varphi + {\rm ln} \left[
(\xi'(z))^{-\frac{1}{n-1}} (\bar{\xi}'(\bar{z}))^{-1} \right],\\
\psi \rightarrow \tilde{\psi} &=&
(\xi'(z))^{-\frac{1}{n-1}} \psi,
\end{eqnarray*}

\noindent where primes denotes derivatives with respect to the arguments.
We see that the variables ${\rm e}^\varphi$ and $\psi$ are
respectively conformal tensors (or ``primary fields'') of dimensions
$(\frac{1}{n-1}, \ \ 1)$ and $(\frac{1}{n-1}, \ \ 0)$.
This left-right asymmetric nature is refered to as heterotic
conformal symmetry in this chapter.

It should be remarked that the heterotic conformal symmetry is not a new
type of conformal symmetry [7]. It is just the usual conformal symmetry
plus some appropriate extension in only one of the two chiral sectors.
Such extension should amount to W-like algebras, which has
nothing to do with the space-time transformation or ``point symmetries''
of the system. Actually, we have worked out the maximum point symmetry
group of the system (2-3), using the standard method of prolongation [5],
which turns out to be just the two-dimensional conformal group. Such result
is not included here because the calculation is rather tedious and the
method is standard. However, from this result, we can conclude that the
system we are considering is integrable because it has long be concluded
that conformal invariance implies integrability [8].

Now let us show that the heterotic Liouville system (2,3) possess
infinite many symmetries generated by some simple, obvious symmetry
and a recursion operator. In order to achieve this let us rewrite the
system (2,3) as follows

\begin{eqnarray}
\partial_{\bar{z}} u &=& n(n-1)\psi^{n-2}{\rm e}^{\partial_z^{-1} u},\\
\partial_{\bar{z}} \psi &=& {\rm e}^{\partial_z^{-1} u},
\end{eqnarray}

\noindent where

\begin{eqnarray*}
u = \partial_z \varphi,
{}~~~~\partial_z^{-1} = \int^z \/ {\rm d}z.
\end{eqnarray*}

\noindent Substututing (5) into (4), we have

\begin{eqnarray*}
\partial_{\bar{z}}u=n(n-1)\psi^{n-2}\partial_{\bar{z}}\psi.
\end{eqnarray*}

\noindent Upon integration with respect to $\bar{z}$, we get

\begin{eqnarray*}
u = n \psi^{n-1},
\end{eqnarray*}

\noindent where we have omitted an integration constant depending
arbitrarily on $z$ without lost of generality. Substitute the last result
into (4) we finally get

\begin{eqnarray*}
\partial_{\bar{z}} u = n(n-1)\left(\frac{u}{n}\right)^{\frac{n-2}{n-1}}
{\rm e}^{\partial_z^{-1} u}.
\end{eqnarray*}

\noindent Rewriting $u$ as $v^{n-1}$ and rescaling $\bar{z}$ as
$\bar{z} \rightarrow n^{1-n} \bar{z}$, we have for $v$ the following equation

\begin{equation}
\partial_{\bar{z}} v = {\rm e}^{\partial_z^{-1} v^{n-1}}.
\end{equation}

By a symmetry of equation (6) we mean an infinitesimal change
of the variable $v$ \index{symmetry}

\begin{eqnarray*}
v \rightarrow \tilde{v}=v + \epsilon \sigma
\end{eqnarray*}

\noindent such that $\tilde{v}$ still satisfy equation (6). It is easy
to see that $\sigma$ is a symmetry of (6) if and only if it is
a solution of the equation

\begin{equation}
D_\Delta \sigma = 0,
{}~~~~D_\Delta \equiv \partial_{\bar{z}} - (n-1) v_{\bar{z}} \partial_z^{-1}
v^{n-2},
\end{equation}

\noindent where we have denoted $\partial_{\bar{z}} v$ as $v_{\bar{z}}$.
Straight forward calculations show that the following quantities are
symmetries,

\begin{eqnarray*}
\sigma_1 = v_z,
{}~~~~ \sigma_2 = v_{\bar{z}}.
\end{eqnarray*}

\noindent Moreover, the operator

\begin{eqnarray*}
\Phi &=& \partial_z(\partial_z - v^{n-1} \partial_z^{-1} v^{n-1})
=\partial_z G \partial_z G^{-2} v^{-(n-1)} \partial_z G \partial_z^{-1}
v^{n-1},\\
G &\equiv& {\rm e}^{\partial_z^{-1}v^{n-1}}
\end{eqnarray*}

\noindent is a recursion operator, \index{recursion operator}
namely, if $\sigma$ is a symmetry
of the system (6), then so is

\begin{equation}
\sigma^{(n)} \equiv \Phi^n \sigma.
\end{equation}

To prove that $\Phi$ is a recursion operator one only needs to check that
[5]

\begin{equation}
D_\Delta \Phi = \tilde{\Phi} D_\Delta
\end{equation}

\noindent for some other operator $\tilde{\Phi}$. Thus it is obvious that
$\Phi^{-1}$ will also be a recursion operator provided the operator
$\tilde{\Phi}$ is invertible. In the present case we find that the operator
$\tilde{\Phi}$ is equal to

\begin{eqnarray*}
\tilde{\Phi} = \Phi - (n-1) G \partial_z^{-1} (v_{zz} + 2 v_z \partial_z
+v^{n-1}v_z) \partial_{\bar{z}}^{-1} (I - \partial_z^{-1} v^{n-1} ).
\end{eqnarray*}

\noindent where $I$ is the identity operator.
To prove that $\tilde{\Phi}$ is invertible, let us check that

\begin{eqnarray*}
\partial_{\bar{z}}( \partial_z^{-1} v^{n-1} - I) &=&
( \partial_z^{-1} v^{n-1} - I) D_{\Delta},\\
\partial_z^{-1} v^{n-1} - I &=& -\partial_z^{-1} ( \partial_z - v^{n-1} )
= - \partial_z^{-1} G \partial_z G^{-1}.
\end{eqnarray*}

\noindent Since now $D_\Delta$ is invertible,

\begin{eqnarray*}
D_\Delta^{-1} &=& ( \partial_z^{-1} v^{n-1} - I)^{-1} \partial_{\bar{z}}
( \partial_z^{-1} v^{n-1} - I)\\
&=& G \partial_z^{-1} G^{-1} \partial_z \partial_{\bar{z}}
\partial_z^{-1} G \partial_z G^{-1},
\end{eqnarray*}

\noindent it follows from equation (9) that $\tilde{\Phi}$ is also
invertible. Finally we can generate infinite many symmetries
of the system (6) starting from the simple symmetries and using the
recursion operators $\Phi^{\pm 1}$. Here are two things needed to be
mentioned. First, one cannot expect to generate new symmetries
from $\sigma_2$ by applying possitive powers of $\Phi$ because
$\Phi \sigma_2 = 0$. Second, we tried but failed to prove the hereditary
property for the recursion operator $\Phi$. According to Refs. [9, 10],
integrability requires heriditary operators. Thus there may exist other
recursion operators which possess the hereditary property in our system.

Now let us return to the heterotic Liouville system (2,3). Besides the
infinite many symmetries found above,
we also find that the system (2,3) has interesting geometrical
implications. Recall that the fact that Liouville equation corresponds to
surfaces of constant Gauss curvature in Euclidean three-space is nothing
but a special case of the more general theorem [3] saying that for a
general two-dimensional surface with local conformal coordinates $
x$, $y$ embeded in Euclidean three-space, the Gauss curvature $K$ can be
represented by the induced metric ${\rm d}r^2 = {\rm e}^\varphi ( {\rm d}x^2
+{\rm d}y^2)$ as

\begin{eqnarray*}
K = -\frac{1}{2}{\rm e}^{-\varphi}( \partial_x^2 +\partial_y^2) \varphi.
\end{eqnarray*}

\noindent Choosing $z=x+iy$ and $\bar{z} =x-iy$, the above equation is turned
into

\begin{eqnarray*}
K = -\frac{1}{2}{\rm e}^{-\varphi} \partial_z \partial_{\bar{z}} \varphi.
\end{eqnarray*}

\noindent One thus realizes that the heterotic Liouville system (2,3)
corresponds to the two-dimensional surfaces of non-constant curvature

\begin{eqnarray*}
K=n(n-1)\psi^{n-2},
\end{eqnarray*}

\noindent and the change of this curvature along the $\bar{z}$-direction
is determined by equation (3).

It seems interesting to mention that the first nontrivial extension of
Liouville in the system (2,3), {\it i.e.} the case of $n=3$, corresponds
to the so-called $(2,\;1)$-Toda model constructed from the zerocurvature
equation for a pair of Lax connections taking values in the Witt algebra
[6],

\begin{eqnarray*}
& &[ \partial_z - A_z,\; \partial_{\bar{z}} - A_{\bar{z}} ] = 0,\\
& &A_z = \partial_\varphi L_0 + 3\psi L_1 + L_2,\\
& &A_{\bar{z}} = -{\rm exp} ({\rm ad} \varphi L_0)L_{-1},\\
& & [ L_n,\;L_m ] = (n-m) L_{n+m}.
\end{eqnarray*}

As the Lax representation is of crucial importance in the theory of classical
integrable systems, we would like to explore the Lax representations for the
Liouville systems with $n>3$. However we have been unable to solve this
problem so far.

In the end of this chapter, we would like to point out some other open
problems.

(1) Although we have found infinite many symmetries of the system (2,3),
the problem of finding the complete set of symmetries is still opening. It
is particularly interesting to see what kind of algebra it would be for this
complete set of symmetries.

(2) As the Liouville system has important applications in two-dimensional
gravitational theories, it would be an interesting problem to see if there
exist physically interesting models of two-dimensional gravity corresponding
to the heterotic Liouville systems.

(3) As far as we know, almost all the well-studied integrable nonlinear
partial differential equations have close relations to Ricatti equations.
In the present case, the role of Ricatti equation is replaced by Bernoulli
equation. Therefore it is interesting to ask whether the heterotic
Liouville systems belong to a new integrable class of nonlinear partial
differential equations, and if so, are there exist any other
integrable systems beloning to this new integrable class?
We hope these problems can be solved in the subsequent investigations.

%% Ber ref %%

\chapter{Two-Extended Toda Fields in
Three-Dimensions}

\section{Introduction}
Two-dimensional integrable field theories have proven to be very fruitful
over the last twenty years. Among such models, Toda field theories
received particular attentions because they are related to most of the
important subjects of modern theoretical physics. For examples, the conformal
Toda models played important roles in the investigations of extended
conformal algebras (W algebras) and W gravities, their affine and
conformal affine analogues have been shown to be interesting models
as solitonic equations, and also as prototypes of critical-offcritical
conformal field theories, and the quantum Toda field theories are among the
important quantum integrable field theories which admit the beautiful
quantum group symmetries and/or factorizable S-matrices. Besides all these,
the study of Toda field theories really helps to establish and understand
the systematic methods for treating two-dimensional integrable systems.

Recently, accompanying the investigations of the so-called W-infinity
($W_\infty$, $W_{1+\infty}$ and $w_\infty$) algebras, there arose a wide
interests of studying certain kind of three-dimensional integrable models,
especially the well-known KP hierarchy and the ``continuous limit''
of Toda theories [1-4, 7-8].
The latter, being related to $w_\infty$ algebra [1-3] and
real Euclidean self-dual Einstein gravity [5] and possessing physically
non-trivial instanton solutions [4], has been studied by numerous authors
from various view points. However, as far as we know, the study of
three-dimensional Toda model has not been put forward to the same extent
as in the two-dimensional case. One of the still opening question asks that,
in the three-dimensional case, whether there exist any structure like the
``fundamental Poisson relation'' in the sense of the {\it St Petersburg}
(former Leningrad) group [6], or, whether one can treat the three-dimensional
integrable models using  the hamiltonian techniques developed for
two-dimensional integrable models.

It seems to us that it may be too ambitious to answer the last question
at the present stage because we have not even made clear enough by what
the term ``integrability'' in three-dimensions is meant. In two-dimensional
case, a system of nonlinear partial differential equations is said to be
integrable  if it can be represented by the following ``zero-curvature''
\index{zero-curvature} equation

\begin{displaymath}
[ \partial_+ - A_+, \ \  \partial_- - A_-] = 0,
\end{displaymath}

\noindent where the potentials $A_\pm$ are usually Lie algebra-valued.
This definition of integrability is certainly different from the
classical concept of Liouville integrability. It is sometimes referred to
as Lax integrability because the ``zero-curvature'' equation is
just the compatibility condition of the Lax pair $\partial_\pm T
= A_\pm T$, and it has been proved that the Lax integrability is reduced
to Liouville integrability for two-dimensional systems if and only if
the Lax operator $A_1 = \frac{1}{2}( A_+ - A_-) $ possesses a
classical $r$-matrix structure [18].
However, for the three-dimensional case,
even the Lie algebra-valued Lax potentials $A_\pm$ are hard to find for most
systems. So whenever we are speaking of integrable three-dimensional models,
we are talking about those models which are solved exactly in some way.
Therefore, there seems to be a very long way to establish a systematic
approach for three-dimensional integrable field theories.

Fortunately, due to the development of the concept of contragradient
continuum Lie algebras by Saveliev and Vershik [7],
one is now able to establish the
Lax pair representations for numbers of three-dimensional models. The
simplest example is just the three-dimensional Toda model mentioned above.
Although the continuum Lie algebra-valued Lax pair representation for
three-dimensional Toda model \index{ Toda ! three-dimensional}
looks rather formal at first sight, it seems to us that
this is precisely the right way to generalize the theories of
two-dimensional integrability to the case of three-dimensions. In this
chapter, we shall study the three-dimensional generalization of the
two-extended Toda model proposed by us sometime earlier [9]. We shall try
to generalize many of the concepts and methods of two-dimensional
integrable models to the three-dimensional case based on this model.
As a by-product, we point out that the model in consideration should
correspond to a new  type of W-infinity algebra, probably may be denoted
by $w_\infty^{(2)}$. The explicit structure of this algebra is planed
to be calculated in the future.

\section{Review of the continuum Lie algebras}
\index{continuum Lie algebra}

Before constructing the three-dimensional two-extended Toda model, let us
first give a brief review of the continuum Lie algebras. Due to Saveliev
and Vershik [7],
the contragradient continuum Lie algebra ${\cal G}(E, {\cal K}, S)$
is defined as the quotient algebra ${\cal G'}(E, {\cal K}, S)/J$,
where $E$ is a vector space over some field $\phi$, ${\cal K}$
and $S$ are bilinear mappings $E \times E
\rightarrow E$, ${\cal G'}$ is the Lie algebra freely generated by the
``local part'' ${\cal G'} \equiv
{\cal G_{-1}} \oplus {\cal G_0} \oplus {\cal G_{+1}}$
through the relations

\begin{eqnarray}
&[X_0(\varphi), \ X_0(\psi)] = 0,
{}~~~~[X_0(\varphi), \ X_{\pm 1}(\psi)] = \pm X_{\pm 1}({\cal K}(\varphi,
\psi)),&
\nonumber\\
&[ X_{+1}(\varphi ), \ X_{-1}(\psi ) ] = X_{0} (S(\varphi , \psi )),
\nonumber&
\end{eqnarray}

\noindent where $\varphi , \ \psi \in E,$ and $J$ is the
largest homogeneous ideal having a trivial intersection with ${\cal G_0}$.
In order that the above relations really defines a Lie algebra structure,
the mappings ${\cal K}$ and $S$ have to subject to some additional
constraints. As a special case one can choose $E$ to be a commutative
associative algebra with the multiplication $\bullet$, and
${\cal K}$ and $S$ have a {\em linear} form, say,

\begin{displaymath}
{\cal K} (\varphi,\;\psi)={\cal K}(\varphi) \bullet \psi,
{}~~~~S(\varphi,\;\psi)=S(\varphi \bullet \psi).
\end{displaymath}

\noindent In these cases one can futher choose $S=id$, which corresponds
to the so-called {\em standard form} [7]. In this chapter, we are
particularly
interested in the standard contragradient continuum Lie algebras with
$E$ being the algebra of $C^\infty$ functions on some one-dimensional
manifold ${\cal M}$ with the local coordinate $t$.
In such cases, one can replace the generating relations by the following
relations between the ``kernel generators'',

\begin{displaymath}
[h(t), h(t')] = 0, \ [h(t), e_\pm(t')] = \pm K(t, t')e_\pm (t'), \
[e_+ (t), e_- (t')] = \delta (t-t') h(t),
\end{displaymath}

\noindent where $K(t, t')$ is called Cartan operator
\index{Cartan operator ! kernel of}, or exactly speaking,
the kernel of the Cartan operator ${\cal K}$ (we shall use the term
{\it Cartan operator} for both $K(t, t')$ and ${\cal K}$ with
abuse of terminology), and
$X_i(\varphi ) \equiv \int dt X_{i}(t)\varphi (t),$ $X_0 (t) \equiv h(t),
\;X_{\pm 1}(t) \equiv e_\pm (t).$
In general, the Cartan operator may be an integral
operator possessing a continuous spectrum
(in contrast to the Kac-Moody algebra in which the Cartan matrix possesses
a ``discrete spectrum'', namely its eigenvalues), and it may or may not
be symmetrizable ({\it $K(t, t')$ is symmetrizable if there exist a function
$v(t)$ such that ${\cal Q} (t, t') \equiv K(t, t')v(t') =
K(t', t)v(t)$. The operator ${\cal Q} (t, t')$ is called the symmetrized
Cartan operator}).
\index{Cartan operator ! symmetrized}
In this chapter, we shall always assume that the Cartan
operator is symmetrizable. As a concrete example,
we can choose $K(t, t') = \partial_t^2
\delta (t-t')$, which is itself symmetric under $t \leftrightarrow t'$
(and this algebra corresponds to the continuous limit of the Lie algebra
$A_\infty$). As we shall see in the following context,
the two-extended Toda model
corresponding to this last special choice of $K$ is actually a generalization
of the three-dimensional Toda model of Refs.[1-5]. Other choices of $K$
can also give three-dimensional generalizations of the two-extended Toda
model, but the corresponding equations often
appear as integro-differential equations.

The general structure theory for the contragradient continuum Lie algebras
is not established yet. Nevertheless, it is enough for us to know the
fact that the Killing form can be appropriately defined according to
concrete choices of $K$ and $S$, and by definition, the contragradient
continuum Lie algebras are naturally Z-graded. In the case of $S=id$
with a symmetrizable Cartan operator $K(t, t')$, the Killing form
can be defined as

\begin{displaymath}
\langle h(t), \; h(t') \rangle = K(t, t')v(t') = {\cal Q} (t, t'),
{}~~~~\langle e_+(t), \; e_-(t') \rangle = v(t)\delta (t-t').
\end{displaymath}

\noindent Particularly, if $K(t, t')=\partial_t^2 \delta (t-t')$, the
function $v(t)$ can be chosen to be the constant 1, and it
was shown in Ref.[8] that there exist highest weight representations for
the corresponding Lie algebra
${\cal G}(C^\infty {\cal M}, {\cal K}, id)$, with the highest weight state
denoted by $\left| \tau \right \rangle$. These materials are all what
is needed for our constructions.

\section{Two-extended Toda model in three-dimensions}

With the above mathematical preparation given, let us now go on to the
central subject of this chapter---the two-extended Toda model in
three-dimensions.

The two-dimensional case of this model is studied by us in a series of
chapters [9,10], in which the W-algebra symmetries,
fundamental Poisson relation,
chiral exchange algebras, general solution and the Wronskian-type special
solution in relation to the WZNW reduction and classical W-surfaces are
made clear. The crucial difference between the two-extended Toda model and
the standard one lies in that, in  the standard case, the Lax connections
$A_\pm$ take values respectively in the subspaces ${\cal G_0} \oplus
{\cal G^{(\pm 1)}}$ of the underlying Lie algebra ${\cal G}$
(where ${\cal G}^{(\pm i)}$
denote the $i$-th graded sector of the Lie  algebra ${\cal G}$), whilst
for the two-extended model, these connections take values in the subspaces
${\cal G_0} \oplus {\cal G^{(\pm 1)}} \oplus {\cal G^{(\pm 2)}}$. It is
precisely this difference that makes the two-dimensional two-extended
Toda model having the extended conformal symmetry algebra $W[{\cal G}, H, 2]$,
in contrast to the standard $W[{\cal G}, H, 1]$ symmetry
algebra for the usual Toda model (we are using the
convention of Ref.[10], where the symbol $W[A_N, principal \; gradation, 1]$
corresponds to the usual $W_{N+1}$ algebra).

Let us be more concrete. The Lax pair of the two-extended Toda model can
be written

\begin{equation}
\partial_\pm T = A_\pm T,
{}~~~~A_\pm = \pm \left[ \frac{1}{2} \partial_\pm \Phi + \exp \left(
\mp \frac{1}{2} {\rm ad} \Phi \right) \bar{\Psi}_\pm + \exp \left(
\mp \frac{1}{2} {\rm ad} \Phi \right) \mu_\pm \right], \label{eq:1}
\end{equation}

\noindent where, in the two-dimensional case, the fields $\Phi$ takes
value in the Cartan subalgebra of the Kac-Moody algebra ${\cal G}$,
$\bar{\Psi}_\pm$ lie in ${\cal G^{(\pm 1)}}$, and $\mu_\pm$ are constant,
regular, representative elements of ${\cal G^{(\pm 2)}}$, respectively.

Now in order to generalize this model into the three-dimensional case,
we require that the fields $\Phi, \; \bar{\Psi}_\pm$ and the constants
$\mu_\pm$ be continuum Lie algebra-valued. That means, the above quantities
can be rewritten in the form

\begin{eqnarray}
&\Phi(x_+, \; x_-) \equiv \int dt h(t) \varphi(x_+, \; x_-, \; t)&
\nonumber\\
&\Psi_\pm (x_+, \; x_-) \equiv \int dt e_\mp (t) \psi_\pm
(x_+, \; x_-, \; t)& \label{eq:2}\\
&\mu_\pm \equiv \pm \frac{1}{2} \int dt dt'
\Omega (t, t') [e_\pm (t), \; e_\pm (t')],
&\nonumber
\end{eqnarray}

\noindent where $\Omega (t, \; t')$ is some antisymmetric function of
$t$ and $t'$, and

\begin{eqnarray}
&\bar{\Psi}_\pm (x_+, \; x_-) \equiv \pm [ \mu_\pm, \; \Psi_\pm ], &
\nonumber\\
&= \frac{1}{2} \int dt dt' dt_1 \Omega (t, t') \psi_\pm (x_+, \; x_-, \; t_1)
[ [ e_\pm (t), \; e_\pm (t')], \; e_\mp (t_1) ]& \label{eq:3}\\
&= \int dt dt' \Omega (t, t') K(t, \; t') \psi_\pm (x_+, \; x_-, \; t)
e_\pm (t').& \nonumber
\nonumber
\end{eqnarray}

With these definitions in mind, we are now ready to write down the equations
of motion for the two-extended Toda model. This can be down by first
calculating the compatibility condition of the Lax pair, which gives the
result

\begin{eqnarray}
&\partial_+ \partial_- \Phi + [ \exp ({\rm ad} \Phi ) \bar{\Psi}_-, \;
\bar{\Psi}_+ ] + [ \exp ({\rm ad} \Phi ) \mu_-, \;
\mu_+ ] =0,&\nonumber \\
&\partial_+ \Psi_- = \exp (-{\rm ad} \Phi ) \bar{\Psi}_+, &\label{eq:4}\\
&\partial_- \Psi_+ = \exp ({\rm ad} \Phi ) \bar{\Psi}_-,& \nonumber
\end{eqnarray}

\noindent and then substituting the definitions (\ref{eq:2})-(\ref{eq:3})
into (\ref{eq:4}). It
finally follows that

\begin{eqnarray}
& &\partial_+ \partial_- \varphi (x_+, x_-, t) - \int dt_1 dt_2
\Omega (t_1, t) \Omega (t_2, t) K(t_1, t) K(t_2, t) \nonumber\\
& &~~~~\times \psi_- (x_+, x_-, t_1) \psi_+ (x_+, x_-, t_2)
\Xi (t) \nonumber\\
& &~~~~+ \int dt_1 \Omega^2 (t_1, t) K(t_1, t) \Xi (t_1) \Xi (t) = 0,
\nonumber\\
& &\partial_+ \psi_- (x_+, x_-, t) = \int dt_1 \Omega
(t_1, t) \psi_+ (x_+, x_-, t_1) K(t_1, t) \Xi (t),\label{eq:5}\\
& &\partial_- \psi_+ (x_+, x_-, t) = \int dt_1 \Omega
(t_1, t) \psi_- (x_+, x_-, t_1) K(t_1, t) \Xi (t),\nonumber
\end{eqnarray}

\noindent where the explicit dependence of $\Xi$ on $x_\pm$ is omitted
for brief of notations,

\begin{equation}
\Xi (t) = \exp \left(- \int d\tilde{t}
K(\tilde{t}, t) \varphi (x_+, x_-, \tilde{t}) \right).\label{eq:6}
\end{equation}

The system (\ref{eq:5}) of integro-differential
equations appears to be rather
complicated at a first glance. However, provided the Cartan operator
$K(t, t')$ and (correspondingly) the function $\Omega (t, t')$ are
appropriately chosen, the three-dimensional two-extended Toda model
can be rewritten in a very neat form. For example, if we choose the
underlying contragradient continuum Lie algebra to be
${\cal G} (C^\infty {\cal M}, \;
\partial_t^2 \delta (t-t'), \; id)$ and let $\Omega (t, t') = t-t'$,
eq.(\ref{eq:5}) can be rewritten

\begin{eqnarray}
& &\partial_+ \partial_- \varphi - 4 \partial_t \psi_- \partial_t \psi_+
\exp (- \partial_t^2 \varphi ) \nonumber\\
& &~~~~+ 2 \exp (- 2 \partial_t^2 \varphi )=0,\nonumber\\
& &\partial_+ \psi_- = 2 \partial_t \psi_+ \exp (- \partial_t^2 \varphi ),
\label{eq:7}\\
& &\partial_- \psi_+ = 2 \partial_t \psi_- \exp (- \partial_t^2 \varphi ),
\nonumber
\end{eqnarray}

\noindent which is exactly an extension of the ``three-dimensional Toda
model'' studied by several authors.

The reason for the above system to be
interesting to study is that, first,
this system of  equations is really a continuous limit of the
two-dimensional $A_\infty$ two-extended Toda model, just as the usual
three-dimensional Toda model

$$\partial_+ \partial_- \varphi +
\exp (- \partial_t^2 \varphi ) =0$$

\noindent is the continuous limit of the
two-dimensional $A_\infty$ Toda model; Second, as a direct extension of
the three-dimensional Toda model, this system is expected to possess
many extended characteristics of the three-dimensional Toda model, such as
a generalized $w_\infty$ symmetry, extended self-dual Einstein spaces,
{\it etc.} Actually, already in the two-dimensional case, the two-extended
Toda model with the underlying Lie algebra $A_\infty$ was suggested [9] to
possess a conformal symmetry algebra which can be denoted $W_\infty^{(2)}$,
which is the generalization to the case of integer-half integer conformal
spectrum of the usual nonlinear $W_\infty$ algebra or, the
large $N$ limit of the $W_N^{(2)}$ algebra. The conformal algebras with
integer-half integer spectra were referred to as ``bosonic superconformal
algebra'' by Fuchs [11] and by us. Using this terminology, the algebra
$W_\infty^{(2)}$ may be called ``bosonic super W-infinity'' algebra. The
$w_\infty^{(2)}$ algebra is supposed to be a linear variant of
$W_\infty^{(2)}$, just as $w_\infty$ is a linear variant of $W_\infty$.
However, the explicit structure of the algebras $W_\infty^{(2)}$ and
$w_\infty^{(2)}$ is still difficult to be constructed. The central
difficulty is that, in the two-extended case, it is not as easy as in the
usual case to choose as a good basis a complete set of chiral conserved
quantities. Nevertheless, there should be no problem on the
existence of such bosonic superconformal algebras. We hope the difficulty
in choosing a basis for such algebras could be overcome in the future.

It might be interesting to note that the system (\ref{eq:7}) really admits
physically interesting solutions. For example, the instanton-like
solution \index{solution ! instanton-like}

\begin{eqnarray}
\displaystyle &\varphi = \int^t dt_1 \int^{t_1} dt_2 \; \ln \left[
\frac{1}{4} (t_2 - a) (t_2 -b) \frac{ \partial_+ f_+ (x_+) \partial_-
f_- (x_-)}{(1-f_+ (x_+) f_- (x_-))^2} \right],&\nonumber\\
&\psi_\pm = g_\pm (x_\pm)& \nonumber
\end{eqnarray}

\noindent explicitly solves eq.(\ref{eq:7}),
where $f_\pm$ and $g_\pm$ are arbitrary
functions of the arguments $x_\pm$. To obtain more solutions of the system,
we have to generalize the techniques for solving two-dimensional
Toda-type models. But in the present section, we would rather introduce
the effective action for the three-dimensional two-extended Toda model and
leave the task of generalizing the techniques for solving two-dimensional
Toda-type models to the next section.

The effective action \index{effective action! 3-d two-extension of Toda}
for the system (\ref{eq:4})-(\ref{eq:5}) read [9]

\begin{eqnarray}
& &I(\Phi, \Psi_\pm) = \frac{1}{4} \int dx_+ dx_- \langle
\partial_+ \Phi \partial_- \Phi + \bar{\Psi}_+ \partial_- \Psi_+ +
\bar{\Psi}_- \partial_+ \Psi_- \rangle \nonumber\\
& &~~~~- \frac{1}{2}
\int dx_+ dx_- \left\langle \exp (-{\rm ad} \Phi) (\bar{\Psi}_+)\bar{\Psi}_-
+ \exp (-{\rm ad} \Phi) (\mu_+)\mu_- \right\rangle \label{eq:8}\\
& &~~~~= \frac{1}{4} \int dx_+ dx_- dt\; v(t) \left\{
\partial_+ \varphi ({\cal K})(\partial_- \varphi) \right. \nonumber\\
& &~~~~+ ({\cal K\Omega})(\psi_+ )\partial_- \psi_+ + ({\cal K\Omega})
(\psi_- )\partial_+ \psi_- \nonumber\\
& &~~~~\left. - 2({\cal K\Omega})(\psi_+) \; ({\cal K\Omega})(\psi_-) \Xi
+ ({\cal K\Omega^2})(\Xi)\Xi \right\},\label{eq:9}
\end{eqnarray}

\noindent where

\begin{eqnarray}
&({\cal K})(f) \equiv \int dt_1 K(t_1, t) f(t_1),&\nonumber\\
&({\cal K\Omega})(f)\equiv \int dt_1 K(t_1, t) \Omega (t_1, t) f(t_1),&
\nonumber\\
&({\cal K\Omega^2})(f) \equiv \int dt_1 K(t_1, t)
\Omega^2 (t_1, t) f(t_1).\nonumber&
\end{eqnarray}

\noindent In the particular case of ${\cal G} (C^\infty {\cal M}, \;
\partial_t^2 \delta (t-t'), \; id)$ model with $\Omega (t, t') = t-t'$,
the above action can be rewritten

\begin{eqnarray}
& &I(\varphi, \psi_\pm) = \frac{1}{4} \int dx_+ dx_- \left\{
\partial_+ \varphi \partial_- \partial_t^2 \varphi \right.\nonumber\\
& &~~~~+ 2\partial_t \psi_+
\partial_- \psi_+ + 2\partial_t \psi_- \partial_+ \psi_-\nonumber\\
& &~~~~\left.
-8\partial_t \psi_+ \partial_t \psi_- \exp (-\partial_t^2 \varphi)
+2 \exp (-2\partial_t^2 \varphi) \right\}.\label{eq:10}
\end{eqnarray}

Notice that the last action is of the fourth-order in derivatives
for the field $\varphi$. Such an action does not describe a
three-dimensional relativistic field theory in the usual sense. Actually,
the extra dimension $t$ is an algebraical dimension
which has different meaning in contrast to the other two space-time
dimensions.

Since we are not experienced in treating fourth-order actions, now we
prefer to define the canonical Poisson brackets
for the original form (\ref{eq:8}) of the action. Remembering that the
fields $\Phi,\;\Psi_\pm$ are just continuum Lie algebra-valued
{\it two-dimensional} fields, and the action (\ref{eq:8}) for these fields
is of the second order, we can define the Poisson bracket for these fields in
the usual way. That means, we can define the canonical conjugate momenta
$\Pi_\Phi,\;\Pi_{\Psi_\pm}$ as

\begin{equation}
\Pi_\Phi \equiv \frac{\delta {\cal L}}{\delta \partial_0 \Phi}
= \frac{1}{2} \partial_0 \Phi,
{}~~~~\Pi_{\Psi_\pm} \equiv \frac{\delta {\cal L}}{\delta \partial_0 \Psi_\pm}
= \frac{1}{2} \bar{\Psi}_\pm,\label{eq:11}
\end{equation}

\noindent where $x_0$ and $x_1$ are defined as $x_\pm = x_0 \pm x_1$,
and introduce the canonical equal-time $x_0$ Poisson brackets

\begin{eqnarray}
&\left\{\Pi_\Phi (x_1)\;\otimes,\; \Phi (x_1')\right\} = \delta (x_1 - x_1')
\int dtdt' {\cal Q^{-1}} (t, t') h(t) \otimes h(t'),&\nonumber\\
&\left\{\Pi_{\Psi_\pm} (x_1)\;\otimes,\;\Psi_\pm (x_1')\right\} =
\delta (x_1 - x_1') \int dt v^{-1} (t) e_\pm(t) \otimes e_\mp (t),&
\label{eq:12}
\end{eqnarray}

\noindent where ${\cal Q^{-1}} (t, t')$ is the formal inverse of the
symmetrized Cartan operator ${\cal Q} (t, t')$,

\begin{displaymath}
\int dt {\cal Q^{-1}} (t_1, t){\cal Q} (t, t_2)
= \int dt {\cal Q} (t_1, t){\cal Q^{-1}} (t, t_2)
= \delta (t_1 - t_2).
\end{displaymath}

\noindent Remembering the definitions (\ref{eq:11}) of the fields
$\Pi_\Phi,\;\Pi_{\Psi_\pm}$ and requiring  the Poisson bracket
for the component fields $\varphi,\;\psi_\pm$ to be consistent with
the above-defined ones, we find that the  Poisson bracket
for the fields $\varphi,\;\psi_\pm$  may be appropriately defined as

\begin{eqnarray}
&\left\{ \frac{1}{2}({\cal Q})(\partial_0 \varphi) (x_1, t),
\varphi (x_1', t') \right\} = \delta (x_1 -x_1') \delta (t-t'),&\nonumber\\
&\left\{ \frac{1}{2}({\cal K\Omega})(\psi_\pm) (x_1, t), \psi_\pm (x_1', t')
\right\} = \delta (x_1 -x_1') \delta (t-t').&\label{eq:13}
\end{eqnarray}

\noindent These Poisson brackets are exactly what will arise if we treat
directly the action (\ref{eq:9}) and define the Poisson brackets in the usual
way.
Therefore, in spite of the different meaning of the variable t, the Poisson
brackets can be defined safely using the usual method.

\section{Fundamental Poisson relation}
As mentioned in the last section, the hamiltonian method of treating
higher-dimensional ($d>2$) integrable systems is still lacking, such
as the fundamental Poisson structure and classical $r$-matrices.
For some particular examples, such as the self-dual Yang-Mills theory
(where the Lax-type linear systems were found for a long
time by the use of prolongation method [12]),
the classical $r$-matrix has been found by L.L.Chau {\it et al} [13] using
the $J$-field formulation. However, up to now, there is
very little knowledge about whether there may be a
fundamental Poisson relation for the transport matrices since the transport
matrix depends on all of the four space-time variables, and the Poisson
bracket for such matrices cannot be obtained easily by integrating
those for the potentials in the linear systems.

Fortunately, in the three-dimensional Toda-type models, the ``transport
operator'' $T$ does not depend on the third variable $t$, and thus we can
get the Poisson bracket for $T$ by directly integrating the Poisson bracket
for the Lax connections.

Let us first calculate the Poisson bracket $\{ A_1 (x_1) \otimes,\;
A_1 (x_1')\}$ with $A_1 \equiv \frac{1}{2}(A_+ - A_-)$. By direct
calculation, we have

\begin{eqnarray}
& & \{ A_1 (x_1) \otimes,\; A_1 (x_1')\} = \frac{1}{2}
\int dtdt' \; K(t, t')\Omega (t, t') \Xi^{1/2} (t') \nonumber\\
& &~~~~\times \left\{ \psi_+ (t) \left[ e_+ (t') \otimes h(t') -
h(t') \otimes e_+ (t') \right] \right.\nonumber \\
& &~~~~+ \left. \psi_- (t) \left[ e_+ (t') \otimes h(t') - h(t') \otimes
e_+ (t') \right] \right\} \delta (x_1 - x_1')\nonumber \\
& &~~~~+ \frac{1}{2}\int dtdt' \; \Omega (t, t') \Xi^{1/2} (t)\Xi^{1/2} (t')
\nonumber\\
& &~~~~\times \left\{ [ e_+ (t), e_+ (t')] \otimes \left( h(t) + h(t')
\right) \right. \nonumber\\
& &~~~~- \left( h(t) + h(t')\right) \otimes [ e_+ (t), e_+ (t')] \nonumber\\
& &~~~~- [ e_- (t), e_- (t')] \otimes \left( h(t) + h(t')\right) \nonumber \\
& &~~~~+ \left( h(t) + h(t')\right) \otimes [ e_- (t), e_- (t')]\nonumber\\
& &~~~~+ 2K(t, t') \left[ e_+(t) \otimes e_+(t')
- e_+(t') \otimes e_+(t) \right.
\nonumber\\
& &~~~~+ \left.
\left. e_-(t) \otimes e_- (t') - e_-(t') \otimes e_-(t) \right]
\right\} \delta (x_1-x_1').\label{eq:14}
\end{eqnarray}

\noindent Equation (\ref{eq:14}) can be rewritten as

\begin{equation}
\{ A_1 (x_1) \otimes,\; A_1 (x_1')\} = [r,
A_1 (x_1) \otimes 1 + 1 \otimes A_1 (x_1')] \delta (x_1-x_1')  \label{eq:15}
\end{equation}

\noindent where $r$ is a ${\cal G}\otimes{\cal G}$ valued constant,
which may be called ``$r$-operator'', \index{$r$-operator}

\begin{eqnarray}
&r = \frac{1}{2}\sum_{a}\sum_{n=1}^{\infty}
{ \int dt_1  ...  \int dt_n } \left\{
e_+^{(a)} (t_1,\; ... ,\;t_n) \otimes
e_-^{(a)} (t_1,\; ... ,\;t_n) \right.&\nonumber\\
&\left. - e_-^{(a)} (t_1,\; ... ,\;t_n) \otimes
e_+^{(a)} (t_1,\; ... ,\;t_n) \right\}
+ \lambda C,~~~~ \lambda \ {\rm arbitrary}&\label{eq:16}
\end{eqnarray}

\noindent where $e_\pm^{(a)} (t_1,\; ... ,\;t_n) \in {\cal G}^{(\pm n)}$,
the superscript $^{(a)}$ indicates different (linearly independent)
elements of ${\cal G}^{(\pm n)}$, which are assumed to be normalized
such that

\begin{displaymath}
\langle e_\pm^{(a)} (t_1,\; ... ,\;t_n), \;
e_\mp^{(b)} (t_1',\; ... ,\;t_n') \rangle = \delta^{ab} \delta (t_1-t_1')
 ...  \delta (t_n-t_n'),
\end{displaymath}

\noindent and the summation over $a$ is taken over
all such elements. The constant $C$ is the ``tensor Casimir operator'' of the
continuum Lie algebra ${\cal G}$ satisfying the conditions

\begin{eqnarray}
&[C, {\cal A}\otimes 1 + 1 \otimes {\cal A}] = 0,&\nonumber\\
&\langle C, \; {\cal A}\otimes 1 \rangle =
\langle C, \; 1 \otimes{\cal A} \rangle = {\cal A},
{}~~~~\forall {\cal A} \in {\cal G}.&\nonumber
\end{eqnarray}

\noindent Explicitly, we can write

\begin{eqnarray}
&C=\int dtdt' {\cal Q^{-1}} (t, t') h(t) \otimes h(t')&\nonumber\\
&+ \sum_{a}\sum_{n=1}^{\infty} { \int dt_1  ...  \int dt_n } \left\{
e_+^{(a)} (t_1,\; ... ,\;t_n) \otimes
e_-^{(a)} (t_1,\; ... ,\;t_n)  \right. &\nonumber\\
&\left. + e_-^{(a)} (t_1,\; ... ,\;t_n) \otimes
e_+^{(a)} (t_1,\; ... ,\;t_n) \right\} . \nonumber
\end{eqnarray}

\noindent Of all the choices of $\lambda$, two special values $\lambda =
\pm \frac{1}{2}$ are particularly important because the corresponding
$r$-operators \index{$r$-operator}

\begin{eqnarray}
&r_\pm = \pm \frac{1}{2} \left\{ \int dtdt' {\cal Q^{-1}} (t, t')
h(t) \otimes h(t') \right.& \nonumber\\
&\left. + 2 \sum_{a}\sum_{n=1}^{\infty}
{ \int dt_1  ...  \int dt_n } e_\pm^{(a)} (t_1,\; ... ,\;t_n) \otimes
e_\mp^{(a)} (t_1,\; ... ,\;t_n) \right\} .& \label{eq:17}
\end{eqnarray}

\noindent satisfy the classical Yang-Baxter equation
\index{Classical Yang-Baxter equation}

\begin{displaymath}
 [ r_\pm^{12}, r_\pm^{13}] + [ r_\pm^{12},  r_\pm^{23}] + [ r_\pm^{32},
 r_\pm^{13}] = 0.
\end{displaymath}

\noindent The fundamental Poisson relation for $T$ is then easily obtained
by integrating the equation $\partial_1 T = A_1 T$, yielding

\begin{equation}
\left\{ T(x_1) \otimes, T(x_1) \right\} = [r_\pm, T(x_1) \otimes T(x_1) ].
\label{eq:18}
\end{equation}

\section{Exchange algebra and dressing transformation}
Let us now proceed in analogy to the two-dimensional case. In the following,
we shall assume that the highest weight representations for ${\cal G}$
exist, and the highest weight vector $\left| \tau \right\rangle$ satisfies

\begin{eqnarray}
&h(t) |\tau \rangle = \tau (t) |\tau \rangle,
{}~~~~e_+(t) |\tau \rangle = 0, ~~~~ \forall t,&\nonumber\\
&\langle \tau | \tau \rangle = 1.&\nonumber
\end{eqnarray}

Notice that the Lax pair (\ref{eq:1})
admits a gauge freedom $T \rightarrow gT$ with
$g \in G$, the underlying continuum Lie group. Using this gauge degree of
freedom, we can transform the Lax connections $A\pm$ such that
one of them takes the form
$A_\pm \in {\cal G}^{(\pm 1)} \oplus {\cal G}^{(\pm 2)}$ or
$A_\pm \in \oplus {\cal G}^{(\pm 2)}$. Then projecting the resulting
transformed Lax pairs onto the states $|\tau \rangle$, $\int dt e_-(t)
|\tau \rangle$ and their dual states, we can get two sets of ``chiral''
vectors which means that they depend on  only one space-time variable,
$x_+$ or $x_-$.

To be explicit, we have the following chiral vectors,

\begin{eqnarray}
&\xi_1^+ (x_+) =\langle \tau | \exp (\frac{1}{2} \Phi )T,
{}~~~~
\xi_1^- (x_-) =T^{-1} \exp (\frac{1}{2} \Phi )| \tau  \rangle,&\nonumber\\
&\xi_2^+ (x_+) =\left\{\langle \tau | \int dt e_+(t) \right\}
\exp (\Psi_+ )\exp (\frac{1}{2} \Phi )T,&\label{eq:x}\\
&\xi_2^- (x_-) =T^{-1} \exp (\frac{1}{2} \Phi )
\exp (\Psi_- )\left\{\int dt e_-(t)| \tau  \rangle\right\},& \nonumber
\end{eqnarray}

\noindent with

\begin{displaymath}
\partial_\pm \xi_a^\mp = 0, ~~~~ a=1,\;2.
\end{displaymath}

\noindent Following the standard method [14], we can show that these chiral
vectors satisfy the exchange algebra [9] \index{exchange algebra}

\begin{eqnarray}
&\left\{ \xi_a^+ (x) \otimes,\; \xi_b^+ (y) \right\} =
\xi_a^+ (x) \otimes \xi_b^+ (y) \left( r_+ \theta (x-y)+ r_-
\theta (y-x) \right),
&\nonumber\\
&\left\{ \xi_a^+ (x) \otimes,\; \xi_b^- (y) \right\} =
-\left(\xi_a^+ (x) \otimes 1 \right) r_-
\left( 1 \otimes \xi_b^- (y)\right),&\nonumber\\
&\left\{ \xi_a^- (x) \otimes,\; \xi_b^+ (y) \right\} =
-\left( 1 \otimes \xi_b^- (y)\right) r_+
\left(\xi_a^+ (x) \otimes 1 \right),&\label{eq:19}\\
&\left\{ \xi_a^- (x) \otimes,\; \xi_b^- (y) \right\} =
\left( r_- \theta (x-y)+ r_+ \theta (y-x) \right)
\xi_a^- (x) \otimes \xi_b^- (y).&\nonumber
\end{eqnarray}

Now let us consider the dressing problem of the three-dimensional
two-extended Toda model. As in the two-dimensional case, the dressing
transformation depends essentially on the factorization of the underlying
Lie group $G$, under which each group element $g$ is factorized as

\begin{equation}
g=g_-^{-1}g_+,\label{eq:20}
\end{equation}

\noindent and the dressing transformation transforms the transport operator
$T$ as

\begin{equation}
T \rightarrow T^g = \Theta_\pm T g_\pm^{-1},
{}~~~~\Theta_-^{-1}\Theta_+ = \Theta \equiv TgT^{-1}.\label{eq:21}
\end{equation}

\noindent At present, the factorization problem is solved by the $r$-operators
 $r_\pm$ as follows,

\begin{equation}
{\cal A}_\pm \equiv {\cal R}_\pm {\cal A} \equiv
\langle r_\pm, \;1 \otimes {\cal A} \rangle_2,
{}~~~~\Rightarrow {\cal A} = {\cal A}_+ -{\cal A}_-,\label{eq:22}
\end{equation}

\noindent which is just the infinitesimal form of the factorization problem.
The fact that the positive and negative transformations of $T$ give rise to
the same $T^g$ implies that the transformed Lax potentials $A_\pm^g$
have the same form compared to the original $A_\pm$. Recalling the concrete
form (\ref{eq:17}) of the $r_\pm$-operators,
we can rewrite the $\Theta_\pm$ operators
in the following Z-graded form,

\begin{equation}
\Theta_\pm = \exp \left(\frac{1}{2} \theta_\pm^{(0)} \right)
\exp (\theta^{(\pm 1)})  ...  \exp (\theta^{(\pm 1)})  ... ,\label{eq:23}
\end{equation}

\noindent where $\theta^{(a)} \in {\cal G}^{(a)}$, and the form-preserving
condition for the Lax connections $A_\pm$ implies that the fields $\Phi,  \;
\Psi_\pm$ must transform as [9]

\begin{eqnarray}
&\Phi^g = \Phi + \theta_+^{(0)} = \Phi - \theta_-^{(0)},
{}~~~~\theta_+^{(0)}+\theta_-^{(0)} = 0,&\nonumber\\
&\Psi_\pm^g = \Psi_\pm \mp \exp \left( \pm \frac{1}{2} {\rm ad} \Phi \right)
\theta^{(\mp 1)}.&\nonumber
\end{eqnarray}

\noindent Note that each element $\theta^{(0)} \in {\cal G}^{(0)}$ can be
written as $\int dt \theta^{(0)} (t) h(t)$, and each $\theta^{(\pm 1)}
\in {\cal G}^{(\pm 1)}$ can be written as
$\int dt \theta^{(\pm 1)} (t) e_\pm(t)$, we can rewrite the above dressing
transformation laws in terms of the component fields,

\begin{eqnarray}
&\varphi^g (x_+,x_-,t) = \varphi (x_+,x_-,t) \pm \theta_\pm^{(0)}
(x_+,x_-,t),&\nonumber\\
&\psi_\pm^g (x_+,x_-,t) = \psi_\pm (x_+,x_-,t) \mp
\Xi^{1/2} (t) \theta^{(\mp 1)} (x_+,x_-,t) .&\nonumber
\end{eqnarray}

It is particularly interesting to note that the chiral vectors $\xi_a^{\pm}$
transform only by a shift of constant group elements,

\begin{displaymath}
(\xi_a^+ )^g = \xi_a^+ g_-^{-1},
{}~~~~(\xi_a^- )^g = g_+ \xi_a^-.
\end{displaymath}

\noindent In order that the above transformations preserve the form of
the chiral exchange algebra (\ref{eq:19}),
the  constant group elements $g_\pm$
must subject to some nontrivial Poisson brackets,

\begin{eqnarray}
&\left\{g_+ \otimes, g_+ \right\} = [r_\pm, \;g_+ \otimes g_+],&\nonumber\\
&\left\{g_- \otimes, g_- \right\} = [r_\pm, \;g_- \otimes g_-],&\nonumber\\
&\left\{g_+ \otimes, g_- \right\} = [r_+, \;g_+ \otimes g_-],&\nonumber\\
&\left\{g_- \otimes, g_+ \right\} = [r_-, \;g_- \otimes g_+].&\nonumber
\end{eqnarray}

\noindent These structures, while considered in the framework of
two-dimensional Toda-type theories, correspond to the semi-classical
limit of the quantum (Kac-Moody) group [15]. Therefore,
it might be interesting
to see whether the above structure can give rise to any structure
like a ``quantum continuum Lie group'' after quantization. We leave
this problem open for later considerations.

\section{Formal solutions via Leznov-Saveliev analysis {\rm [16]} }
\index{solution ! formal}
As in the two-dimensional case, the chiral vectors $\xi_a^\pm$ are also
useful for constructing formal solutions of the two-extended Toda
system (\ref{eq:4})-(\ref{eq:5}).
Recalling the definitions of these chiral vectors, we have

\begin{eqnarray}
&\langle \tau | \exp (\Phi) |\tau \rangle = \xi_1^+ (x_+) \xi_1^- (x_-),&
\nonumber\\
&\langle \tau | \exp (\Phi) \exp (\Psi_-)
\left\{\int dt e_-(t) |\tau \rangle\right\} = \xi_1^+ (x_+) \xi_2^- (x_-),&
\nonumber\\
&\left\{\int dt \langle \tau |e_+(t) \right\}
\exp (\Psi_+)\exp (\Phi)| \tau \rangle = \xi_2^+ (x_+) \xi_1^- (x_-).
\nonumber&
\end{eqnarray}

\noindent In terms of the component fields, the above equations read

\begin{displaymath}
\exp \int dt \varphi (x_+, x_-, t) \tau(t) = \xi_1^+ (x_+) \xi_1^- (x_-),
\end{displaymath}
\begin{equation}
\int dt \psi_- (x_+, x_-, t) \tau (t) = \frac {\xi_1^+ (x_+) \xi_2^- (x_-)}
{\xi_1^+ (x_+) \xi_1^- (x_-)},\label{eq:y}
\end{equation}
\begin{displaymath}
\int dt \psi_+ (x_+, x_-, t) \tau (t) = \frac {\xi_2^+ (x_+) \xi_1^- (x_-)}
{\xi_1^+ (x_+) \xi_1^- (x_-)}.
\end{displaymath}

Let us consider in more detail the above relations and show how one can
obtain formal solutions out of these relations.

Define $T_{L/R} = \exp (\pm \frac{1}{2} \Phi) T$ and decompose them as

\begin{displaymath}
T_L = e^{K_+}N_- M_+,
{}~~~~T_R = e^{K_-}N_+ M_-,
\end{displaymath}

\noindent where $K_\pm \in G_0,\;N_\pm,\;M_\pm \in G_\pm$, we can get
from eq.(\ref{eq:x}) that

\begin{displaymath}
\xi_1^+ (x_+) = \langle \tau | e^{K_+}M_+,
{}~~~~\xi_1^- (x_-) = M_-^{-1} e^{-K_-} |\tau \rangle,
\end{displaymath}

\noindent which shows that $K_\pm$ and $M_\pm$ are chiral objects,

\begin{displaymath}
\partial_\pm K_\mp = \partial_\pm M_\mp =0.
\end{displaymath}

\noindent Furthermore, writing $N_\pm = \exp (\chi^{(\pm 1)})
\exp (\chi^{(\pm 2)})  ... $ as we did in eq.(\ref{eq:23}), we obtain [9]

\begin{displaymath}
\xi_2^+ (x_+) = \left\{\int dt \langle \tau | e_+ (t)\right\}
e^{K_+} \left[1+ e^{-{\rm ad} K_+} P_+ \right]M_+,
\end{displaymath}
\begin{displaymath}
\xi_2^- (x_-) = M_-^{-1}\left[1+ e^{-{\rm ad} K_-} P_- \right] e^{-K_-}
\left\{ \int dt e_- (t) | \tau \rangle \right\},
\end{displaymath}

\noindent where

\begin{displaymath}
P_\pm \equiv \Psi_\pm \pm \exp ({\rm ad} K_\pm) \chi^{(\mp 1)},
{}~~~~ \partial_\pm P\mp =0.
\end{displaymath}

\noindent More detailed calculations [9] show that the operators $M_\pm$ are
not independent of $K_\pm$ and $P\pm$,

\begin{eqnarray}
&\partial_+ M_+M_+^{-1} = \exp (-{\rm ad} K_+) (\bar{P}_+ + \mu_+ ),&
\nonumber\\
&M_-\partial_- M_-^{-1} = \exp (-{\rm ad} K_-) (\bar{P}_- + \mu_- ).&
\nonumber
\end{eqnarray}

\noindent with $\bar{P}_\pm \equiv \pm [ \mu_\pm,\; P_\pm ]$.
Therefore, eqs.(\ref{eq:y}) become

\begin{displaymath}
\exp \int dt \varphi (x_+, x_-, t) \tau(t) =
\langle \tau | e^{K_+}M_+M_-^{-1} e^{-K_-} |\tau \rangle,
\end{displaymath}
\ben
& &\int dt \psi_- (x_+, x_-, t) \tau (t) \\
& &~~~~~~~~= \frac
{\langle \tau | e^{K_+}M_+M_-^{-1} [1+ \exp (-{\rm ad}K_-)P_-]
e^{-K_-} \int dt e_-(t) |\tau \rangle}
{\langle \tau | e^{K_+}M_+M_-^{-1} e^{-K_-} |\tau \rangle},
\een
\ben
& &\int dt \psi_+ (x_+, x_-, t) \tau (t) \\
& &~~~~~~~~= \frac
{\int dt \langle \tau | e_+(t) e^{K_+}[1+ \exp (-{\rm ad}K_+)P_+]M_+M_-^{-1}
e^{-K_-} |\tau \rangle}
{\langle \tau | e^{K_+}M_+M_-^{-1} e^{-K_-} |\tau \rangle}.
\een

\noindent Thus provided we know enough about the highest weight
representations of the underlying continuum Lie algebra ${\cal G}$,
we may be able to obtain the solution of the system (\ref{eq:4})-(\ref{eq:5})
using the above relations.

\section{Concluding remarks}
Let us end this chapter by presenting some concluding remarks.

In this chapter, we studied the three-dimensional two-extended Toda model
by generalizing various techniques for treating two-dimensional models.
The models studied here are usually integro-differential equations, however,
there is one special case, say, eqs.(\ref{eq:7}), which is completely
a system of {\it differential} equations. Actually, this system is
just the $(B_2, C_2)$ flow of the so-called semiclassical or continuous
Toda hierarchy recently proposed by Takasaki and Takebe [17].
The $(B_1, C_1)$
flow of this hierarchy is just the well-known ``continuous Toda''
model, which possess the $w_\infty$ symmetry, and corresponds
to real Euclidean Einstein spaces with at least one rotational Killing
vector [5]. Therefore, it is very interesting to ask whether
eqs.(\ref{eq:7}) correspond to any similar structure. If this
is true, then it might be as well interesting to made it clear the roles
of the more general system, eqs.(\ref{eq:5}), in the gravitational theories.

Although we have given some hint of deriving formal solutions to the
system (\ref{eq:4})-(\ref{eq:5}), we have to say that such constructions are
really {\it formal} since we have not enough knowledge about the highest
weight representations of the continuum Lie algebras to determine
whether our construction can really give rise to explicit, physically
nontrivial and interesting solutions. Thus is seems necessary to
study more about the structures and representations of the
continuum Lie algebras themselves.

\chapter{Connections of \wnl algebnras and exchange algebra}

\section{Introduction}

Both integrable and conformal field theories are important subjects
of study in modern theoretical physics. In view of underlying symmetries,
integrable quantum field theories are characterized by quantum group
symmetries, whilst the conformal field theories are characterized
by $W$ algebras. Quantum groups and $W$ algebras are, however, although
not well understood, not independent objects because any conformal
invariant field theory is automatically integrable.

At the classical level, quantum groups become the well known chiral
exchange algebras (CEA), and $W$ algebras tend to their natural classical
limits. The question therefore arises: are there any intimate relations
between classical exchange algebras and $W$ algebras?
It cannot be true that there is one to one correspondence between
CEA and W algebras. However, as is shown in Ref. \cite{OB1}, all classical
$W$ algebras of the series $W_N$ can be reconstructed from the $sl(N)$
CEAs. In this chapter we are motivated to show that not only the $W_N$ series
but also all the \wnl algebras \cite{poly} can be reconstructed from CEAs.
This can be viewed as an alternative approach of $W$ algebra constructions,
since the standard hamiltonian reduction construction \cite{feher}
of $W$ algebras is based on the
Kac-Moody current algebra, whilst the CEAs are quadratic Poisson algebras
based on the Lie groups which is in some sense ``dual'' to the Kac-Moody
algebra.

\section{$W$ algebra versus CEA: the general scheme}

Let us recall the basics of \wnl algebras. By definition, \wnl algebras is
the reduction of $sl(N)$ current algebra under various $sl(2)$ embeddings.
To be exact, let us consider the vector representation of $sl(N)$. In this
representation, \wnl is nothing but the Poisson bracket algebra of the
gauge independent remnants of the current algebra under the constraints

\bd
J_{i,i+l+k}=\left\{
\ba{cc}
$$-1$$ & $$k=0$$ \cr
$$0$$  & $$k>0$$
\ea
\right. ~~,
\ed

\noindent where $J$ is the $sl(N)$-valued Kac-Moody current. After
appropriate fixing of the gauge degrees of freedom induced by the
above constraints, the matrix $J$ can be put into the form

\bd
J_{\rm fix}=
\begin{array}{c}
\\
\\
\\
N-l+1 \\
\\
\end{array}
\stackrel{
\begin{array}{cccccc}
&  &  & l+1 &  &
\end{array}
}{\left(
\begin{array}{cccccc}
0 & ... & 0 & -1 &  &  \\
& \ddots  &  & \ddots  & \ddots  &  \\
&  & 0 & ... & 0 & -1 \\
J_{N-l+1,1} &  & ... & ... &  & J_{N-l+1,N} \\
\vdots  &  &  &  &  & \vdots  \\
J_{N,1} &  & ... & ... &  & J_{N,N}
\end{array}
\right) } ,
\ed

\noindent where, of cause, ${\rm Tr} J_{\rm fix}=0$.
The matrix elements $J_{a,b}$ with $1 \leq a \leq N$ and $N-l+1
\leq b \leq N$ form a basis for \wnl algebra which we shall adopt in
this chapter.

CEA is a type of quadratic Poisson bracket algebra whose structure is
determined by the classical $r$-matrix. Such algebras are particularly
important in integrable systems such as (conformal) Toda field theories
and WZNW models. For these models the ${\cal G} \otimes {\cal G}$-valued
$r$-matrices take the standard triangular form \index{$r$-matrix}

\bd
r_\pm= \pm \left( \sum_{ij} (K^{-1})^{ij} H_i \otimes H_j
+ 2 \sum_{ \alpha > 0} E_{ \pm \alpha} \otimes E_{ \mp \alpha} \right),
\ed

\noindent where $K$ is the Cartan matrix of the Lie algebra ${\cal G}$,
$H_i$ are Cartan generators of ${\cal G}$, and $E_{ \pm \alpha}$ are
(normalized) step operators of ${\cal G}$ corresponding to the
positive/negative root $\pm \alpha$. In what follows we shall abuse the
Lie algebra generators and their representation matrices in the defining
representation of the Lie algebra. The CEA takes the form

\be
\{ \xi (x) \otimes, \; \xi (y) \} = \xi (x) \otimes \xi (y)
\left[ r_+ \theta_+ (x-y) + r_- \theta_- (y-x) \right], \label{x}
\ee

\noindent where $\xi (x)$ is a row vector of $N$ components, and

\bd
\theta(x-y)= \left\{
\ba{cc}
1&$$x>y$$\cr
$$\frac{1}{2}$$ & $$x=y$$\cr
0 & $$x<y$$
\ea
\right. .
\ed

\noindent In this chapter we shall use $l$ linearly independent
realizations of the same algebra (\ref{x}) based on $sl(N)$ to construct
\wnl algebra. That means, we assume the existence of the algebra
\index{exchange algebra}

\bea
& &\{ f_a (x) \otimes, \; f_b (y) \} = f_a (x) \otimes f_b (y)
\left[ r_+ \theta_+ (x-y) + r_- \theta_- (y-x) \right],\label{xx}\\
& &~~~~a,\;b=1,\;...,\;l \nonumber
\eea

\noindent with all $f_a (x)$ being $N$-component row vectors
\footnote{This assumption is actually realizable: suppose $f_a$
is the $a$-th row of the holomorphic patch of WZNW field $g(x,\;\bar{x})=
g(x)\bar{g}(\bar{x})$ \cite{BY}. }.

Now let us consider the following problem.
Let $f_a^i$ denote the $i$-th component of
$f_a$. Define

\be
f_{kl+a}^i=(-\p )^k f_a^i (x) \label{fkl}
\ee

\noindent for all $k \in {\bf Z}_+$. Then the determinants

\bd
R_a \equiv \left|
\ba{ccccc}
$$f_1^1$$ & $$f_1^2$$ & ... & $$f_1^N$$ & $$f_1^i$$ \cr
$$f_2^1$$ & $$f_2^2$$ & ... & $$f_2^N$$ & $$f_2^i$$ \cr
$$\vdots$$& $$\vdots$$&     & $$\vdots$$& $$\vdots$$\cr
$$f_N^1$$ & $$f_N^2$$ & ... & $$f_N^N$$ & $$f_N^i$$ \cr
$$f_{N+a}^1$$ & $$f_{N+a}^2$$ & ... & $$f_{N+a}^N$$ & $$f_{N+a}^i$$
\ea
\right|
\ed

\noindent vanish identically for $a=1,\;2,\;...,\;l$ and $i=1,\;...,\;N$.
Expanding $R_a$ according to the last column we get

\bd
\sum_{b=1}^{N+1} R_a(b,\;N+1) f_b^i = 0,
\ed

\noindent where $R_a(m,\;n)$ is the algebraic co-minor of $J_a$ with
respect to the $(m,\;n)$-th entry. In particular we denote $\Delta_N =
J_a(N+1,\;N+1)$, which is the same for all $a$. Thus the above equation
can be rewritten into the form

\be
f_{N+a}^i = - \sum_{b=1}^{N} \frac{R_a(b,\;N+1)}{\Delta_N} f_b^i.
\label{fna}
\ee

On the other hand, consider the matrix linear equation

\bd
( \p + J_{\rm fix} ) h(x) = 0,
\ed

\noindent where $h(x)$ is an $N$-component column vector, whose components
are denoted $h_i(x)$. Using the explicit form of $J_{\rm fix}$ we can easily
get

\bea
& & h_p(x) = - \p h_{p-l} (x),~~~~~~~~~~~~~~( l< p \leq N ) \label{hkl}\\
& & \p h_{N-l+a}(x) = \sum_{j=1}^{N} J_{N-l+a,\;j} h_j(x)~~~~~
(a=1,\;...,\;l). \label{hna}
\eea

\noindent This is a coupled system of linear differential equations
for the functions $h_i(x),~~i=1,\;...,\;N$. If one perform the process of
decoupling the equations for different $h_i(x)$, it would lead to
$l$ independent linear $N$-th order differential equations for $h_a(x),~~
a=1,\;...,\;l$. Thus by denoting the independent solutions of $h_a(x)$ as
$h_a^i(x)$, we can easily see that the form of equations
(\ref{hkl}-\ref{hna}) coincide
exactly with equations (\ref{fkl}) and (\ref{fna}),
provided we make the identification of variables

\bea
& &h_a^i(x) = f_a^i(x), \nonumber\\
& &J_{N-l+a,\;j} = \frac{R_a(j,\;N+1)}{\Delta_N}. \nonumber
\eea

\noindent Thus a connection between the components of CEA generators and \\
\wnl
generators is established.

Notice that, however, in order to make the above identification meaningful,
the summation

$$
\sum_{a=1}^{l} \frac{R_a(N-l+a,\;N+1)}{\Delta_N}
$$

\noindent should vanish
because this equals the trace of the matrix $J_{\rm fix}$. Remembering the
definition of $\Delta_N$, the last requirement can be equivalently stated

$$
\p {\rm ln} \Delta_N = 0.
$$

\noindent So $\Delta_N$ should be a constant. In other words, $\Delta_N$ should
at least be a center in the Poisson bracket algebra of the original CEA.
This is automatically ensured if we take the CEA generators $f_a(x)$
as the first $l$ rows of the holomorphic patch of WZNW field. So, given the
CEA (\ref{xx}), we can explicitly calculate the generating relations of \wnl.

\section{\w42 as an example}

Having sketched the general connection between CEAs and \wnl algebras, let
us now study a concrete example to check if the above scheme works correctly.

Consider the \w42 algebra, which was first explicitly calculated in
\cite{Dep}. In this case, $N=4$, $l=2$, and the CEA (\ref{xx})
can be written in the component form

\bea
& &\{ f_a^i(x),\;f_b^j(y) \} = - \frac{1}{4} f_a^i(x) f_b^j(y)
\left[ \theta(x-y) - \theta(y-x) \right] \nonumber\\
& &~~~~+ 2 f_a^j(x)f_b^i(y)
\left[ \theta(i-j)\theta(x-y) - \theta(j-i)\theta(y-x) \right]. \label{xxx}
\eea

\noindent By straightforward calculations one can show that $\Delta_4$
is indeed a center under the Poisson brackets (\ref{xxx}). Therefore, one can
choose normalizations for $f_a$ such that $\Delta_4=1$. However, in what
follows, we choose an alternative method, {\it i.e.} renormalize
the \w42 generators as follows,

\bea
& &v_2=\Delta_4 J_{3,1},~~~v_{3/2}=\Delta_4 J_{3,2},~~~
v_1=\Delta_4 J_{3,3},~~~v_{1/2}=\Delta_4 J_{3,4},\\
& &u_{5/2}=\Delta_4 J_{4,1},~~~u_{2}=\Delta_4 J_{4,2},~~~
u_{3/2}=\Delta_4 J_{4,3},~~~u_1=\Delta_4 J_{4,4},
\eea

\noindent where the suffices of $v$ and $u$'s denotes their conformal
dimensions, and $v_1 +u_1=\p \Delta_4$.

After rather tedious calculations we get the following Poisson brackets,

\begin{eqnarray*}
& &\{v_{1/2}(x),~v_{1/2}(y) \} = 0,\\
& &\{v_1(x),~v_1(y) \} = -\Delta_4(x) \Delta_4(y) \delta'(x-y),\\
& &\{v_1(x),~v_{1/2}(y) \}=\Delta_4 v_{1/2} \delta (x-y),\\
& &\{v_{3/2}(x),~v_{3/2}(y) \} = - \frac{3}{4} v_{1/2}(x) v_{1/2}(y)
\delta'(x-y),\\
& &\{v_{3/2}(x),~v_1(y) \} = - \Delta_4 v_{3/2} \delta (x-y) \\
& &~~~~+ \frac{1}{2} v_{1/2}(x) \Delta_4 (y) \delta'(x-y),\\
& &\{v_{3/2}(x),~v_{1/2}(y) \} = 0,\\
& &\{u_{3/2}(x),~u_{3/2}(y) \}=0,\\
& &\{u_{3/2}(x),~v_{3/2}(y) \} =\Delta_4 (u_2 -v_2) \delta (x-y) -
\Delta_4(x) v_1(y) \delta'(x-y) \\
& &~~~~- \Delta_4(x) \Delta_4(y) \delta''(x-y),\\
& &\{u_{3/2}(x),~v_{1}(y) \} =\Delta_4 u_{3/2} \delta (x-y),\\
& &\{u_{3/2}(x),~v_{1/2}(y) \}= \Delta_4 (u_1 -v_1) \delta (x-y) \\
& &~~~~-2\Delta_4(x) \Delta_4(y) \delta'(x-y),\\
& &\{v_2(x),~v_2(y) \}=(v_2(x) \Delta_4(y) + \Delta_4(x) v_2(y) \\
& &~~~~- \frac{3}{4}v_1(x)v_1(y))\delta'(x-y) \\
& &~~~~+ \frac{3}{4}(\Delta_4(x) v_1(y) -v_1(x) \Delta_4(y) )
\delta''(x-y) \\
& &~~~~+ \frac{3}{4}\Delta_4(x) \Delta_4(y) \delta'''(x-y),\\
& &\{v_2(x),~v_{3/2}(y) \}=(v_{3/2} v_1 - v_{1/2} v_2 ) \delta (x-y) \\
& &~~~~+ (v_{3/2}(x) \Delta_4(y) + \frac{1}{4} v_1(x) v_{1/2}(y) \\
& &~~~~- v_{1/2}(x) v_1(y) ) \delta'(x-y) \\
& &~~~~-( v_{1/2}(x) \Delta_4(y) + \frac{1}{4}
\Delta_4(x) v_{1/2}(y)) \delta''(x-y),\\
& &\{v_{2}(x),~v_1(y) \} = - \frac{1}{2} v_1(x) \Delta_4(y) \delta'(x-y)\\
& &~~~~+ \frac{1}{2}\Delta_4(x) \Delta_4(y) \delta''(x-y),\\
& &\{v_2(x),~v_{1/2} (y) \}= \Delta_4 v_{3/2} \delta (x-y)
- v_{1/2}(x) \Delta_4 (y) \delta'(x-y),\\
& &\{v_2(x),~u_{3/2} (y) \}= -\Delta_4 u_{5/2} \delta (x-y),\\
& &\{ u_2(x),~u_2(y) \} = ( u_2(x) \Delta_4(y) + \Delta_4(x) u_2(y) \\
& &~~~~- \frac{3}{4} u_1(x) u_1(y)) \delta'(x-y) \\
& &~~~~+ \frac{3}{4}
(\Delta_4(x) u_1(y) - u_1(x) \Delta_4(y)) \delta''(x-y) \\
& &~~~~+ \frac{3}{4} \Delta_4(x) \Delta_4(y) \delta'''(x-y),\\
& &\{u_2(x),~u_{3/2}(y) \} = \Delta_4 u_{5/2} \delta(x-y) \\
& &~~~~-u_{3/2}(x) \Delta_4(y) \delta'(x-y),\\
& &\{u_2(x),~v_2(y) \} = (u_{5/2} v_{1/2} - u_{3/2} v_{3/2}) \delta(x-y)\\
& &~~~~+ (\frac{1}{4} u_1(x) v_{1}(y)
- u_{3/2} (x)v_{1/2}(y)) \delta'(x-y)\\
& &~~~~+ \frac{1}{4}( u_1(x) \Delta_4(y)
- \Delta_4(x) v_1(y)) \delta''(x-y)\\
& &~~~~- \frac{1}{4} \Delta_4(x) \Delta_4(y) \delta'''(x-y),\\
& &\{u_2(x),~v_{3/2}(y) \} = (u_{2} v_{1/2} - u_{1} v_{3/2}) \delta (x-y)\\
& &~~~~+( \Delta_4(x) v_{3/2}(y) - \frac{3}{4} u_{1}(x) v_{1/2}(y))
\delta'(x-y)\\
& &~~~~+ \frac{3}{4} \Delta_4(x) v_{1/2}(y) \delta''(x-y),\\
& &\{u_2(x),~v_1(y) \} = \frac{1}{2} u_{1} (x) \Delta_4(y) \delta'(x-y)\\
& &~~~~-  \frac{1}{2} \Delta_4(x) \Delta_4(y) \delta''(x-y),\\
& &\{u_2(x),~v_{1/2}(y) \} = - \Delta_4 v_{3/2} \delta(x-y),\\
& &\{u_{5/2}(x),~u_{5/2}(y) \} = -\frac{3}{4} u_{3/2} (x) u_{3/2} (y)
\delta'(x-y),\\
& &\{u_{5/2}(x),~u_2(y) \} = (u_{2} u_{3/2} - u_1 u_{5/2}) \delta(x-y)\\
& &~~~~+ (\frac{1}{4} u_{3/2} (x) u_1(y) - u_1(x) u_{3/2}(y) \\
& &~~~~+ \Delta_4(x) v_{5/2}(y)) \delta'(x-y)\\
& &~~~~+(\Delta_4(x) u_{3/2}(y)
+ \frac{1}{4} u_{3/2}(x) \Delta_4 (y)) \delta''(x-y),\\
& &\{u_{5/2}(x),~v_2(y) \} = (u_{5/2} v_1 - u_{3/2} v_2) \delta (x-y)\\
& &~~~~+( u_{5/2}(x) \Delta_4(y) - \frac{3}{4} u_{3/2}(x) v_1(y) )
\delta'(x-y)\\
& &~~~~- \frac{3}{4} u_{3/2}(x) \Delta_4(y) \delta''(x-y),\\
& &\{u_{5/2}(x),~v_{3/2}(y) \} = (u_{2} v_1 - u_{1} v_2) \delta (x-y)\\
& &~~~~+( u_{2}(x) \Delta_4(y) + \frac{1}{4} u_{3/2}(x) v_{1/2}(y) \\
& &~~~~- u_1(x) v_1(y) + \Delta_4(x) v_2(y)) \delta'(x-y)\\
& &~~~~+ ( \Delta_4(x) v_1(y) - u_{1}(x) \Delta_4(y) )
\delta''(x-y) \\
& &~~~~+ \frac{1}{4} \Delta_4(x) \Delta_4(y) \delta'''(x-y),\\
& &\{u_{5/2}(x),~v_1(y) \} = u_{5/2} \Delta_4 \delta (x-y)\\
& &~~~~- \frac{1}{2} u_{3/2}(x) \Delta_4(y) \delta'(x-y),\\
& &\{u_{5/2}(x),~v_{1/2}(y) \} = (u_{2}-v_2) \Delta_4 \delta (x-y)\\
& &~~~~- u_1(x) \Delta_4(y) \delta'(x-y)
+ \Delta_4(x) \Delta_4(y) \delta''(x-y),\\
& &\{u_{5/2}(x),~u_{3/2}(y) \} = 0.
\end{eqnarray*}

We see that this is exactly the \w42 algebra presented in \cite{Dep} if we
set $\Delta_4 =1$. The Virasoro element of \w42 in this presentation is not
explicit. However, there is a general principle to find this element
out of the Drinfeld-Sokolov gauge
\index{Drinfeld-Sokolov ! gauge}, {\it i.e.} bye taking the trace of the
square of the matrix $J_{\rm fix}$,

\bea
& &T={\rm Tr} \left( J_{\rm fix}^2 \right)\nonumber\\
& &~~~\stackrel{{\rm for}~W_4^{(2)}}{=} \frac{1}{\Delta_4^2}\left(
v_1^2 + u_1^2 + 2 u_{3/2} v_{1/2} \right) -
\frac{2}{\Delta_4} \left( v_2 + u_2 \right). \nonumber
\eea

\section{Summary}

In this chapter we presented a general connection between classical \wnl
algebras and CEAs. This connection reveals a remarkable hidden relation
between the CEAs and $W$ algebras. Especially, when one finds a set of
CEA in some integrable models with conformal symmetry, he might at once
conclude the type of the corresponding $W$ algebra according to the
number of independent CEA generators.

It is interesting to mention that, for each $W$ algebra in the series \wnl,
there corresponds at least one integrable hierarchy which takes this algebra
as the Poisson structure in its second Hamiltonian formalism. For \w42
algebra, one of such hierarchies is constructed in \cite{QPL} recently.
However, it is quite possible that to each \wnl type algebra there associates
several different integrable hierarchies. For example, equation (\ref{fna})
can be viewed as a multicomponent generalization of the well-known
Gelfand-Dickey \cite{Dic} equation. Associated with this equation,
one can construct
a multicomponent KP-KdV type hierarchy \cite{CL}. This is especially
easy when $N$ is an integral multiple of $l$, in which case the
corresponding hierarchy becomes simple matrix extension of the ususl
KP-KdV hierarchies \cite{QC}. Such hierarchies are quite different from
the one obtained in \cite{QPL} when the underlying $W$ algebra is \w42.

The conection between \wnl algebras and CEAs provides with us an easy way
of constructing lattice versions of $W$ algebras, the latter has been
considered as important object in the theories of integrable lattice
models \cite{LD}.
The point is that CEA can be easily put into the lattice form, and
the construction performed in this chapter can be carried out without
difficult in the lattice case. Actually going to the lattice is precisely
the traditional method for the quantization of CEAs. We believe that
provided the quantum determinants are appropriately defined, the
quantized version of \wnl algebras will play an important role in the
theories of quantum groups and/or noncommutative geometries. This may
also be the right way to understand the connections between quantum groups
and quantum $W$ algebras \cite{PU}.

\newpage

%% w42 ref %%

\addcontentsline{toc}{chapter}{Index}
\printindex

\end{document}